\documentclass[aps,pra,eqsecnum,twocolumn,amsmath,amssymb,10pt]{revtex4-2}

\usepackage{graphicx}
\usepackage{bm}
\usepackage[usenames,dvipsnames]{color}
\definecolor{altncolor}{rgb}{0,0,0.8}
\usepackage[colorlinks=true, linkcolor=green, anchorcolor=altncolor,
citecolor=altncolor, filecolor=altncolor, menucolor=altncolor,
urlcolor=altncolor]{hyperref}

\begin{document}

\title{Doppler sensitivity and resonant tuning of Rydberg atom-based antennas}

\author{Peter B. Weichman}

\affiliation{FAST Labs$^{TM}$, BAE Systems, 600 District Avenue, Burlington, MA 01803}

\date{\today}

\begin{abstract}

Radio frequency antennas based on Rydberg atom vapor cells can in principle reach sensitivities beyond those of any conventional wire antenna, especially at lower frequencies where very long wires are needed to accommodate the growing wavelength. They also have other desirable features such as nonmetallic, hence lower profile, elements. This paper presents a detailed theoretical investigation of Rydberg antenna sensitivity, elucidating parameter regimes that could cumulatively lead to 2--3 orders of magnitude sensitivity increase beyond that of currently tested configurations. The key insight is to optimally \emph{combine} the advantages of two well-studied approaches: (i) three laser ``2D star configuration'' setups that, enhanced as well with increased laser power, to some degree compensate for atom motion-induced Doppler broadening, and (ii) resonant coupling between a pair of near-degenerate Rydberg levels, tuned via a local oscillator to the incident signal of interest. The advantage of the star setup is subtle because it only restores overall sensitivity to the \emph{expected} Doppler-limited value, compensating for additional significant off-resonance reductions where differently moving atom sub-populations actually destructively interfere with each other in the net signal. The additional unique advantage of the local oscillator tuning is that it leads to vastly narrower line widths, as low as $\sim$\,10 kHz set by the intrinsic Rydberg state lifetimes, rather than the typical $\sim$\,10 MHz scale set by the core state lifetimes. Intuitively, with this setup the two Rydberg states may be tuned to act as an independent high-q cavity, a point of view supported through a study of the frequency-dependence of the antenna resonant response. There are a number of practical experimental advances, especially larger $\sim$\,1 cm laser beam widths, required to suppress various extrinsic line broadening effects and to fully exploit this cavity response.

\end{abstract}

\maketitle

\section{Introduction}
\label{sec:intro}

Rydberg atoms are formed by exciting a (typically alkali atom, Rb or Cs) ground state electron into a very high orbit. The resulting hydrogen-like atomic state can have (depending on the exact excited state) an electric dipole moment scaling with principal quantum number $n$ as $p_\mathrm{Ry} \sim e a_0 n^2$ where $a_0 = 0.529$ \AA\ is the Bohr radius. For $n = O(10^2)$ this implies a potential $O(10^4)$ increase in the coupling strength of the atom with an incident external electric field ${\bf E}_\mathrm{in}(t)$. Designing practical devices based on this concept requires a number of advances that have been steadily accumulating over the past 30+ years \cite{HFI1990,FIM2005,RBTEY2011,CTSSAW2012,Sedlacek2012,NIST2014,SK2018,Michigan2019,Waterloo2021,MITRE2021}---these will be discussed in more detail in Sec.\ \ref{sec:exptsetup}. Detection sensitivity for the most advanced setups is actually based on two nearby Rydberg states, with energy difference close to the resonance condition $\Delta E_\mathrm{Ry} = \hbar \omega_\mathrm{in}$ where $\omega_\mathrm{in}$ is the incident field center frequency \cite{NIST2019a,NIST2019b}, constrained also by the photon unit angular momentum selection rule. The ``frequency tuner'' part of the setup is hence enabled by choosing different state pairs to obtain the near-resonance. There are many such pairs available, enabling unprecedented coverage from near-DC to THz regimes \cite{Durham2017,JC2020,ARL2020}. Further tuning can be obtained through Stark shifts associated with locally applied DC electric fields \cite{BVBZ2022}.

\begin{figure}

\includegraphics[width=3.0in,viewport = 0 0 410 390,clip]{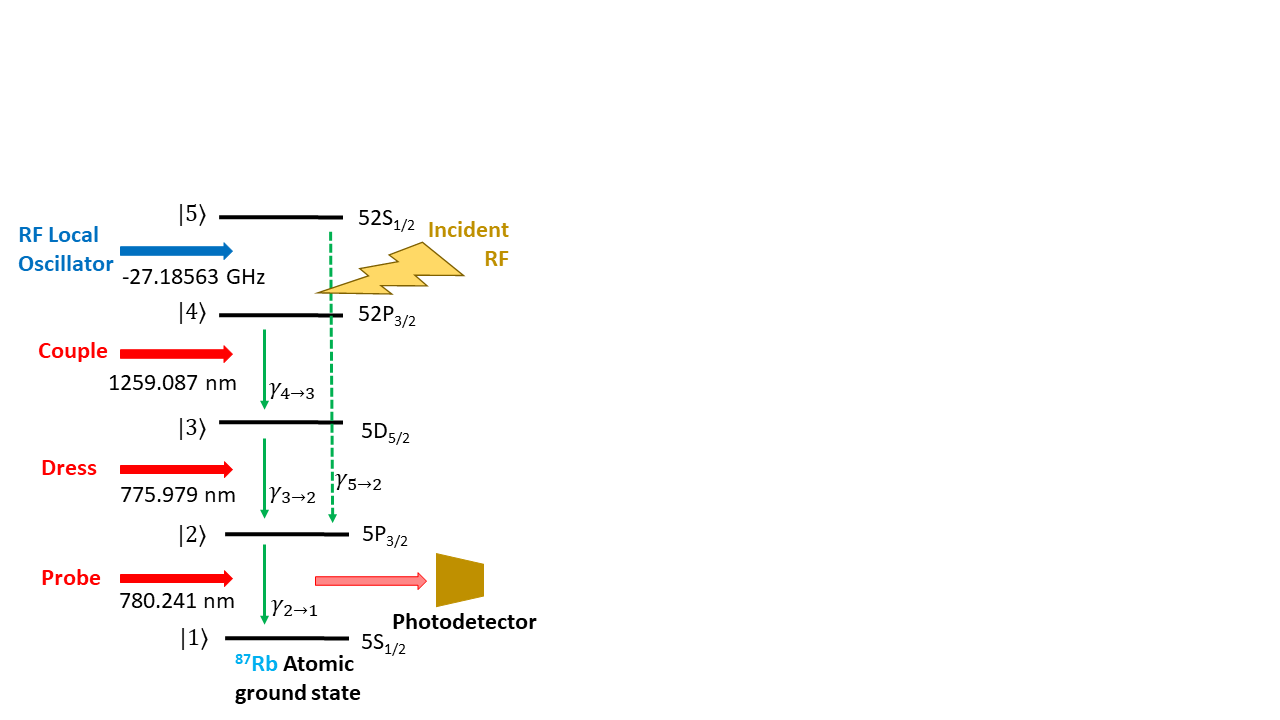}

\caption{Example three-laser, five-state $^{87}$Rb Rydberg atom setup. In combination, Probe, Dress, and Coupling beams promote the atom from its ground state $|1\rangle$ to a selected $n=52$ Rydberg state $|4\rangle$. An additional vapor cell RF local oscillator \cite{NIST2019a,NIST2019b} couples the latter to a second Rydberg state $|5\rangle$ (which happens to be lower in energy in this case). The incident field, with small frequency difference $|\omega_\mathrm{in} - \omega_\mathrm{LO}| \ll \omega_\mathrm{LO}$, is superposed upon the latter, and through resonant tuning influences the Probe beam transmission ${\cal P}_\mathrm{EIT}$. The dominant single photon decay paths are also shown. The values for this example are $(\gamma_{2 \to 1}, \gamma_{3 \to 2}, \gamma_{4 \to 3}, \gamma_{5 \to 2})/2\pi = (6606.50, 646.18, 1.64, 1.09)$ kHz.}

\label{fig:rysetup}
\end{figure}

Sensors capable of operation at ambient temperature are of huge benefit as well. For Rydberg sensors there are actually several competing effects at work. Alkali atom vapor densities, in equilibrium with the solid, increase exponentially with temperature, benefiting sensitivity to a point through increasing the number of participating atoms. However, the mean interatomic distance must be kept significantly larger than the $O(n^2 a)$ Rydberg atom diameter to avoid strong collision-broadening effects. Increasing atomic thermal velocity $v_\mathrm{th} = O(10^2)$ m/s also leads to increased Doppler broadening---intuitively, only a sub-population of sufficiently slow-moving atoms, $v \ll v_\mathrm{th}$, remain on-resonance with the corresponding laser frequencies and contribute to the sensor output signal. For some applications one may approximately compensate for atom motion through ``recoil free'' laser setups enforcing vanishing of the vector sum of their wavevectors, equation (\ref{3.20}) and Fig.\ \ref{fig:starconfig} below \cite{RBTEY2011,SKAW2016}. However for the Rydberg sensor, as seen in Fig.\ \ref{fig:rysetup}, each laser is tuned to a different transition which is unavoidably detuned by atom motion along the beam direction. Although not at all obvious, there turns out still to be a significant advantage to enforcing the recoil free condition. It will be seen that although the overall excitation level of the vapor is not strongly affected, the contribution of differently moving sub-populations to the net sensor signal can be both positive and negative, and increasing violation of the recoil free condition turns out to lead, among other effects, to increasing signal cancellation. Given satisfaction of this condition, it will be seen that there are benefits as well to increasing one or more laser powers, not only through increased probe beam photon count rates, but also through expansion of the effective slow moving population. The latter scales as $\Omega/\Delta_\mathrm{Dopp}$ where $\Omega$ is the laser Rabi frequency and $\Delta_\mathrm{las}^\mathrm{Dopp}/2\pi = v_\mathrm{th}/\lambda = O(10^2)$ MHz is the thermal Doppler shift associated with a given laser wavelength $\lambda \sim 1\ \mu$m.

The earlier approaches were enabled by optical mixing of atomic core levels with a single Rydberg target state, with sensor operation based on probe laser transmission---electromagnetically induced transparency (EIT) effect---as a function of probe or coupling laser detuning (extensively reviewed in Ref.\ \cite{ARL2020}). The more recent two Rydberg state approach imposes additional experimental complexity, but also introduces significant new degrees of signal tuning capability, greatly increasing potential sensitivity. In addition to directly identifying (or actively creating) a target Rydberg state with a partner that is near-resonant with the incident RF signal, a vapor cell RF local oscillator (LO) can be applied to further match that of a known or expected signal frequency (to within 100 kHz, say). Physically, one might view the Rydberg pair as a kind of resonant cavity, with quality, hence inverse line width, controlled by the $\sim$\,1 ms state lifetimes, vastly longer than the $< 1\ \mu$s core excited state lifetimes (see the example in Fig.\ \ref{fig:rysetup}). The latter limits earlier approaches based on the $> 1$ MHz Autler--Townes splitting peak width. It will be shown that this ``Rydberg cavity'' effect can indeed be exploited by using very small local oscillator amplitudes, with Rabi frequency $\Omega_\mathrm{LO}/2\pi \sim 10$ kHz comparable to the dissipation/dephasing linewidth. It is important here as well that the RF Doppler shift $\Delta^\mathrm{Dopp}_\mathrm{RF}/2\pi = v_\mathrm{th}/\lambda_\mathrm{RF} \alt 10$ kHz is typically smaller than this linewidth.

For resonance-based sensors sensitivity typically falls off sharply with frequency difference. Thus, maximum sensitivity occurs in what we will call the ``adiabatic limit,'' in which the device is perfectly tuned to the expected signal frequency. However, signals of interest will typically be imperfectly predictable and/or have finite bandwidth, and it is therefore critical to characterize the sensor under such more realistic conditions. Most of our results will be aimed at evaluating optimal performance in the adiabatic limit, but at the end we will also quantify the sensor resonant bandwidth using a full dynamic linear response theory. This analysis verifies that the sensitivity bandwidth may indeed may be interpreted in terms of the quality of the Rydberg cavity.

The key conclusion of this work is that a combination of Doppler compensation, laser amplitude enhancement, and LO control of the Rydberg cavity can in principle enable 2--3 orders of magnitude increase in RF signal electric field sensitivity. Not surprisingly, there are a number of practical experimental advances necessary to taking full advantage of this. For the 2D star setup these especially include broader laser beams (say width $w \sim 1$ cm, compared to the typical $w < 1$ mm) to increase the interaction volume (multiple laser beam intersection region) and reduce transit time broadening. Associated with the latter is a linewidth $f_\mathrm{tr} = v_\mathrm{th}/w \sim 100$ kHz for $w = 1$ mm, completely dominating the intrinsic $\sim$\,10 kHz Rydberg cavity linewidth and hugely degrading sensitivity. For $w \sim 1$ cm the two are comparable, and the advertised sensitivity enhancement is mostly restored. As a final point, for the high sensitivity setup the sensor application is restricted to the weak field regime, $|\Omega_\mathrm{in}| < |\Omega_\mathrm{LO}|$, beyond which the response becomes highly nonlinear. For stronger signals one would need to switch back to a more conventional setup.

\subsection{Quantum vs.\ classical sensitivity limits}
\label{sec:qvsclasssense}

The intense interest in Rydberg sensors is driven by their theoretical potential to exceed the sensitivity of any conventional classical antenna. It is not actually straightforward to compare the two since their operational characteristics and noise limitations are so different---atomic transition dynamics vs.\ induced currents in a conductor. For example, conventional antennas are typically limited by thermal noise while Rydberg sensors are limited by photon count statistics, hence shot noise. The latter is quantified by a combination of laser illumination power, the number of participating atoms, and the efficiency of the light--RF signal--atom coupling scheme. Exploring the limits on this coupling efficiency is the focus of this paper, elucidating parameter regimes that could cumulatively lead to the cited 2--3 orders of magnitude sensitivity increase over current experimental setups. If experimentally confirmed, this should indeed produce the desired record breaking sensitivities.

To motivate this conclusion, consider first the minimum detectable electric field for a conventional antenna, which is governed by the signal-to-noise ratio
\begin{equation}
\mathrm{SNR} = \frac{P_R}{P_N},
\label{1.1}
\end{equation}
in which $P_R$ is the received power and $P_N$ the noise power. The numerator
\begin{equation}
P_R = \frac{|{\bf E}|^2}{2 Z_0} A_\mathrm{eff}
\label{1.2}
\end{equation}
is the power incident on the receiver determined by the Poynting vector (here $Z_0 = \sqrt{\mu_0/\epsilon_0} = 377\ \Omega$ is the vacuum impedance). The effective receiver area may be expressed in the form \cite{Balanis2016}
\begin{equation}
A_\mathrm{eff} \approx \frac{G_R L_R^2}{4\pi} \left\{\begin{array}{ll}
1, & \lambda < L_R \\
(L_R/\lambda)^2, & \lambda > L_R
\end{array} \right.
\label{1.3}
\end{equation}
in which $\lambda = c/f$ is the wavelength, $L_R$ is the physical antenna diameter, and $G_R$ is a geometry-dependent gain factor (normalized, e.g., so that $G_R = 1$ for a 1D line antenna). Thus, at high frequencies the power is limited by the physical collection area, while at low frequencies the efficiency drops rapidly due to the reduction in induced voltage $V_R \sim E L_R (L_R/\lambda)$ across the antenna ends. It is for this reason that the most efficient individual antenna elements are designed to have $L_R \approx \lambda/2$ which becomes an increasingly challenging problem as the frequency drops (e.g., HF band and below, for which $\lambda > 10$ m). As detailed below, Rydberg sensors are not limited by this condition, pointing to low frequency applications as perhaps the most promising.

The noise power may be expressed in the form
\begin{equation}
P_N = \frac{F_N}{\tau}
\label{1.4}
\end{equation}
in which $F_N$ is the noise spectral density (with units of power per unit frequency, hence energy) in the neighborhood of the signal frequency and $\tau$ is the averaging/integration time. One typically expresses
\begin{equation}
F_N = \phi_N k_B T_0
\label{1.5}
\end{equation}
in which $T_0$ is the physical receiver temperature and $\phi_N$ is a dimensionless noise figure. In many cases $\phi_N > 1$ is a fundamental constraint, and one often sees $\phi_N = 3$ quoted as a representative value characterizing the internal noise of high quality devices. For the case of external (especially nonthermal) noise sources $\phi_N$ may be many orders of magnitude larger. As will be seen, rising temperature does degrade Rydberg sensitivity, but in a very different way through atom motion-induced Doppler shifts of the laser frequencies.

We define the E-field sensitivity by ${\cal E}_R = E_R \sqrt{\tau}$ with $E_R$ determined by setting $\mathrm{SNR} = 1$. This produces
\begin{eqnarray}
{\cal E}_R &=& \frac{1}{L_\mathrm{eff}} \sqrt{\frac{8 \pi Z_0 k_B T_0 \phi_N}{G_R}}
\nonumber \\
&\approx& \frac{1}{L_\mathrm{eff}[\mathrm{cm}]} \mathrm{\frac{\mu V}{m\sqrt{Hz}}},
\label{1.6}
\end{eqnarray}
in which the numerical value exhibited in the second line is obtained by choosing $G_R = 1$, $\phi_N = 3$, and $T_0 = 300$ K. The notation indicates that the antenna size is measured in cm, and from (\ref{1.3}), we define the effective length parameter
\begin{equation}
L_\mathrm{eff} = \sqrt{\frac{4\pi A_\mathrm{eff}}{G_R}} = L_R \min\{1, L_R/\lambda\}.
\label{1.7}
\end{equation}
Note that ${\cal E}_R$ is distinct from a ``minimum useable field,'' often taken as $10 {\cal E}_R$ (20 dB SNR), which is large enough to greatly suppress statistical measurement errors. As a practical comparison, GPS signal powers (with $f = 1.23$ GHz, $\lambda = 25$ cm) lie in the $10^{-16}$ W range (e.g., 100 W effective isotropic source at $d = 20,000$ km impinging on a 50 cm$^2$ effective area), and employ averaging times in the 10 ms range (filtered by the GPS code sequence). The electric field sensitivity of a standard $L_R \simeq 10$ cm (hence $L_\mathrm{eff} \simeq 4$ cm) cell phone antenna must therefore satisfy, via (\ref{1.2}),
\begin{equation}
{\cal E}_R < (25\ \mu\mathrm{V/m}) \times (0.1\ \mathrm{s}^{1/2})
= 2.5\ \frac{\mu\mathrm{V}}{\mathrm{m\sqrt{\mathrm{Hz}}}},
\label{1.8}
\end{equation}
consistent by an order of magnitude with the lower bound (\ref{1.6}).

Rydberg sensor sensitivity varies tremendously with the type of measurement performed. Perhaps the simplest conceptually is direct measurement of the Autler--Townes frequency splitting (AC Stark effect) of the vapor absorption peak \cite{Michigan2019,Waterloo2021}, linear in the RF electric field amplitude. This has the advantage of producing a calibrated result, with proportionality constant depending only on universal atomic parameters (e.g., Rydberg level transition dipole moment). However, resolution of the splitting is limited by the peak width, and the minimum quoted E-field detection is in the $\sim$ 30 mV/m range \cite{Waterloo2021}, four orders of magnitude short of (\ref{1.8}).

The focus of the present work is on the much higher resolution achievable if one relaxes the degree of universality of the calibration feature---focusing instead on strong atomic responses that generally depend on the full atomic vapor dissipative dynamics. Along these lines, the modeling results in \cite{ARL2020} (to be improved upon in Sec.\ \ref{sec:rysnr}) predict ${\cal E}_R \sim 1\ \mu$V/m/$\sqrt{\mathrm{Hz}}$ for Rydberg antenna setups at 15 GHz---comparable to (\ref{1.6}), though with tremendous, highly frequency-dependent, variability around this. In reality it is very difficult achieve these theoretical limits. This same reference quotes a state-of-the-art Rydberg antenna sensitivity of 100 $\mu$V/m/$\sqrt{\mathrm{Hz}}$, and Ref.\ \cite{NIST2019b} quotes a value 45 $\mu$V/m/$\sqrt{\mathrm{Hz}}$. However, more recent work has achieved significant improvements: through very careful setup optimization, Ref.\ \cite{Shanxi2020} quotes a record 5.5 $\mu$V/m/$\sqrt{\mathrm{Hz}}$, and more recent work has improved this to 3 $\mu$V/m/$\sqrt{\mathrm{Hz}}$ \cite{foot:cqcom}. It follows that demonstrated Rydberg antenna sensitivities may be on the verge of meeting standard GPS signal detection requirements.

On the other hand, minimum E-field sensitivity may be contrasted with minimum E-field \emph{calibration} standards. The latter are designed to measure low field levels without distorting the source field (which would otherwise perturb the measured field itself). The EM boundary conditions on metal antennas are strongly distorting and one must work very hard to mitigate this. The result is that the minimum calibration field is far larger than the thermal limit. For example, Refs.\ \cite{Sedlacek2012,NIST2014,Waterloo2021}, quote values in the 30--100 mV/m range, but do note that 3 mV/m can be achieved using optical measurements of the antenna output (as opposed, e.g., to direct rectified current measurements through an antenna output load).

In contrast, the Rydberg sensor is non-metallic, and hence automatically much lower profile. The minimum E-field values quoted above may therefore also be used directly as a calibration standard---aided by the fact that its measurement characteristics are known in terms of fundamental atomic spectroscopy properties \cite{ARL2020})---and hence improve on the best conventional sensor by an order of magnitude or more. Although not yet officially adopted by NIST, and representing a rather limited application, it is likely that the Rydberg sensor does represent the current best E-field calibration standard.

\subsection{Atom motion-induced Doppler and other broadening effects}
\label{sec:atomdoppeff}

Practical sensor applications benefit enormously from room temperature operation. However, as indicated above, atomic motion-induced Doppler shifts can greatly degrade sensitivity by moving a transition off the laser tuned resonance. Only sub-populations of atoms that happen to be moving at the correct velocity will optimally contribute to the desired signal. Nevertheless, a number of studies have been aimed at reducing such effects to the degree possible for various applications.

\begin{figure}

\includegraphics[width=2.0in,viewport = 20 20 360 365,clip]{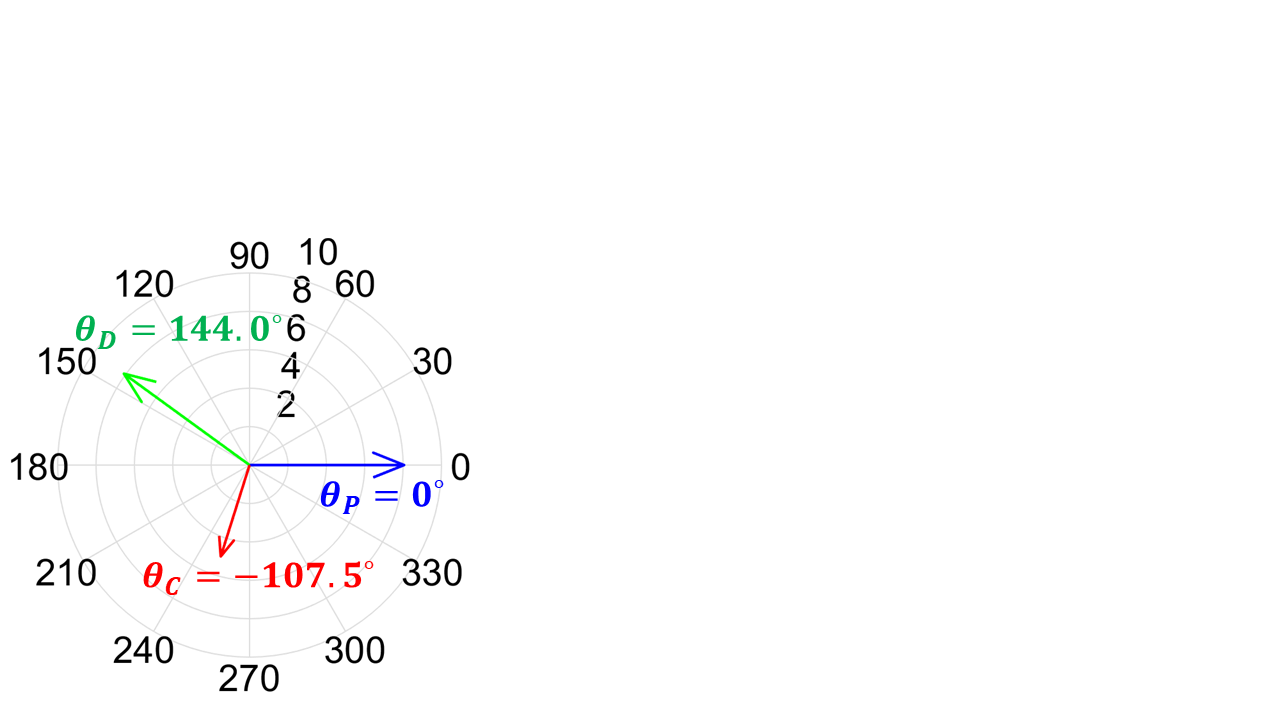}

\caption{Probe (blue), Dressing (green), and Coupling (red) laser orientation ``2D star'' configuration enforcing the Doppler condition (\ref{3.20}) in the form ${\bf k}_P + {\bf k}_D + {\bf k}_C = 0$ for the $^{87}$Rb setup described in Fig.\ \ref{fig:rysetup}. In support of the five-state model reduction, all lasers and RF fields are chosen also to have vertical linear polarization relative to this 2D plane. This way the Rabi frequencies involve only the $z$-component of the transition dipole moments [see (\ref{2.2}) and (\ref{2.3})], and transitions between states with different vertical angular momentum values $m_F$ are suppressed.}

\label{fig:starconfig}
\end{figure}

For example, the 2D 3-laser layout pictured in Fig.\ \ref{fig:starconfig}, with relative angles tuned to minimize the total atom recoil momentum, has been proposed to generate spatially uniform spin waves for light storage applications (though not necessarily focused on Rydberg atoms) \cite{SKAW2016}, and as a route to enhanced Rydberg atom qubit fidelity (Ref.\ \cite{RBTEY2011} and references therein). Even with the simpler 1D setup, Doppler degradation of the absorption signal linewidth and amplitude for a single Rydberg state can be reduced by tuning the ratio of the dressing and coupling relative laser intensities \cite{CTSSAW2012}. However, both of these applications involve only the four state version of Fig.\ \ref{fig:rysetup}, hence do not consider the enhanced RF signal detection enabled by resonant coupling between two Rydberg states that is the focus of this paper.

A very specific 1D RF sensing setup was considered in Ref.\ \cite{Waterloo2021} that very nearly satisfies the zero atom recoil condition through careful selection of the atomic states. Only the Autler--Townes splitting measurement was considered (yielding 28.2 mV/m resolution), and this setup of course greatly reduces sensor flexibility since it limits the choice of RF frequencies.

This paper will also highlight the 2D ``star'' setup, but now focused on the RF sensing problem. The laser orientation angles may be thought of as being adjusted according to the Rydberg state $|4 \rangle$ selected by the coupling laser frequency, in turn chosen so that there exists a $|4\rangle$--$|5\rangle$ transition near-resonant with the incident RF signal. The RF sensing problem involves coherent transitions between the two different Rydberg states, coherently coupled to the three core states as well. The general signal optimization problem, not limited by the Autler--Townes splitting measurement, will be seen to lie in a very different physical regime from the previously cited applications, and as detailed below introduces a variety of new exploitable degrees of freedom.

As alluded to above, an additional experimental challenges for 2D setups is the wider laser beams required to (i) properly illuminate a reasonable vapor volume within the intersecting region, and (ii) increase the dwell time of the moving atom within this region (reduce transit time broadening). Thus, probe beam transit lengths in the 1 cm range through the excited volume of the vapor are required to obtain sufficient transmitted amplitude (exponential decay) variability with tuning parameters. For 1D setups the required beam overlap volumes can be accomplished with the commonly used $\sim$\,1 mm wide beams. For 2D setups this is no longer the case and additional work will be required to fully exploit its predicted advantages. Similarly, $\sim$\,1 cm interaction region diameters are needed to avoid the effect of transit time broadening overwhelming the intrinsic Rydberg cavity linewidth---the atom actually needs to reside in the interaction volume long enough to detect and exploit the existence of a high-q cavity.

\subsection{Outline}
\label{sec:outline}

The remainder of this paper is outlined as follows. Experimental designs of interest (pictured in Fig.\ \ref{fig:rysetup}) are described in Sec.\ \ref{sec:exptsetup}. There are substantial sensing advantages to using laser cooled and trapped atoms, but the focus here will be on the much simpler, hence more broadly applicable setup exploiting an equilibrium room temperature vapor cell despite performance degradation due to thermal motion. Theoretical models, based on the Lindblad formalism, are described in Sec.\ \ref{sec:lindblad}. The assumptions underlying the widely used reduction to the five-state model illustrated in Fig.\ \ref{fig:rysetup} are discussed, especially the proper inclusion of the various previously described linewidth broadening effects. Results leading to 2--3 orders of magnitude increase in signal sensitivity are described in Sec.\ \ref{sec:senseresults}. The finite frequency response and associated sensor bandwidth is discussed in Sec.\ \ref{sec:Ryfinitefresp}, with details of the linear response mathematical formalism relegated to App.\ \ref{app:dynresp}. We note that it has recently been proposed that significant further sensitivity enhancements, by another order of magnitude or more, are possible using resonant or confining microwave structures to amplify the incident field within the vapor cell \cite{BVBZ2022}. The impact of these improvements on (shot noise-limited) signal SNR are described in Sec.\ \ref{sec:rysnr} and compared to the previously discussed classical antenna results. The paper is concluded in Sec.\ \ref{sec:conclude}.

\section{Experimental setup}
\label{sec:exptsetup}

The Rydberg state population is maintained by a finely tuned laser setup, exciting the atom via a sequence of transitions between a selected set of states $\{|n\rangle \}$ with energies $\{\varepsilon_n \}$. A particular transition $|m\rangle \to |n\rangle$ is efficiently excited only when one of the laser frequencies $\omega_\alpha$ lies sufficiently close to the energy difference $(\varepsilon_n - \varepsilon_m)/\hbar$, allowing one to ignore, to a good approximation, all unconnected states. Given the energy degeneracies produced by conservation of total atomic angular momentum (and indexed by quantum number $m_F$), identification of the set $\{|n\rangle \}$ may be subtle. This will be discussed further in Sec.\ \ref{sec:lindblad} where the theoretical model is developed.

The focus here is the 3-laser setup shown in Fig.\ \ref{fig:rysetup}, significantly extending a recent 2-laser setup analysis \cite{ARL2021}. As detailed below, the laser layout geometry, a 2D example of which is pictured in Fig.\ \ref{fig:starconfig}, enables some degree of Doppler control \cite{foot:3laser}, hence improved atomic resonance enhancement.

\subsection{EIT signal}
\label{sec:eitsig}

The combination of laser, local oscillator (LO) \cite{NIST2019a,NIST2019b} (in the form, e.g, of a carefully designed dielectric cavity surrounding the atomic vapor cell \cite{BVBZ2022}), and incident field illumination creates a coherent superposition of core and Rydberg states. The resonant tuning of this state is the basis for the sensing capability through the electromagnetically induced transparency (EIT) effect. Thus, the perturbation of the Rydberg states by the incident field induces a variation in the core state amplitudes, and this affects the absorption characteristics of the Probe beam. Specifically, the measured signal is the photon count corresponding to the transmitted fraction
\begin{equation}
{\cal P}_\mathrm{EIT}(t) = \frac{P_\mathrm{ph}(t)}{P_0}
\label{2.1}
\end{equation}
where $P_0$ is the laser power entering the vapor cell and $P_\mathrm{ph}$ the power at the photodetector \cite{HFI1990,FIM2005,SK2018,Michigan2019,MITRE2021}. For small incident field the variation of ${\cal P}_\mathrm{EIT}$ will be correspondingly small, and the fundamental noise floor will be determined by the Poisson statistics (shot noise) of the photon count perturbation (see Sec.\ \ref{sec:rysnr}). Thermal noise effects internal to the photodetector are present, but are negligible in comparison. To maximize the perturbation the setup, especially cell length, is designed to produce moderate overall absorption, ${\cal P}_\mathrm{EIT} \sim 0.5$.

\begin{table}

\begin{tabular}{ll}
Vapor cell dimensions                  & (1 cm)$^3$ \\
Vapor number density $N_A$             & $10^{10}$ atoms/cm$^3$ \\
Laser beam width $w$                   & 1 cm \\
Probe transition hyperfine levels      & $F_1 = 2$, $F_2 = 3$ \\
Probe transition dipole moment $|{\bf d}_{12}|$ & $1.7314 \times 10^{-29}$ C-m \\
Temperature $T$                        & 300 K \\
Thermal velocity $v_\mathrm{th} = \sqrt{k_BT/m_A}$  & 169.3 m/s
\end{tabular}

\caption{Example $^{87}$Rb experimental and atomic parameters. As explained in the text, the ${\bf d}_{12}$ value is a commonly used effective value derived from an average over the rotationally degenerate quantum numbers $m_{F1}, m_{F2}$ consistent with the Probe beam polarization. The quoted beam width value is substantially larger than the current $\sim$\,1 mm typical for 1D setups, but is needed to fully illuminate the desired $\sim$\,1 cm$^3$ cell volume within the three intersecting laser beams (Fig.\ \ref{fig:starconfig}) and reduce transit time broadening. Longer cell lengths will be considered as well to explore optimal, though less practical, sensitivity limits (Sec.\ \ref{sec:inhomogprobe}).}

\label{tab:Ryparms}
\end{table}

\subsection{Electric field coupling and Rabi frequencies}
\label{sec:efieldrabi}

The coupling of the illumination fields to the atom is quantified by the Rabi frequencies
\begin{equation}
\Omega_{mn} = \frac{1}{\hbar} {\bf E}_{mn} \cdot {\bf d}_{mn}
\label{2.2}
\end{equation}
with transition dipole moment \cite{FIM2005}
\begin{equation}
{\bf d}_{mn} = \langle m| e{\bf R} |n \rangle,
\label{2.3}
\end{equation}
where ${\bf R} = \sum_i ({\bf r}_i - {\bf r}_N)$ is the electron charge displacement operator relative to the nuclear coordinate ${\bf r}_N$. Here, ${\bf E}_{mn}$ is the amplitude of the time-dependent field ${\bf E}_{mn} e^{-i\omega_\alpha t}$ corresponding to the illumination frequency $\omega_\alpha$ tuned close to the transition between states $|m\rangle$ and $|n\rangle$: $\hbar \omega_\alpha \simeq \varepsilon_n - \varepsilon_m$. Whenever convenient, the subscripts $(P, D, C, \mathrm{LO})$ will be used interchangeably with $(12, 23, 34, 45)$. Some example experimental parameter values are listed in Table \ref{tab:Ryparms}.

For the special case of the transition between Rydberg states there are two contributions
\begin{equation}
\Omega_\mathrm{Ry}(t)
= \frac{1}{\hbar} {\bf E}_\mathrm{Ry}(t) \cdot {\bf d}_{45}
= \Omega_\mathrm{LO} + \Omega_\mathrm{in}(t)
\label{2.4}
\end{equation}
in which the total field ${\bf E}_\mathrm{Ry}(t) = {\bf E}_\mathrm{LO} + {\bf E}_\mathrm{in}(t)$ by design includes both the LO and spectrally ``nearby'' incident field. The LO multiplier $e^{-i\omega_\mathrm{LO} t}$ has been factored out, and $\Omega_\mathrm{in}(t)$ may then be thought of as a narrow-band communication signal centered on the (small) frequency difference $\Delta \omega_\mathrm{in} = \omega_\mathrm{in} - \omega_\mathrm{LO}$ \cite{NIST2019c,NIST2022,BBNTS2022}.

For the purposes of the present discussion it may be assumed that $\omega_\mathrm{in}$ is to some degree known, and leads to a choice of Rydberg states for which the energy level difference $|\varepsilon_5 - \varepsilon_4|$ indeed lies sufficiently close to $\hbar \omega_\mathrm{in}$ (e.g., frequency within a few MHz). The Coupling laser is tuned to populate the state $|4 \rangle$, while the LO introduces a controlled coupling to state $|5 \rangle$. As will be seen, tuning both the frequency and the amplitude of the LO can be used to enormously enhance the sensitivity of ${\cal P}_\mathrm{EIT}$ to $\Omega_\mathrm{in}$---the Rydberg cavity effect alluded to in the Introduction.

If $\omega_\mathrm{in}$ is not accurately known, one would need to tune the sensor across a sufficiently large bandwidth to locate the signal. This is in principle no different from the action of a classical frequency tuner, but the process in this case is clearly more complicated as one must hop between different selected pairs of Rydberg levels, enabled through the combination of tunable Coupling laser and LO.

\subsection{Signal linear response}
\label{sec:siglinresp}

The signal sensitivity may be characterized by the frequency domain linear response $S_\mathrm{EIT}$ defined by
\begin{eqnarray}
{\cal P}_\mathrm{EIT}[\Omega_\mathrm{Ry}(t)]
&=& {\cal P}_\mathrm{EIT}(\Omega_\mathrm{LO})
+ S_\mathrm{EIT}(\Omega_\mathrm{LO},\omega)
\Omega^0_\mathrm{in} e^{-i \omega t}
\nonumber \\
&&+\ O(|\Omega^0_\mathrm{in}|^2)
\label{2.5}
\end{eqnarray}
defined by a harmonic signal $\Omega_\mathrm{in}(t) = \Omega^0_\mathrm{in} e^{-i \omega t}$, hence fixed $\Delta \omega_\mathrm{in} = \omega$. This paper will first focus (Sec.\ \ref{sec:senseresults}) on the adiabatic limit $|\omega|/\omega_\mathrm{LO} \ll 1$, for which
\begin{equation}
S^\mathrm{ad}_\mathrm{EIT}(\Omega_\mathrm{LO})
\equiv S_\mathrm{EIT}(\Omega_\mathrm{LO},0)
= \frac{\partial {\cal P}_\mathrm{EIT}}{\partial \Omega_\mathrm{LO}}
\label{2.6}
\end{equation}
reduces to the response to a steady state perturbation of the LO amplitude. As will be discussed in Sec.\ \ref{sec:Ryfinitefresp}, based on the full dynamical linear response theory developed in App.\ \ref{app:dynresp}, $|S_\mathrm{EIT}(\Omega_\mathrm{LO},\omega)|$ degrades with nonzero frequency, so the adiabatic limit represents maximum achievable sensitivity. The more general time domain linear response signal
\begin{eqnarray}
\Delta {\cal P}_\mathrm{EIT}[\Omega_\mathrm{Ry}(t)]
&\equiv& {\cal P}_\mathrm{EIT}[\Omega_\mathrm{Ry}(t)]
- {\cal P}_\mathrm{EIT}(\Omega_\mathrm{LO}) \ \ \ \ \ \
\label{2.7} \\
&=& \int \frac{d\omega}{2\pi} S_\mathrm{EIT}(\Omega_\mathrm{LO},\omega)
\Omega^0_\mathrm{in}(\omega) e^{-i \omega t}
\nonumber
\end{eqnarray}
will be strongly distorted if the signal spectrum $\Omega^0_\mathrm{in}(\omega)$ is supported on a bandwidth larger than this. To avoid distortion of wider band signals one needs to tune $\omega_\mathrm{LO}$ some distance, much larger than the bandwidth, away from the signal center frequency where $S_\mathrm{EIT}(\Omega_\mathrm{LO},\omega) \simeq S_\mathrm{EIT}(\Omega_\mathrm{LO},\Delta \omega_\mathrm{in})$ may be approximated as constant across the band---at the expense of reduced amplitude response \cite{NIST2019c,NIST2022,BBNTS2022}.

Note that since $S_\mathrm{EIT}$ is defined as a linear response coefficient it can be either sign. Maximum response corresponds to the magnitude, hence to the steepest part of the ${\cal P}_\mathrm{EIT}(\Omega_\mathrm{LO})$ curve, whether ascending or descending. Both positive and negative cases will be seen in the results to follow.

\section{Equation of motion: Lindblad operator formalism}
\label{sec:lindblad}

The strong illuminations permit a semiclassical modeling approach in which the electric field is treated as classical, coupling to the atom via the standard Stark term, while the spontaneous decays are treated statistically \cite{FIM2005}. Thus, the full atom--photon field problem is replaced by one involving atomic states alone, described by a density matrix $\hat \rho$, projected here onto the (five, in this case) illumination-driven states. The equation of motion
\begin{equation}
\partial_t \hat \rho = i [\hat \rho, \hat H] + \hat D[\hat \rho]
\label{3.1}
\end{equation}
(setting $\hbar = 1$) includes both unitary evolution via the Stark Hamiltonian $\hat H$, and nonunitary relaxation via the Lindblad operator $\hat D[\hat \rho]$ that incorporates the spontaneous decay rates in the ``quantum master equation'' form \cite{FIM2005}:
\begin{equation}
D_{mn} = \left\{\begin{array}{ll}
-\frac{1}{2} \rho_{mn} \sum_p (\gamma_{m \to p} + \gamma_{n \to p}), &  m \neq n \\
\sum_p (\rho_{pp} \gamma_{p \to n} - \rho_{nn} \gamma_{n \to p}), & m = n.
\end{array} \right.
\label{3.2}
\end{equation}

Due to the tree-like structure seen in Fig.\ \ref{fig:rysetup}, in which there are no closed excitation loops (e.g., no additional laser directly coupling states $|3\rangle$, $|5\rangle$ or $|2\rangle$, $|4\rangle$) one may consistently transform to a ``rotating'' frame in which all driving frequencies are absorbed into the states:
\begin{equation}
|n \rangle \to e^{i(\varepsilon_n + \Delta_n) t} |n\rangle,\ \
\rho_{mn} \to e^{i(\varepsilon_m - \varepsilon_n + \Delta_m - \Delta_n) t}\rho_{mn},
\label{3.3}
\end{equation}
where $\varepsilon_n$ are the bare atom energy levels and
\begin{equation}
\Delta_n = \varepsilon_n - \varepsilon_m - \omega_{mn},
\label{3.4}
\end{equation}
is the detuning of the illumination frequency $\omega_{mn}$ coupling states $|m\rangle$, $|n\rangle$. The Lindblad operator is unaffected by this transformation and in the absence of the incident field ($\Omega_\mathrm{in} = 0$) the transformed Hamiltonian is time-independent with off-diagonal elements $H_{mn} = -\Omega_{mn}/2 = -\Omega_{nm}^*/2$ and diagonal elements given by tree-branch partial sums of detunings \cite{FIM2005}---see (\ref{3.6}) below.

\subsection{Reduction to five-level system model}
\label{sec:5levelmodel}

As alluded to above, the choice of the number of states $\{|n\rangle\}$ to explicitly keep in the model is not entirely obvious due to the $2F+1$ degeneracy implied by the finite total angular momentum quantum number (hyperfine level) $F$. The $^{87}$Rb and $^{133}$Cs ground state hyperfine splittings are in the several GHz range, hence essentially equally populated at room temperature ($k_BT/h \simeq 6$ THz). The lasers, having linewidths in the kHz range, are in fact tuned to specific hyperfine levels, which specifies selected $F$ values for each state in Fig.\ \ref{fig:rysetup} (not shown).

The near-universal simplification, which will be followed here, is to represent each rotationally degenerate subset, $-F_n \leq m_{Fn} \leq F_n$, with a single state $|n\rangle$ \cite{FIM2005}. The effective transition dipole moment, hence Rabi frequency $\Omega_{mn}$, specified in the model then represents a rotational average defined by the laser polarization. For planar experimental setups considered here (Fig.\ \ref{fig:starconfig}) all lasers and RF fields have vertical linear polarization relative to the plane. Thus, only the $Z$ component of the transition dipole moment (\ref{2.3}) survives, leading to conservation of $m_F$. The representative probe transition values shown in Table \ref{tab:Ryparms} reflect this choice \cite{foot:cqcom}. The Rabi frequency values for the other transitions, used to guide the model choices below, are similarly informed by experimental measurements using specific laser intensities \cite{foot:cqcom}. It is quite possible that detailed comparisons between the predictions in this paper and experiment will require a more careful exploration of these assumptions, and perhaps even an expansion of the atomic state basis, to obtain high accuracy agreement. However, the general conclusions, constrained by the basic physics of the light--atom interactions, are expected to be robust.

\begin{table}

\begin{tabular}{ll}
Transit time broadening coeff.\ $c_\mathrm{Tr} = \frac{w}{v_\mathrm{th}} \gamma_\mathrm{Tr}$  & 2.71 \\
Intrinsic laser linewidth $\gamma_\mathrm{La}/2\pi$        & 6 kHz \\
Rydberg collisional broadening $\gamma_\mathrm{Col}/2\pi$  & 5 kHz \\
Non-lifetime decoherence rate $\gamma_\mathrm{nlt}/2\pi$   & 20 kHz
\end{tabular}

\caption{Example $^{87}$Rb non-intrinsic decay parameters. Here $\gamma_\mathrm{Tr} = c_\mathrm{Tr} v_\mathrm{th}/w$ represents the thermal mean of the inverse travel time across the beam and adds to the rate of decay to ground $\gamma_{n \to 1}$ of all levels. We use here the Gaussian beam geometrical factor $c_\mathrm{Tr} = 4/\sqrt{\pi\ln(2)}$. Using $v_\mathrm{th} = 169.3$ m/s one obtains the values $\gamma_\mathrm{Tr}/2\pi = 7.3$, 73 kHz for $w = 10$, 1 mm, respectively. The other three account for various dephasing processes $\gamma_{n \to n} = (\gamma_{n \to}^2 + \gamma_\mathrm{nlt}^2)^{1/2} + \gamma_\mathrm{La} + \gamma_{\mathrm{Col},n}$ that do not change the atomic state.  Here $\gamma_{n \to}$ is shorthand for the intrinsic decay rate out of state $|n \rangle$ (see Fig.\ \ref{fig:rysetup}). The value of $\gamma_\mathrm{La}$ here is characteristic of current typical lab quality lasers \cite{foot:cqcom}. The collision contribution $\gamma_{\mathrm{Col},n}$ is taken nonzero only for the Rydberg levels and in a full model would depend strongly on temperature and vapor density. The value $\gamma_\mathrm{nlt}$ is similarly a representative experimental fitting parameter accounting for other unmodeled stochastic processes \cite{foot:cqcom}.}

\label{tab:nonintrinsdecay}
\end{table}

\subsubsection{Non-intrinsic decay mechanisms}
\label{subsec:nonintrinsdecay}

It should be mentioned that Lindblad operator decay parameters (\ref{3.2}), in addition to standard atomic level decay, may also include contributions from effects such as beam transit, collisional, and laser linewidth broadening. Collisional (limited to the Rydberg states) and linewidth broadening lead to wavefunction dephasing without changing the atomic state, hence contribute to $\gamma_{n \to n}$. Transit time broadening is treated as an additional decay mechanism $\gamma_{n \to 1} \propto v_\mathrm{th}/w$ for all levels direct to ground state, where $w$ is an effective beam width (see Table \ref{tab:Ryparms}): excited atoms leaving the beam region are replaced by unexcited atoms entering the beam. As discussed in the Introduction, values $w \sim 1$ cm that will be argued for in this paper are substantially larger than the current 1 mm typical for 1D setups, but is required to both reduce transit time broadening and to fully illuminate the 1 cm$^3$ cell volume within the three intersecting laser beams (Fig.\ \ref{fig:starconfig}). Such increased illumination volume is a challenge, but is an active experimental pursuit for a number of different applications relying on increased sensitivity.

Representative values for these linewidth contributions, used in most of the numerical calculations to follow, are listed Table \ref{tab:nonintrinsdecay}. Clearly the values given there, intended only to be representative of some recent model--experimental comparisons \cite{foot:cqcom}, are far from unique, and would need to be adjusted for detailed comparison to a given experiment. However, the major conclusions to follow are all found to be robust against reasonable variations. The main conclusion is that the values here only weakly affect the core state linewidths but strongly influence the Rydberg state linewidths. The latter will be seen below to dominate optimal sensor design, consistent with the high-q Rydberg cavity picture.

\subsubsection{Five-level effective Hamiltonian}
\label{subsec:5levelhameff}

Given the above assumptions, the five-level system focused on here yields the $5 \times 5$ Hamiltonian
\begin{equation}
H = -\left(\begin{array}{ccc|cc}
0 & \Omega_P/2 & 0 & 0 & 0 \\
\Omega_P^*/2 & \Delta_2 & \Omega_D/2 & 0 & 0 \\
0 & \Omega_D^*/2 & \Delta_3 & \Omega_C/2 & 0 \\ \hline
0 & 0 & \Omega_C^*/2 & \Delta_4 & \Omega_\mathrm{LO}/2 \\
0 & 0 & 0 & \Omega_\mathrm{LO}^*/2 & \Delta_5
\end{array} \right)
\label{3.5}
\end{equation}
with detuning parameters
\begin{eqnarray}
\Delta_2 &=& \Delta_P
\nonumber \\
\Delta_3 &=& \Delta_P + \Delta_D
\nonumber \\
\Delta_4 &=& \Delta_P + \Delta_D + \Delta_C
\nonumber \\
\Delta_5 &=& \Delta_P + \Delta_D + \Delta_C + \Delta_\mathrm{LO}.
\label{3.6}
\end{eqnarray}
The lines in (\ref{3.5}) emphasize the core and Rydberg subspaces. All of the results to follow will be based on this space of parameters. As stated, precise translation of these to physical illumination parameters may be subtle, but this is beyond the scope of this paper---awaiting future experimental implementation.

For atom velocity ${\bf v}$, the energy levels are Doppler shifted relative to the stationary illuminators, with resulting detuning shifts
\begin{equation}
\Delta_\alpha \to \Delta_\alpha + {\bf k}_\alpha \cdot {\bf v},
\label{3.7}
\end{equation}
where ${\bf k}_\alpha$ is the wavevector of illuminator $\alpha$. Note that even as the nominal local oscillator Doppler shift $v/\lambda_\mathrm{LO}$ is at least $O(10^4)$ smaller than the laser values, ${\bf E}_\mathrm{LO}({\bf x})$ will also not be plane-wave like, being strongly influenced by the vapor cell geometry \cite{BVBZ2022}. The Doppler shift of $\Delta_\mathrm{LO}$ is therefore not cleanly defined and hence will be neglected.

\subsubsection{Super-vector representation}
\label{subsec:supervecrep}

Since the equation of motion (\ref{3.1}) is linear in $\hat \rho$, by listing its elements as a column `supervector' ${\bm \rho}$, one obtains
\begin{equation}
\partial_t {\bm \rho} = {\bf G} {\bm \rho},\ \
{\bf G} = i{\bf H} + {\bf D}
\label{3.8}
\end{equation}
with commutator matrix
\begin{equation}
H_{mn,pq} = H_{qn} \delta_{pm} - H_{mp} \delta_{qn}.
\label{3.9}
\end{equation}
The elements of ${\bf D}$ are similarly obtained by writing (\ref{3.2}) in the form
\begin{equation}
D[{\bm \rho}]_{mn} = \sum_{p,q} D_{mn,pq} \rho_{pq}
\label{3.10}
\end{equation}
and identifying
\begin{equation}
D_{mn,pq} = \left\{\begin{array}{ll}
-\frac{1}{2} \delta_{pm} \delta_{qn}
\sum_r (\gamma_{m \to r} + \gamma_{n \to r}), & m \neq n \\
\delta_{pq} \gamma_{p \to n}, & m = n \neq p \\
-\delta_{pq} \sum_{r (\neq n)} \gamma_{n \to r}, & m = n = p,
\end{array} \right.
\label{3.11}
\end{equation}
where, to obtain the third line, we note that the $\rho_{nn} \gamma_{p \to n}$ term cancels between the two terms in the second line of (\ref{3.2}).

\subsection{Adiabatic limit}
\label{sec:adialim}

It follows from the equation of motion (\ref{3.8}) that the steady state, or adiabatic, density matrix ${\bm \rho}_\mathrm{ad}$ satisfies
\begin{equation}
{\bf G} {\bm \rho}_\mathrm{ad} = 0.
\label{3.12}
\end{equation}
Existence of a nonempty kernel follows from conservation of probability,
\begin{equation}
{\bf t} \cdot {\bm \rho} \equiv \mathrm{tr}[\hat \rho] = 1,\ \
t_{mn} \equiv \delta_{mn}.
\label{3.13}
\end{equation}
Applying the transpose (row vector) ${\bf t}^T = {\bf t}^\dagger$ to both sides of (\ref{3.8}) one obtains the constraint
\begin{equation}
{\bf t}^\dagger {\bf G} = 0
\label{3.14}
\end{equation}
which guarantees reduced rank of ${\bf G}$.

The linear response sensitivity derivative defined in (\ref{2.5}), as seen below, will involve the density matrix derivative $\partial {\bm \rho}/\partial\Omega_\mathrm{LO}$. One obtains
\begin{equation}
\partial_a {\bm \rho}_\mathrm{ad} = -{\bf G}^{-1}
(\partial_a {\bf G}) {\bm \rho}_\mathrm{ad},
\label{3.15}
\end{equation}
where $a$ is any parameter. The restricted inverse, on the space orthogonal to the kernel, may be computed from the eigen-decomposition of ${\bf G}$,
\begin{eqnarray}
{\bf G} {\bf u}_n^R &=& \lambda_n {\bf u}_n^R,\ \
{\bf u}_n^{L \dagger} {\bf G}\ =\ \lambda_n {\bf u}_n^{L \dagger}
\nonumber \\
{\bf u}^{L \dagger}_m {\bf u}_n^R &=& \delta_{mn},
\label{3.16}
\end{eqnarray}
in which the second line is a normalization condition, leading to the form
\begin{equation}
{\bf G}^{-1} \equiv \sum_{\lambda_n \neq 0}
\frac{1}{\lambda_n} {\bf u}^L_n {\bf u}^{R \dagger}_n.
\label{3.17}
\end{equation}
Only the nonzero eigenvalues $\lambda_n$ appear in the sum. Since ${\bf G}$ is generally not symmetric, both the left and right eigenvectors ${\bf u}^{L,R}_n$ appear. For any illumination parameter (Rabi frequency or detuning value), one obtains the rather sparse matrix
\begin{equation}
\partial_a {\bf G} = i\partial_a {\bf H}.
\label{3.18}
\end{equation}

The Rydberg sensor performance results presented in Sec.\ \ref{sec:senseresults} will all be based on (\ref{3.15}). Extension to finite frequency solutions to (\ref{3.8}), within the dynamic linear response formalism developed in App.\ \ref{app:dynresp}, will be discussed in Sec.\ \ref{sec:Ryfinitefresp}.

\subsection{Thermal averages}
\label{sec:thermave}

Thermal averages are performed using the classical Maxwell distribution (appropriate to finite $T$ dilute vapors):
\begin{equation}
\hat \rho^\mathrm{th} = \left(\frac{m_A}{2\pi k_B T} \right)^{3/2}
\int d{\bf v} e^{-m_A{\bf v}^2/2k_BT} \hat \rho({\bf v})
\label{3.19}
\end{equation}
with atomic mass  $m_A$. For $T = 300$ K the $^{87}$Rb thermal velocity is $v_\mathrm{th} = \sqrt{k_B T/m_A} = 169$ m/s. For $\lambda = 1\ \mu$m this leads to Doppler shifts $v_\mathrm{th}/\lambda \simeq 170$ MHz, enormous compared to the $\sim$\,1 MHz, or less, resonant linewidths encountered below. A naive estimate is hence that, respectively, fewer than 1\% and 0.01\% of atoms that happen to be moving slowly, will contribute to $S^\mathrm{th}_\mathrm{EIT}$ for 1D and 2D setups. In fact, it is generally much worse than this. It will be seen below that the ${\bf v}$ dependence causes different atom populations to give opposite-sign, near-canceling contributions to $S^\mathrm{th}_\mathrm{EIT}$ unless one enforces the condition
\begin{equation}
\sum_\alpha {\bf k}_\alpha = 0.
\label{3.20}
\end{equation}
Enforcing this experimentally for the three laser system is accomplished by appropriately orienting the beams in a 2D ``star'' configuration. The $^{87}$Rb example shown in Fig.\ \ref{fig:rysetup} leads to the configuration shown in Fig.\ \ref{fig:starconfig}. It should be emphasized that the condition (\ref{3.20}) \emph{does not} lead to true Doppler compensation (e.g., effectively smaller $T$), it simply restores the above naive Doppler-reduced estimate.

\subsection{EIT response}
\label{sec:eitresp}

The EIT response is derived from $\hat \rho^\mathrm{th}$ in the form \cite{FIM2005}
\begin{equation}
{\cal P}^\mathrm{th}_\mathrm{EIT}(L)
= \frac{\Omega_P(L)^2}{\Omega_P(0)^2} = e^{-\alpha_P R_P(L)}
\label{3.21}
\end{equation}
in which $\Omega_P(s)$, $0 \leq s \leq L$ represents the probe beam amplitude a distance $s$ into the vapor cell of length $L$ and inhomogeneity across the beam is neglected. The exponential decay argument $R_P$ is derived from the probe transition component of the density matrix
\begin{equation}
R_P(L) = \int_0^L \frac{\mathrm{Im}[\rho^\mathrm{th}_{21}(s)]}{\Omega_P(s)} ds
\label{3.22}
\end{equation}
with coefficient
\begin{equation}
\alpha_P = \frac{2 k_P N_0 |{\bf d}_{12}|^2}{\epsilon_0 \hbar}
\label{3.23}
\end{equation}
in which $N_0$ is the vapor atomic number density. The linear response sensitivity derivative follows in the form
\begin{equation}
S^\mathrm{th}_\mathrm{EIT}(L) = -\alpha_P
R^\mathrm{LO}_P(L) {\cal P}^\mathrm{th}_\mathrm{EIT}(L),\ \
R_P^\mathrm{LO}(L) \equiv \frac{\partial R_P(L)}{\partial \Omega_\mathrm{LO}}.
\label{3.24}
\end{equation}
Note that $\mathrm{Re}[\rho^\mathrm{th}_{21}]$ corresponds to an index of refraction, perhaps detectable via an alternative interference measurement. Note also that absorption is a consequence of decay processes: $\mathrm{Im}[\hat \rho^\mathrm{th}]$ is nonzero only due to the presence of $\hat D[\hat \rho^\mathrm{th}]$, mainly through the relatively large value $\gamma_{2 \to 1}$. Spectral properties of $\hat H$ dominate the resonant behavior, but $\hat D[\hat \rho^\mathrm{th}]$ generates the EIT signal.

For sufficiently small $\Omega_P$, $\hat \rho^\mathrm{th}_{21}$ is linear in $\Omega_P$ and $\hat \rho^\mathrm{th}_{21}/\Omega_P$ is a constant, independent of $s$, proportional to the vapor linear polarizeability tensor whose imaginary (absorptive) part produces the beam attenuation. However, the result (\ref{3.22}) is actually quite general, including nonlinear atomic polarization effects present for larger $\Omega_P$ \cite{foot:polarize}. For the setup shown in Fig.\ \ref{fig:rysetup}, it will be seen that the linear regime corresponds roughly to $\Omega_P/2\pi < 3$ MHz. However, larger values may be desired to increase signal SNR through increased photon count, even with the added complication of an inhomogeneous vapor. This will be analyzed in some detail below.

If one assumes that only $\Omega_P$ varies along the beam, then (\ref{3.21})--(\ref{3.24}) are computed by solving the ODE pair
\begin{eqnarray}
\partial_s \Omega_P(s) &=& -\frac{1}{2} \alpha_P
\mathrm{Im}[\rho^\mathrm{th}_{21}[\Omega_P(s)]]
\nonumber \\
\partial_s R_P^\mathrm{LO}(s) &=& \frac{1}{\Omega_P(s)}
\mathrm{Im}\left[\frac{\partial\rho^\mathrm{th}_{21}}{\partial \Omega_{LO}}
[\Omega_P(s)] \right]
\label{3.25} \\
&-& \frac{\alpha_P}{2} \Omega_P(s) R_P^\mathrm{LO}(s)
\mathrm{Im}\left[\frac{\partial}{\partial \Omega_P(s)}
\frac{\rho^\mathrm{th}_{21}[\Omega_P(s)]}{\Omega_P(s)} \right]
\nonumber
\end{eqnarray}
with all other parameters taken as fixed. The two density matrix derivatives are computed via (\ref{3.15}). Depending on setup details, these equations could be extended to treat simultaneous inhomogeneity of the remaining $\Omega_\alpha$, generated by exponentials of other components of $\hat \rho^\mathrm{th}$, but this lies beyond the scope of the present work.

\begin{figure*}

\includegraphics[width=2.2in,viewport = 0 0 560 540,clip]{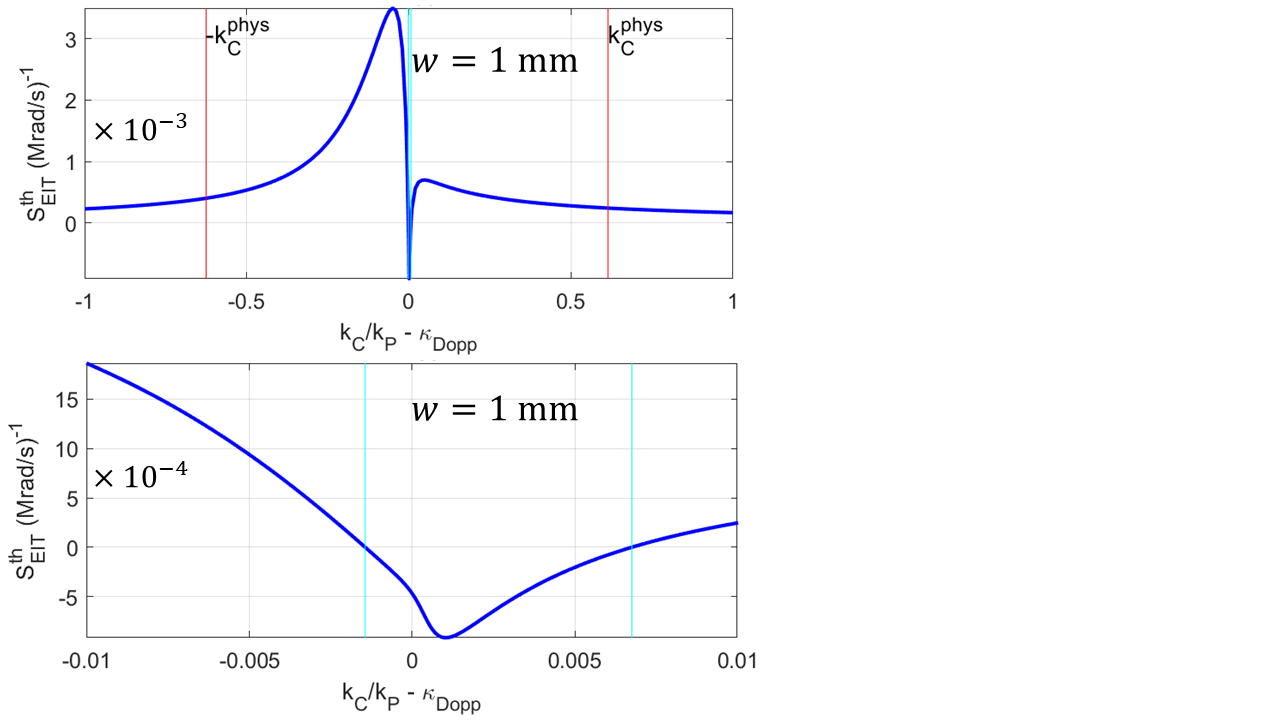}
\includegraphics[width=2.5in,viewport = 0 0 570 510,clip]{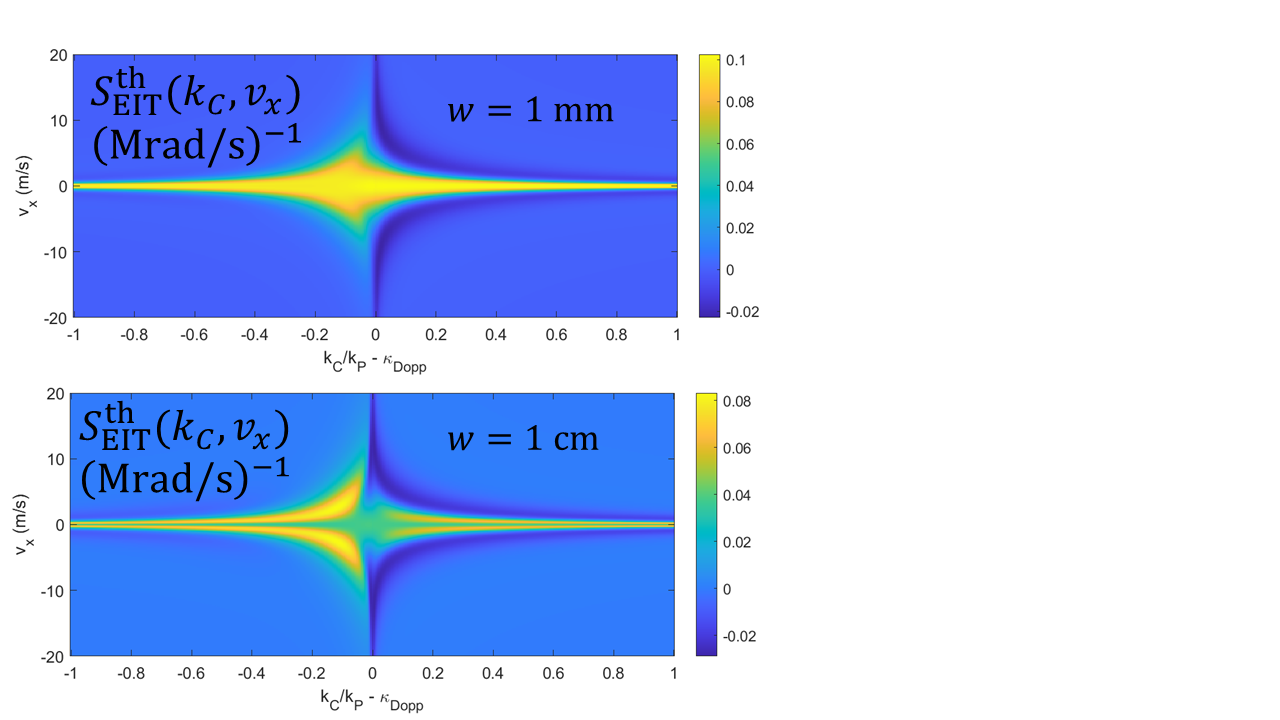}
\includegraphics[width=2.2in,viewport = 0 0 560 540,clip]{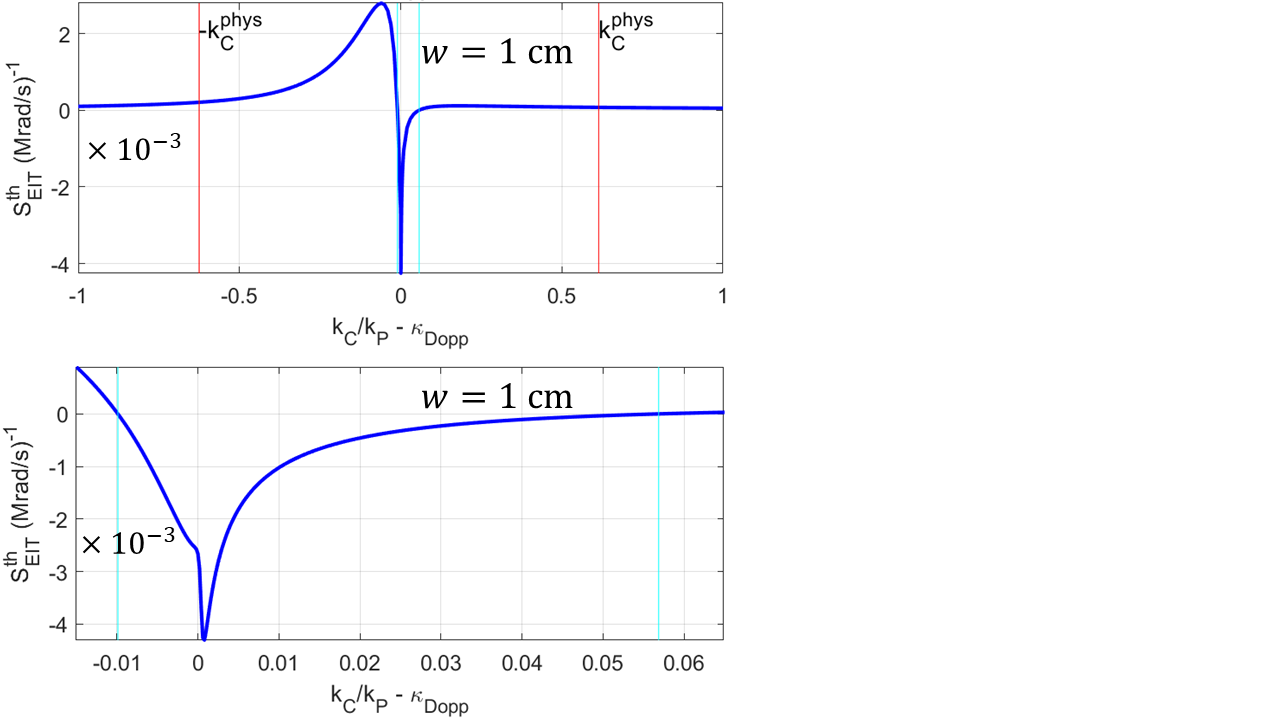}

\caption{\textbf{Top left:} Linear response sensitivity $S_\mathrm{EIT}^\mathrm{th}(T=300\ \mathrm{K})$ vs.\ Doppler detuning for the 1D setup: ${\bf k}_D$ antialigned with ${\bf k}_P$, in turn either aligned or antialigned with ${\bf k}_C$, and for a $w = 1$ mm beam width. Vertical cyan lines mark zero crossings. Here $k_C$ is varied through the Coupling parameter $\Delta_C + k_C v_x$ at fixed $\Delta_C$. Using the parameters in Fig.\ \ref{fig:rysetup}, the true physical sensitivity ($k^\mathrm{phys}_C/k_P = 0.6197$; vertical red lines) is $< 10$\% of its value at the Doppler point $\kappa_\mathrm{Dopp} \equiv k_D/k_P - 1 = 0.00549$. Experimentally motivated parameters are $L = 1$ cm, $(\Omega_P,\Omega_D,\Omega_C,\Omega_\mathrm{LO})/2\pi = (2.7, 3.9, 3.1, 0.85)$ MHz and all $\Delta_\alpha = 0$. \textbf{Bottom left:} Expanded view of the very narrow negative-going main peak. The $O(10^{-3})$ dimensionless scale here corresponds to roughly 0.4 THz in Coupling laser frequency. \textbf{Top and bottom right:} Identical plots using the larger $w = 1$ cm beam width. \textbf{Middle:} Velocity spectra $S_\mathrm{EIT}(v_x) = -\alpha_P R^\mathrm{LO}_P(v_x) {\cal P}_\mathrm{EIT}^\mathrm{th}$ for the 1 mm (top) and 1 cm (bottom) beam widths. The decrease of its thermal average $S_\mathrm{EIT}^\mathrm{th}$ away from $\kappa_\mathrm{Dopp}$ is seen to be due to a combination of cancelation between $v_x$ populations and spectral peak narrowing. For this 1D setup it is seen that the effects of transit time broadening are fairly subtle. In fact, although the narrower beam slightly degrades the peak sensitivity near the Doppler point $k_C/k_P = \kappa_\mathrm{Dopp}$, the sensitivity is significantly \emph{increased} at the physical points $\pm k_C^\mathrm{phys}$.}

\label{fig:doppler1D}
\end{figure*}

\begin{figure}

\includegraphics[width=3.0in,viewport = 0 0 690 550,clip]{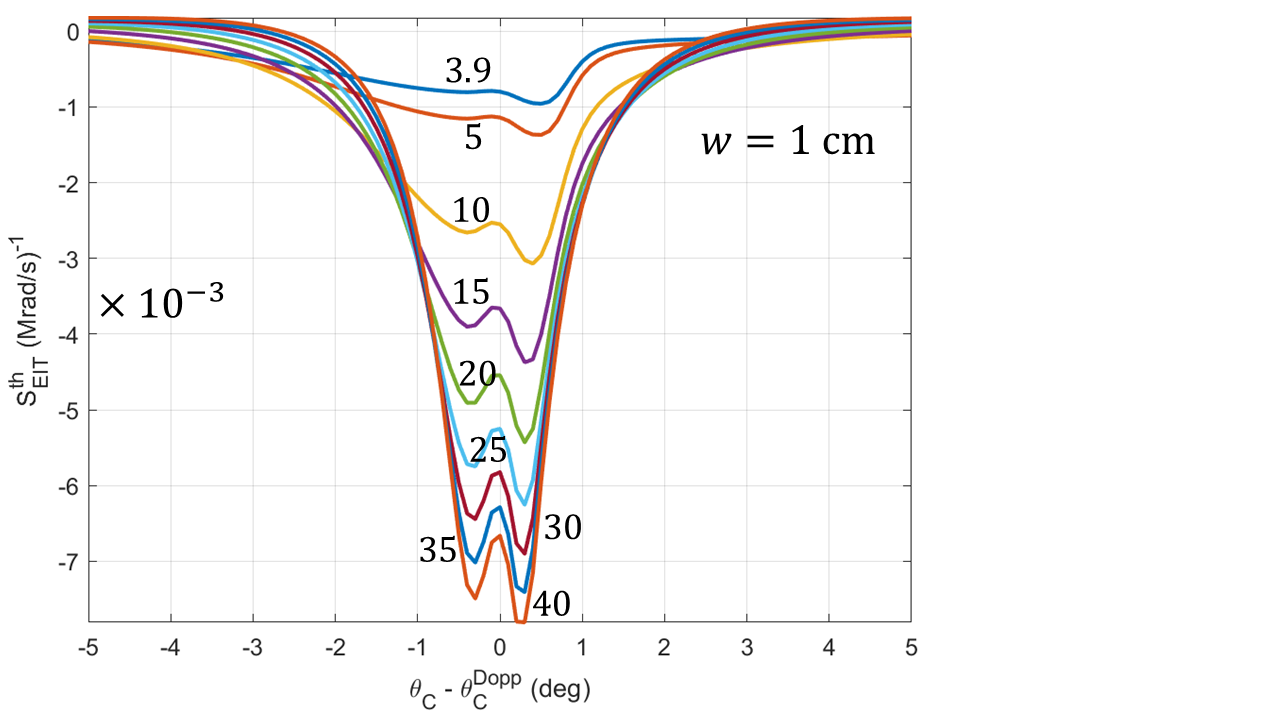}
\includegraphics[width=3.0in,viewport = 0 0 730 540,clip]{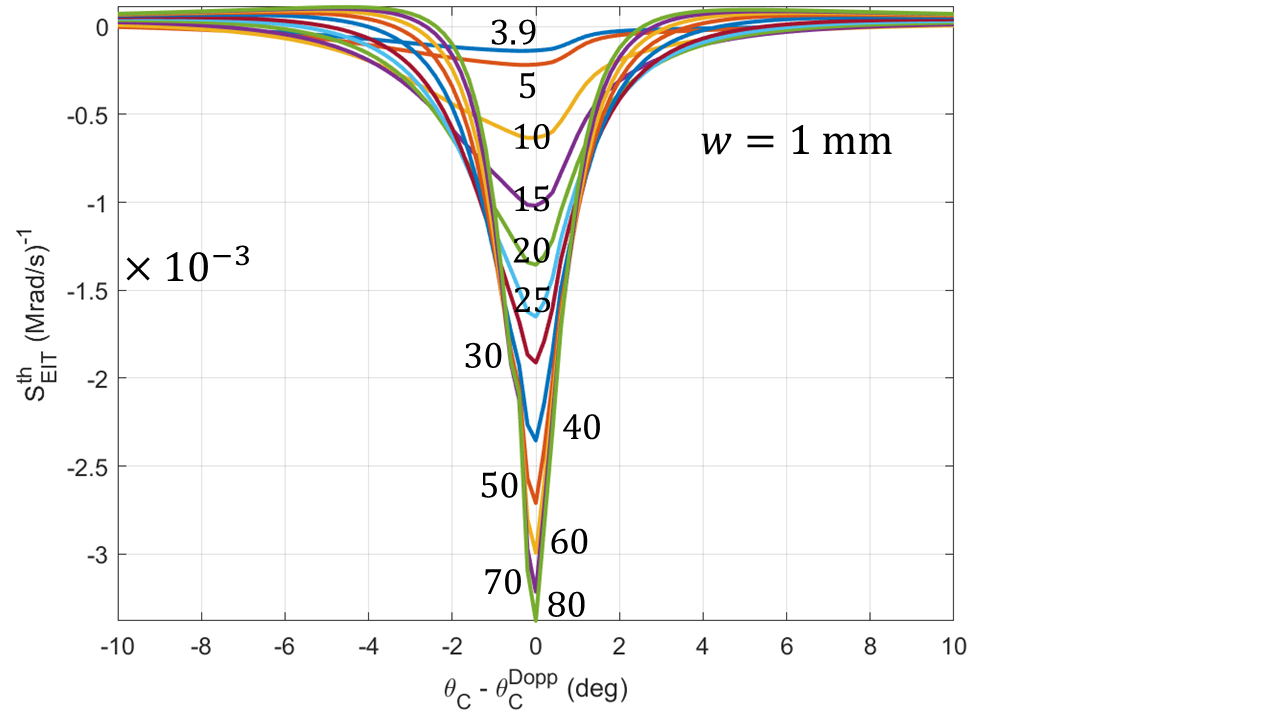}

\caption{\textbf{Top:} EIT linear response sensitivity vs.\ Doppler for the 2D ``star'' setup, pictured in the Fig.\ \ref{fig:starconfig} for $w = 1$ cm beam diameter and for a sequence of Dressing laser amplitudes $\Omega_D/2\pi = (3.9, 5, 10, 15, 20, 25, 30, 35, 40)$ MHz, from top to bottom as indicated. The Coupling detuning $\Delta_C + {\bf k}_C \cdot {\bf v}$ is varied through $\theta_C$ relative to $\theta_C^\mathrm{Dopp} = -107.5^\circ$ with fixed $|{\bf k}_C|$. The sensitivity curve maximum magnitudes increase monotonically with Dressing laser amplitude, initially linearly but saturating at $|S_\mathrm{EIT}^\mathrm{th}| \sim 10^{-2}$ for $\Omega_D/2\pi \sim 40$ MHz. Parameters are otherwise the same as in Fig.\ \ref{fig:doppler1D}. \textbf{Bottom:} Corresponding plot for $w = 1$ mm beam diameter and for a somewhat larger range $\Omega_D/2\pi = (3.9, 5, 10, 15, 20, 25, 30, 40, 50, 60, 70, 80)$ MHz. For smaller $\Omega_D$ the linewidth is about twice as large and the maximum sensitivity about five times smaller. For larger $\Omega_D$ the linewidth narrows and the maximum it is only about 3 times smaller.}

\label{fig:doppler2D}
\end{figure}

For $\Omega_P = 0$ there is no excitation out of the ground state, and any initial state will ultimately relax to the ground state. The adiabatic solution is therefore simply the ground state, $\rho_{lm} = \delta_{l1} \delta_{m1}$; in particular $\rho_{21} = 0$. As alluded to above, for small $\Omega_P$ the leading linear correction produces a constant value (independent of both $\Omega_P$ and depth $s$) for the ratio $\mathrm{Im}[\rho_{21}(\Omega_P)]/\Omega_P$ appearing in (\ref{3.22}), and one obtains the simple result
\begin{equation}
R_P(L) = L \frac{\mathrm{Im}[\rho^\mathrm{th}_{21}(\Omega_P(0))]}{\Omega_P(0)}.
\label{3.26}
\end{equation}
The linear response sensitivity also simplifies to the form
\begin{equation}
S^\mathrm{th}_\mathrm{EIT}(L) = -\frac{\alpha_P L}{\Omega_P}
 \mathrm{Im}\left[\frac{\partial \rho^\mathrm{th}_{21}(\Omega_P)}
{\partial\Omega_\mathrm{LO} }\right]
e^{-\alpha_P R_P(L)}.
\label{3.27}
\end{equation}
This same result holds for sufficiently thin cells for which $\Omega_P \simeq \Omega_P(0)$ may be taken as uniform. We find that this approximation is indeed valid for $L \alt 1$ cm.  For numerical simplicity, most examples below will use this limit. At the end we consider nontrivial solutions of (\ref{3.25}) for larger $\Omega_P$ and/or $L$, addressing questions of optimal cell length.

\begin{figure*}

\includegraphics[width=2.3in,viewport = 0 0 660 560,clip]{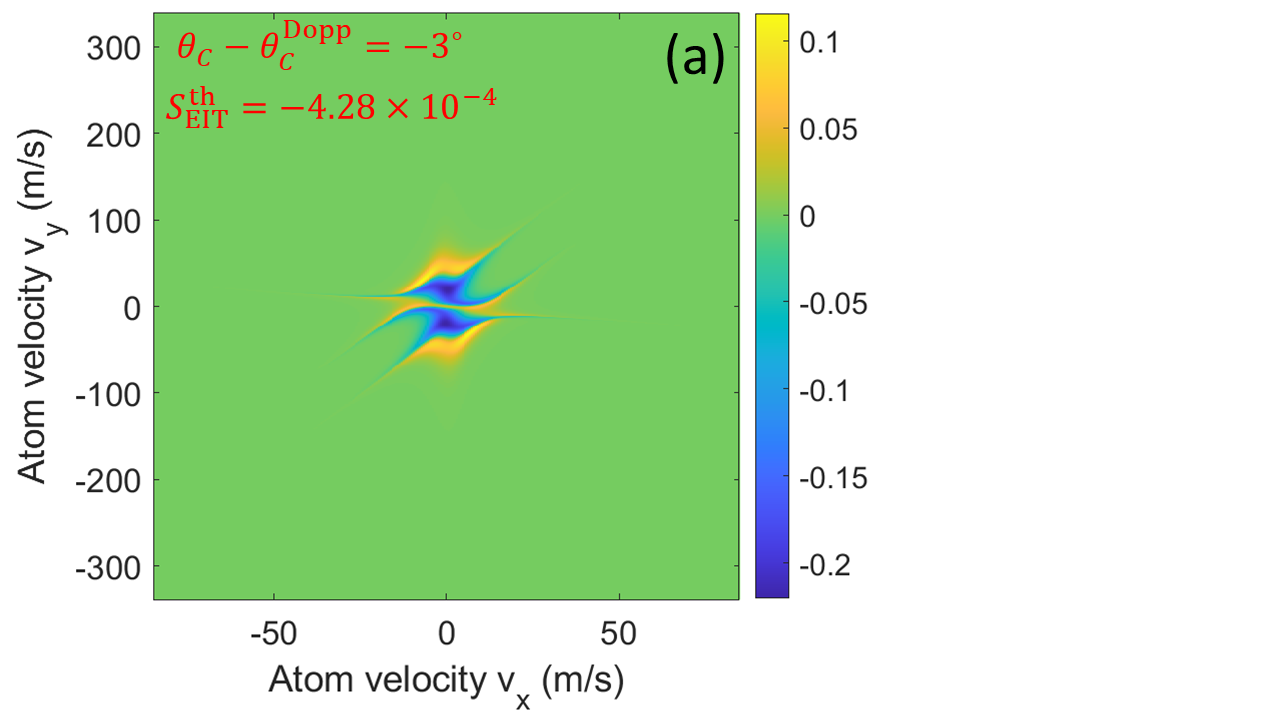}
\includegraphics[width=2.3in,viewport = 0 0 660 560,clip]{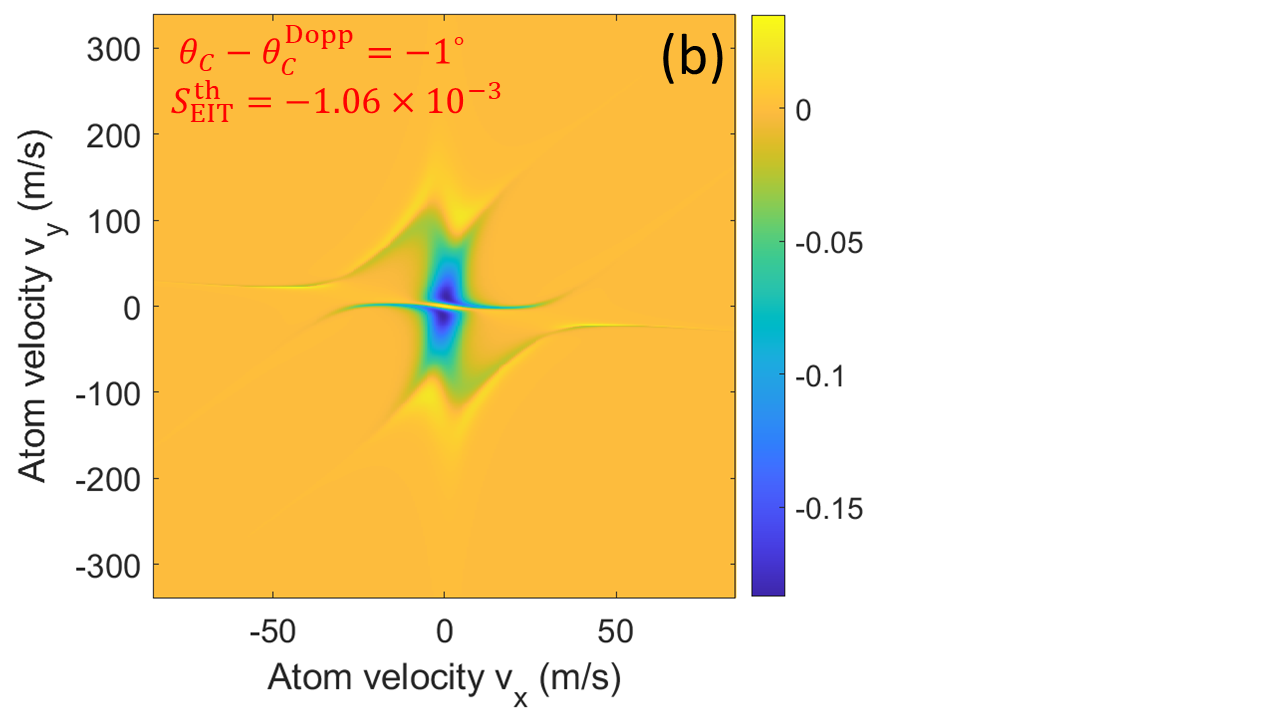}
\includegraphics[width=2.3in,viewport = 0 0 660 560,clip]{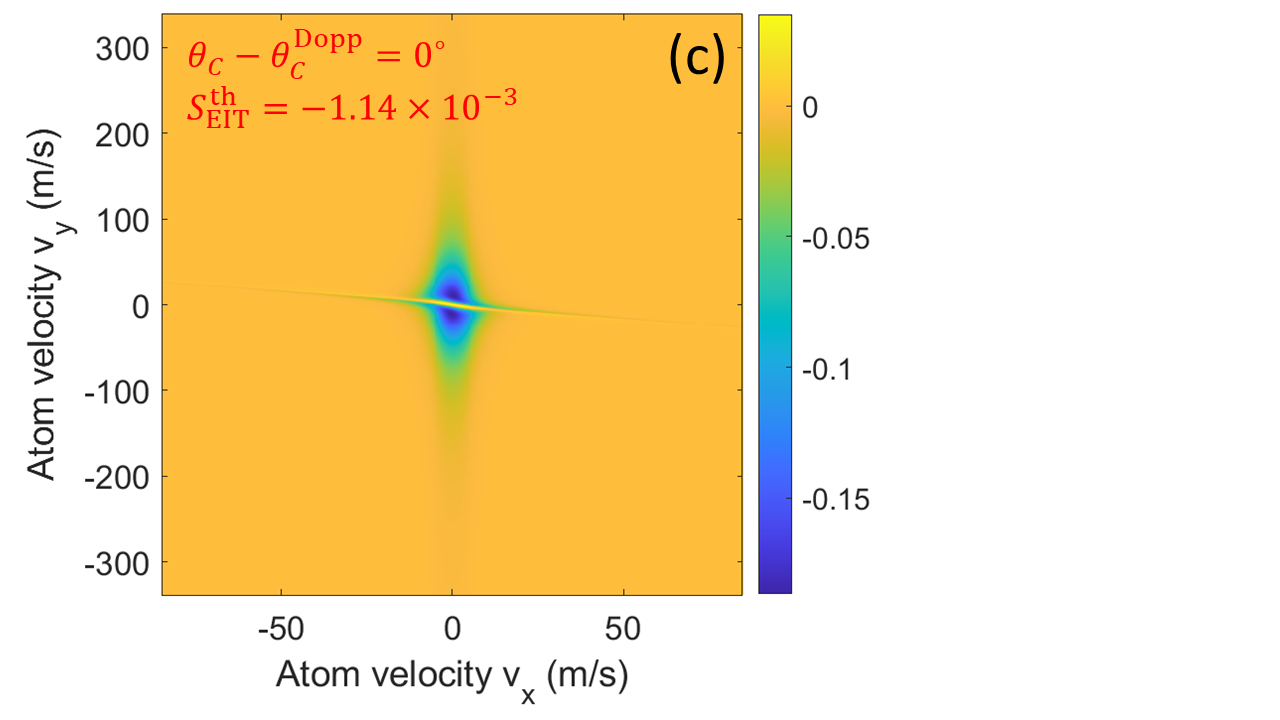}

\includegraphics[width=2.3in,viewport = 0 0 660 560,clip]{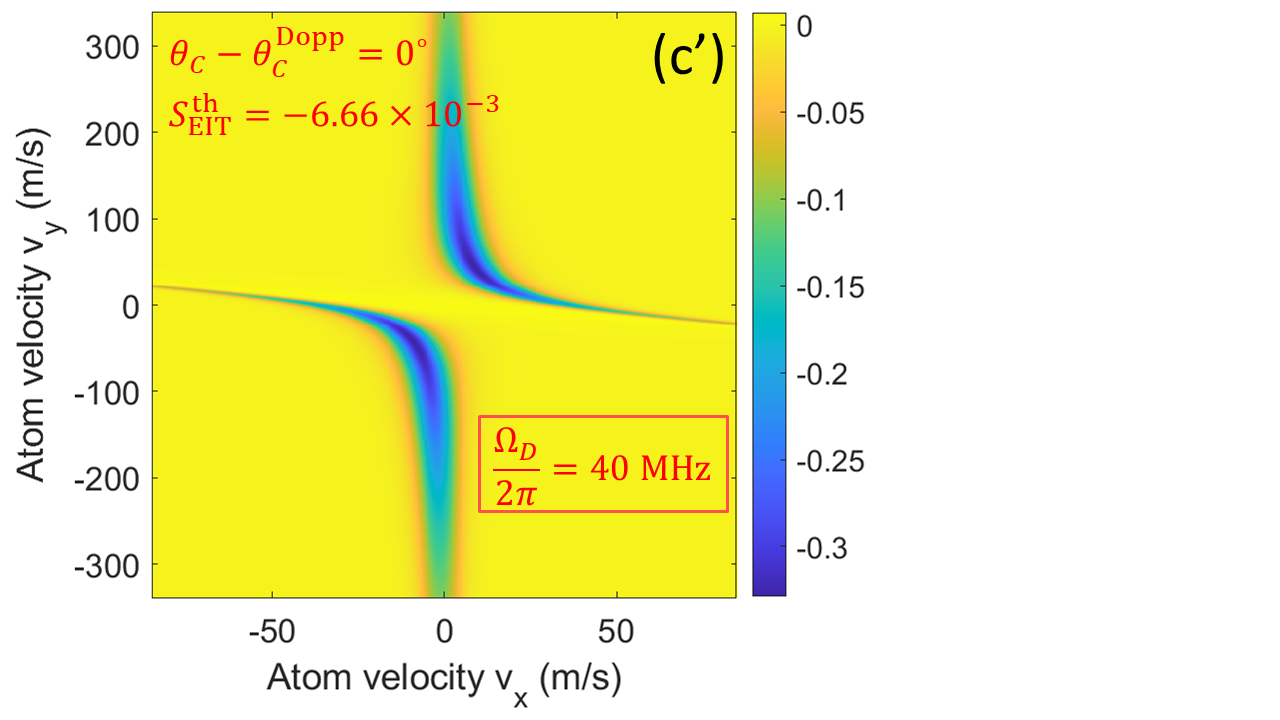}
\includegraphics[width=2.3in,viewport = 0 0 660 560,clip]{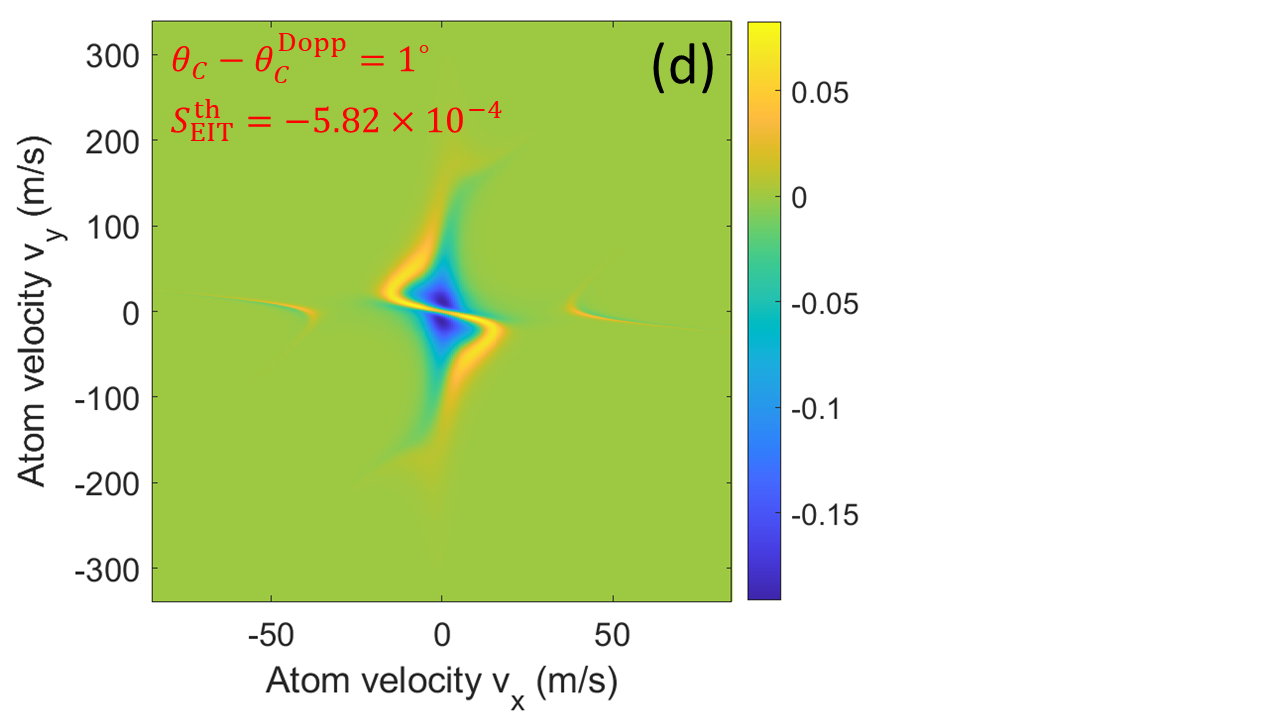}
\includegraphics[width=2.3in,viewport = 0 0 660 560,clip]{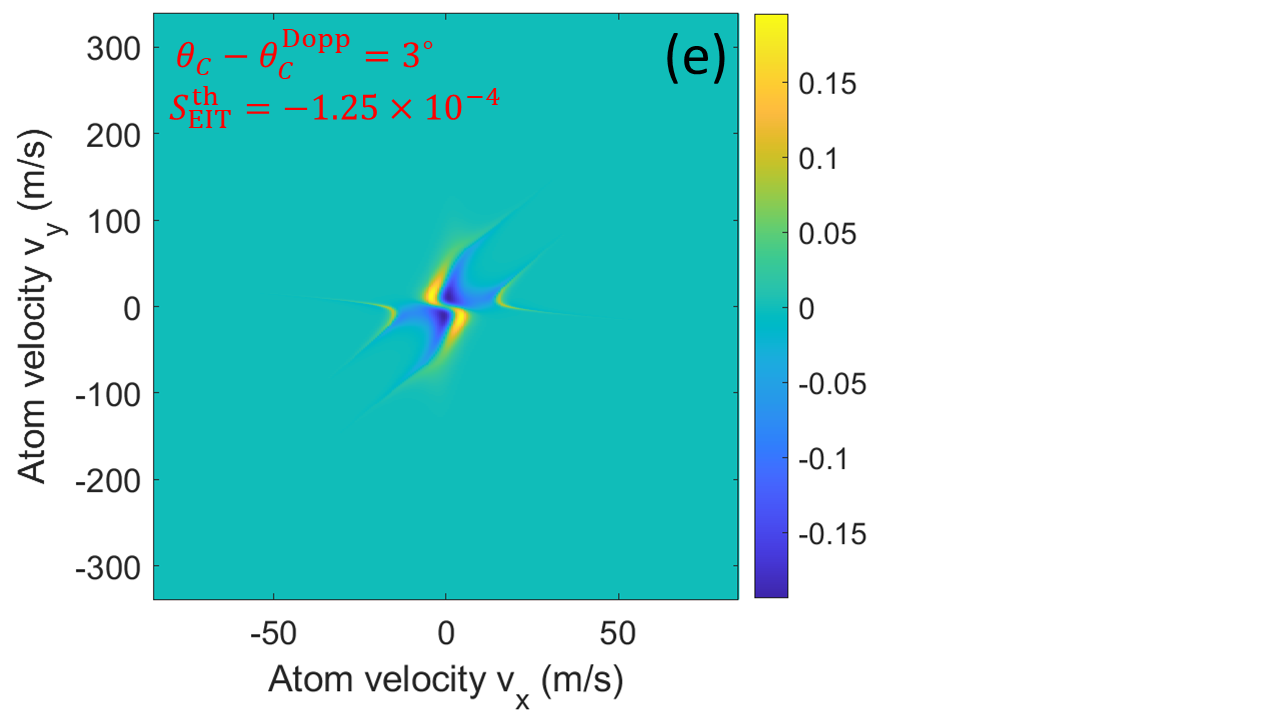}

\caption{EIT linear response sensitivity velocity spectra $S_\mathrm{EIT}({\bf v}) = -\alpha_P R^\mathrm{LO}_P({\bf v}) {\cal P}_\mathrm{EIT}^\mathrm{th}$, in units of (Mrad/s)$^{-1}$, corresponding mainly to the $\Omega_D/2\pi = 5$ MHz case in the upper panel ($w = 1$ cm) of Fig.\ \ref{fig:doppler2D}. The magnitude of the thermal averages (\ref{3.5}), $S^\mathrm{th}_\mathrm{EIT} = [-4.28 \mbox{ \textbf{(a)}}, -10.6 \mbox{ \textbf{(b)}}, -11.4 \mbox{ \textbf{(c)}}, -5.82 \mbox{ \textbf{(d)}}, -1.25 \mbox{ \textbf{(e)}}] \times 10^{-4}$ (Mrad/s)$^{-1}$, decrease rapidly away from the Doppler angle $\theta_C^\mathrm{Dopp} = -107.5^\circ$ (Fig.\ \ref{fig:starconfig}) even while the color scale varies rather weakly. The special panel $({\bf c^\prime})$ demonstrates the alternate Doppler suppression effect of much larger $\Omega_D/2\pi = 40$ MHz, with greatly broadened resonance region and six-fold sensitivity enhancement of $S^\mathrm{th}_\mathrm{EIT} = -66.6 \times 10^{-4}$ (Mrad/s)$^{-1}$. For $^{87}$Rb the thermal standard deviation $v_\mathrm{th} = 169.3$ m/s easily encompasses the nonzero central regions for \textbf{(a--e)}, but not so for $({\bf c^\prime})$ which explains the saturation effect seen in Fig.\ \ref{fig:doppler2D}. Note the distinctly different patterns for $\theta_C > \theta_C^\mathrm{Dopp}$, $\theta_C = \theta_C^\mathrm{Dopp}$, and $\theta_C < \theta_C^\mathrm{Dopp}$ reflecting the rapidly evolving illumination-induced ``dressed'' atomic level resonances. At these laser wavelengths, each 1 m/s velocity change represents a Doppler-induced detuning shift $\Delta v/\lambda \sim 1$ MHz.}

\label{fig:vspec2d}
\end{figure*}

\section{Optimal sensor performance results}
\label{sec:senseresults}

We now describe a set of results, based on the previously described adabatic limit theoretical approach, aimed at maximizing signal sensitivity. At the end (Sec.\ \ref{sec:rysnr}) we will use these results to estimate the correspondingly enhanced SNR and compare to the classical antenna results based on (\ref{1.1})--(\ref{1.5}). in Sec.\ \ref{sec:Ryfinitefresp} we will expand the analysis to include non-adiabatic (finite frequency) effects. We focus first on Doppler effects and the advantages of the star configuration (Fig.\ \ref{fig:starconfig}), and then demonstrate the ultra-narrow linewidths (Rydberg cavity effect) achievable in the small $\Omega_\mathrm{LO}$ regime (Sec.\ \ref{sec:weakLORyCavity}). The critical role of beam width and associated transit time broadening is demonstrated as well. In all cases we demonstrate increasing sensitivity with Dressing laser power, an effect shown to be attributable to a different form of Doppler compensation.

For most of the results, the $L = 1$ cm cell length and $\Omega_P/2\pi = 2.7$ MHz values are sufficiently small that inhomogeneity effects along the beam path are not important and the $s$-independent value (\ref{3.26}) for the exponential argument may be used. Only at the end (Sec.\ \ref{sec:inhomogprobe}) are inhomogeneity effects explored for larger values of both these parameters through full ODE solutions to (\ref{3.25}).

\begin{figure}

\includegraphics[width=3.3in,viewport = 10 0 510 360,clip]{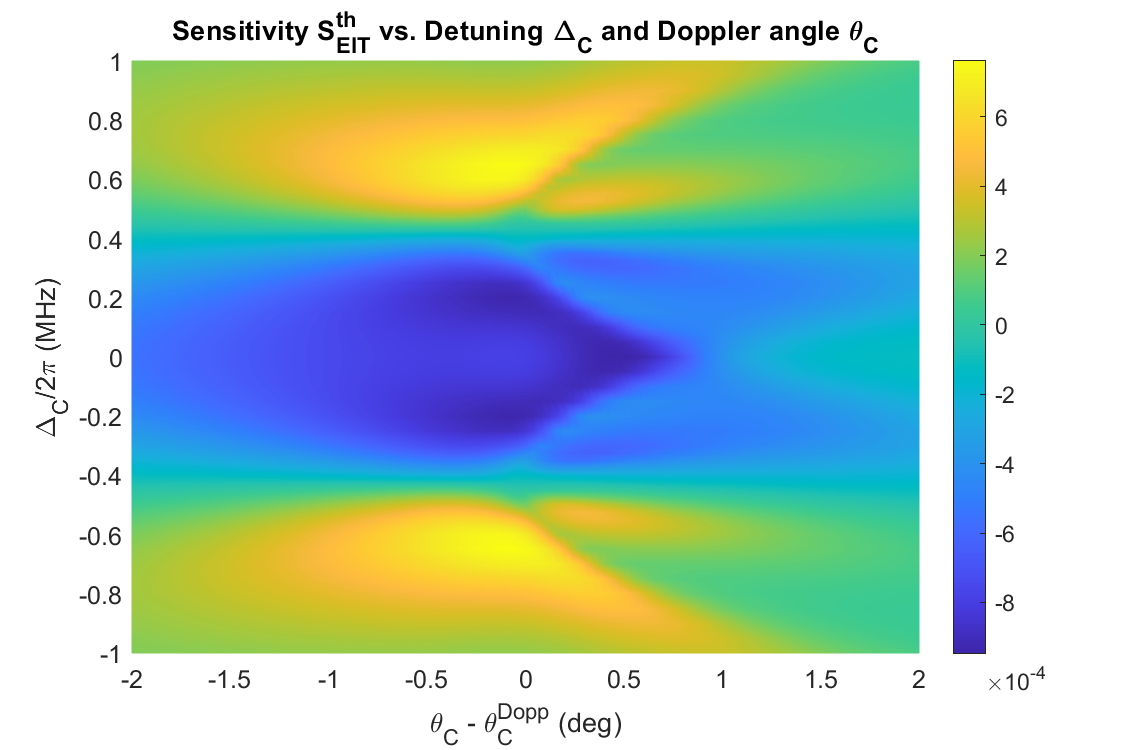}

\caption{EIT linear response sensitivity vs.\ Doppler and Coupling laser detuning for the 2D ``star'' setup, pictured in Fig.\ \ref{fig:starconfig}. Units are (Mrad/s)$^{-1}$ and parameters are otherwise the same as in Fig.\ \ref{fig:doppler1D}. The $\Delta_C = 0$ transect is the same as the topmost curve ($\Omega_D/2\pi = 3.9$ MHz) in the upper panel of Fig.\ \ref{fig:doppler2D}. The detailed transect profile is seen to depend strongly on $\Delta_C$, reaching maximum overall magnitude in windows around $\Delta_C/2\pi \simeq 0$ and $|\Delta_C|/2\pi \simeq 0.6$ MHz, dropping off rapidly for $|\Delta_C|/2\pi > 1$ MHz, and including a local minimum accompanying the sign reversal in the neighborhood of $|\Delta_C|/2\pi = 0.4$ MHz.}

\label{fig:doppler2D_DeltaC}
\end{figure}

\subsection{1D setup Doppler effect demonstration}
\label{sec:1ddopp}

To motivate the importance of Doppler compensation, we consider first the simpler 1D setup (defined here as the $x$-axis), with Dressing laser antialigned with the Probe laser. The Doppler condition corresponds therefore to $k_P + k_C = k_D$. Within the model defined by the Hamiltonian (\ref{3.5}), to illustrate the effects of this condition we artificially vary the coefficient $k_C$ in the Doppler shift term $k_C v_x$ in (\ref{3.7}) while maintaining fixed values of the detuning and Rabi frequency parameters. Using the laser wavelengths shown in Fig.\ \ref{fig:rysetup}, the physical values are $k_C/k_P = \pm 0.6197$ (corresponding to Coupling beam aligned or anti-aligned with the Probe beam) while the Doppler point corresponds to $(k_C/k_P)^\mathrm{Dopp} \equiv k_D/k_P - 1 = 0.00549$.

The signal linear response sensitivity results are shown in Fig.\ \ref{fig:doppler1D} for 1 mm (left) and 1 cm (right) beam widths, with remaining experimentally motivated parameters listed in the caption. As seen in the left and right panels, the maximum sensitivity $|S^\mathrm{th}_\mathrm{EIT}| \sim 4 \times 10^{-3}$ (Mrad/s)$^{-1}$, occurs very close to the Doppler point, and drops by at least an order of magnitude at the physical value. The bottom left and right panels show a magnified view of the ``Doppler resonance'' whose central feature is $\Delta f_C \sim 0.4$ THz wide (compared to the physical $f_C = 238$ THz). However, it corresponds to a much narrower ``Doppler linewidth'' $\Delta f_C v_\mathrm{th}/c \sim 200$ kHz when expressed in terms of thermal variation of the detuning $\Delta_C$.

The middle panel shows the underlying velocity spectra of the quantity
\begin{equation}
S_\mathrm{EIT}(v_x) = -\alpha_P R^\mathrm{LO}_P(v_x)
{\cal P}_\mathrm{EIT}^\mathrm{th}
\label{4.1}
\end{equation}
whose thermal average (integral against the 1D Maxwell distribution) produces the linear response sensitivity (\ref{3.24}). Here $R^\mathrm{LO}_P(v_x)$ is obtained from (\ref{3.22}) and the right hand side of (\ref{3.24}) prior to the thermal average, i.e., by evaluating $\mathrm{Im}[\rho_{21}(v_x)]$ at a fixed value of $v_x$. The thermally averaged transmission coefficient ${\cal P}_\mathrm{EIT}^\mathrm{th} \sim 0.8$ is required here to obtain the correct normalization, but is a factor of order unity for the cell length $L = 1$ cm considered here. As one moves away from the Doppler compensated region, one sees that the peak sensitivity (color scale) does not change very much, but the spectrum strongly narrows. This, in combination with some cancelation between positively and negatively sensitive atom populations (especially to the right of the Doppler point) is what leads to the rapid drop in the thermal average seen in the upper panel. The latter cancelation effect will be seen below to be more important for the 2D setup.

Finally, we note that the effects of transit time broadening (see Table \ref{tab:nonintrinsdecay} and its caption) are quite subtle and varied. The resonant structure near the Doppler point, reflecting the fine scale structure of the velocity spectra, varies strongly, with extremal sensitivity flipping sign ($3.4 \times 10^{-3}$  (Mrad/s)$^{-1}$ for $w = 1$ mm, $-4.2 \times 10^{-3}$ (Mrad/s)$^{-1}$ for $w = 1$ cm), but the magnitude changes little. On the other hand, the sensitivity at the physical points significantly \emph{increases} for the narrower beam: $(2.4, 4.0) \times 10^{-4}$ (Mrad/s)$^{-1}$ for $w = 1$ mm vs.\ $(0.75, 2.1) \times 10^{-4}$ (Mrad/s)$^{-1}$ for $w = 1$ cm (with the two figures for $k_C = k_C^\mathrm{phys}$ and $k_C = -k_C^\mathrm{phys}$, respectively). From the middle panel of Fig.\ \ref{fig:doppler1D}, it appears that this effect originates in broadening of the underlying resonant lineshapes leading to reduced cancelation between different speed atoms.

Another interesting feature of the $w = 1$ cm velocity spectrum is the contrast between the slow moving atom value $S_\mathrm{EIT}(v_x = 0) \sim 0.04$ (Mrad/s)$^{-1}$, which might be thought of as the $T = 0$ sensitivity, and the significantly larger values for finite velocity atoms ($v_x \sim \pm 5$ m/s, in the central region of the plot). The thermal mean picks up both contributions, and this explains the thermal reduction factor of only $\sim$\,10\% near the Doppler point vs.\ the previously estimated 1\% for the 1D setup.

\subsubsection{Key takeaways}
\label{subsec:keytakeaway}

The 1D example clearly demonstrates the sensitivity advantage of the Doppler compensated setup (artificially generated, in this case), and it is interesting that there is no intrinsic advantage to increasing beam diameter other than to increase photon count. Away from the Doppler point, narrower beams in fact lead to higher sensitivity values through increased weight in the tails of the underlying resonance. There are many practical experimental advantages to the 1D setup, and this appears to be yet another (not previously recognized).

In the analysis of the 2D star setup to follow, however, the opposite conclusion will be found. The Doppler advantage is much more sensitive to linewidth broadening, and the $\sim$\,1 cm beam diameter is necessary to achieve the promised orders of magnitude sensitivity increase. The increased photon count from such a beam is completely separate from this aspect of the analysis, and is an added benefit.

\begin{figure}

\includegraphics[width=3.2in,viewport = 0 0 940 530,clip]{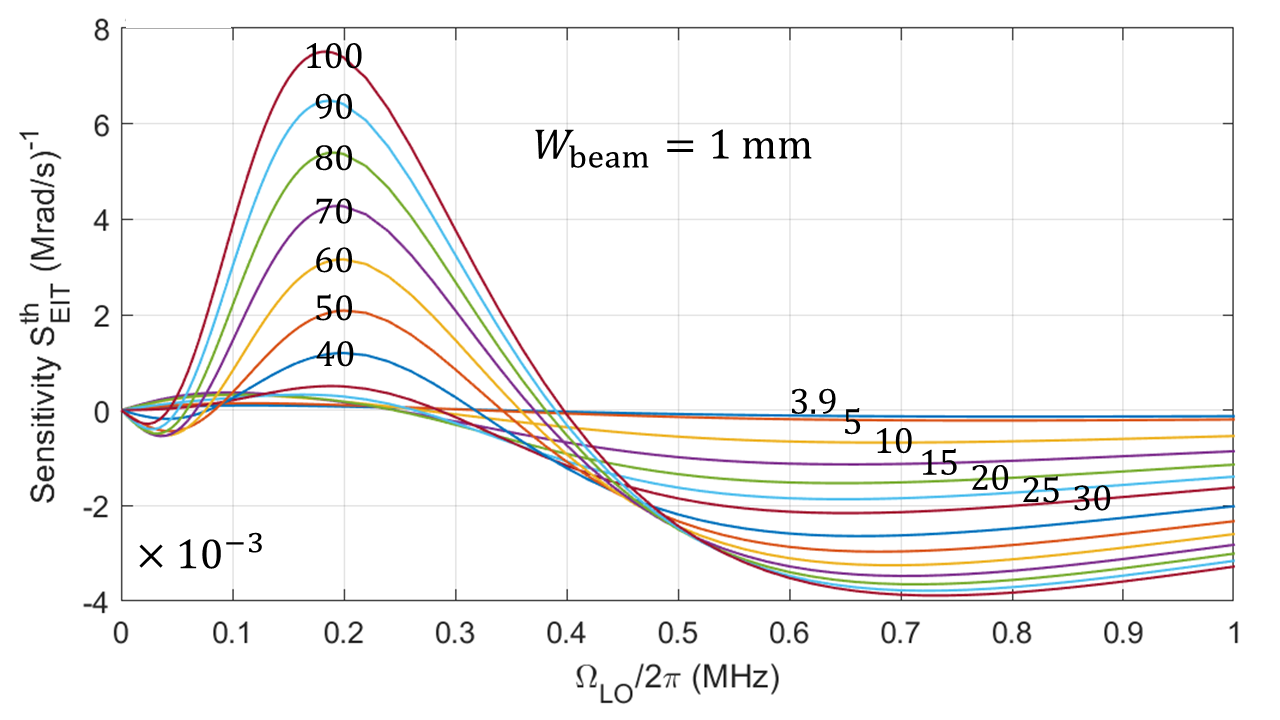}
\includegraphics[width=3.2in,viewport = 0 0 940 530,clip]{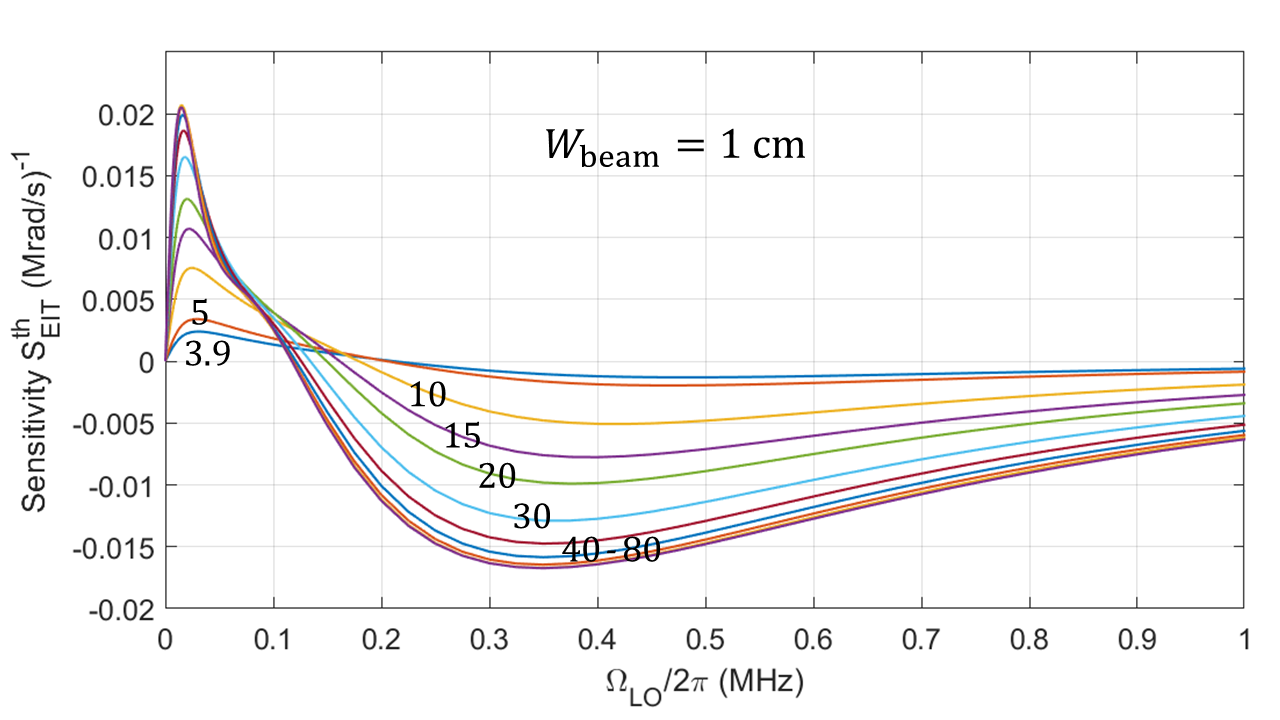}
\includegraphics[width=3.2in,viewport = 0 0 950 530,clip]{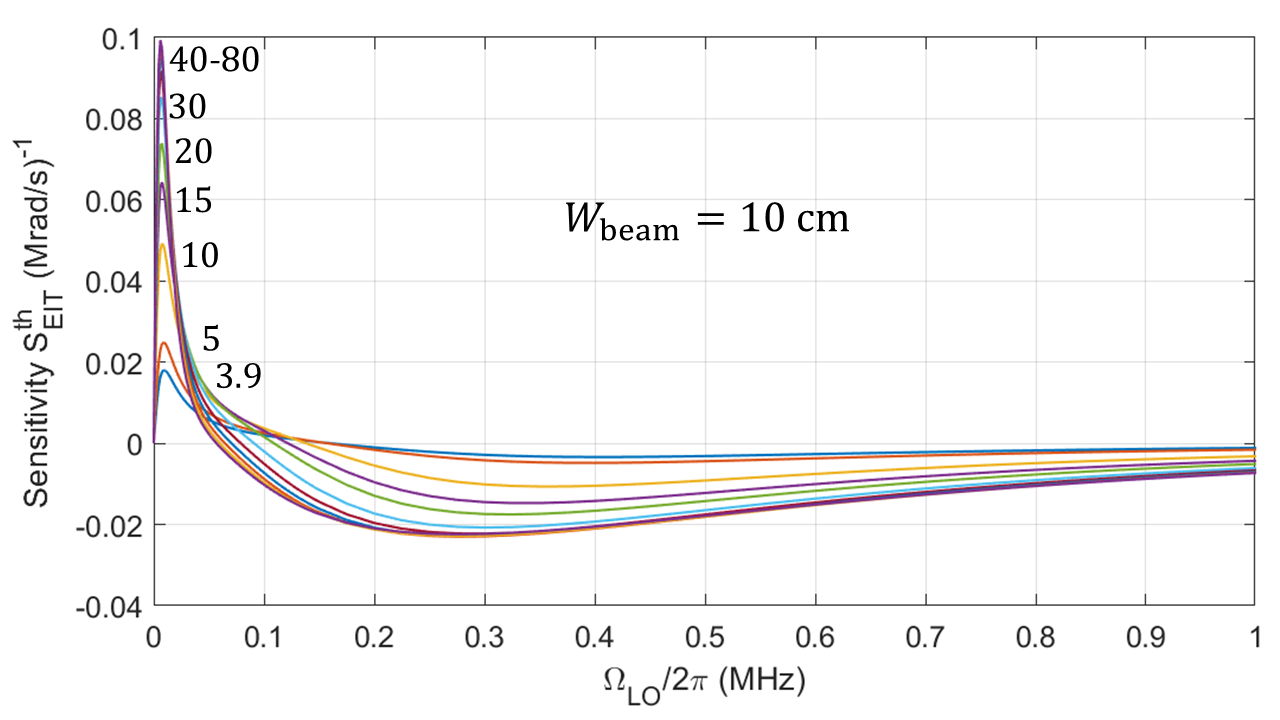}

\caption{EIT linear response sensitivity $S_\mathrm{EIT}^\mathrm{th}$ including the regime of very small local oscillator strength $\Omega_\mathrm{LO}$ for three different beam widths $w = 1, 10 , 100$ mm, and for sequences of Dress laser amplitudes $\Omega_D$ similar to those used in Fig.\ \ref{fig:doppler2D}. The $40 \leq \Omega_D/2\pi \leq 80$ MHz ranges in the middle and lower plots are in steps of 10 MHz. As usual, parameters are otherwise the same as in Fig.\ \ref{fig:doppler1D}. As discussed in the main text (Sec.\ \ref{sec:weakLORyCavity}), the interesting nonmonotonic structure of the curves reflects a combination of the underlying Rydberg level spectral resonances and line broadening mechanisms. For larger $w$, peak sensitivities at small $\Omega_\mathrm{LO}$ are evident. The sensitivity enhancement with increasing $\Omega_D$ seen in Fig.\ \ref{fig:doppler2D} is clearly preserved.}

\label{fig:SensOmegaLO}
\end{figure}

\subsection{2D star configuration Doppler effect demonstration}
\label{sec:2dstardopp}

The 2D setup enables proper physical control of the Doppler condition ${\bf k}_P + {\bf k}_D + {\bf k}_C = 0$ by way of the ``star'' configuration shown in Fig.\ \ref{fig:starconfig}. We will consider perturbations of this condition by varying the angle $\theta_C$ of ${\bf k}_C$. Figure \ref{fig:doppler2D} shows the thermally averaged linear response sensitivity vs.\ $\theta_C$ for various Dressing laser amplitudes $\Omega_D$ for beam widths $w = 1$ mm and 1 cm. The sensitivity magnitude in all cases drops by a factor of $O(10)$ on $\Delta \theta_C \sim 1^\circ$ scales, corresponding to $f_C \Delta\theta_C v_\mathrm{th}/c \sim 2$ MHz thermal variation of the detuning $\Delta_C$. The fine-scale features, on the other hand, are on $0.1^\circ$ scales, which serve to define a Doppler linewidth of $\sim$\,200 kHz, consistent with the 1D result above.

For the 1 mm Probe beam case, the $O(10^{-4})$ peak sensitivity value for $\Omega_D/2\pi = 3.9$ MHz is consistent with the naive $\sim$\,0.01\% estimate above. This value is also similar to the 1D setup physical values seen in Fig.\ \ref{fig:doppler1D}, so there is clearly no significant advantage to the 2D configuration. Reality is actually even worse than this: we use here cell length $L = 1$ cm for both cases, with the entire Probe beam region of the vapor considered as an active volume. If all three beams are taken as 1 mm, the sensitivity would drop by another order of magnitude due to the proportionally smaller change in transmission $P^\mathrm{th}_\mathrm{EIT}$ through the active region---an effective cell length $L \sim 1$ mm would now appear in the prefactor of equation (\ref{3.27}). Thus, in absence of any beam broadening the 1D setup has the clear advantage.

On the other hand, if all beams are broadened to 1 cm, one observes an order of magnitude larger $O(10^{-3})$ peak sensitivity, a consequence, as we shall see below, of reduced transit time broadening in addition to the Doppler enhancement. For both beam widths, the increase with $\Omega_D$ (and with $\Omega_C$, but not shown here) occurs because increasing the off-diagonal components of the Hamiltonian (\ref{3.5}) reduces the influence of the Doppler-dominated diagonal components. The effect begins to saturate above $\Omega_D/2\pi \sim 40$ MHz because the increasing velocities involved are suppressed by the Maxwell distribution (\ref{3.19}).

Insight into these effects is again obtained from the underlying velocity spectrum $S_\mathrm{EIT}({\bf v})$ (now depending on both velocity components) shown in Fig.\ \ref{fig:vspec2d} for five different values of $\theta_C$. Different atom populations clearly contribute very differently. For example, for $\theta_C = \theta_C^\mathrm{Dopp}$, the stationary atom value $S_\mathrm{EIT}({\bf v} = 0) \simeq 0.05$ is actually dominated by neighboring large negative sensitivity regions $S_\mathrm{EIT}({\bf v}) \sim -0.3$. The special panel ${\bf (c^\prime)}$ of the figure demonstrates the effect of large $\Omega_D$, verifying additional effective Doppler suppression via spectral broadening.

As in the 1D case, away from the Doppler point, the scale of signal sensitivity variation with ${\bf v}$ is unchanged, but the geometry of the positive and negative regions changes dramatically, with $S_\mathrm{EIT}^\mathrm{th} \to 0$ through a combination of cancellation between regions and shrinkage of the near-resonant region. It is observed numerically that the cancellation effect operates more strongly here than in the 1D case (especially compared the left side of the bottom panel of Fig.\ \ref{fig:doppler1D} where it is essentially absent).

Note that even as the 1D configuration has a built in $O(10^{-2})$ Doppler disadvantage, it is simpler to set up and has geometric advantages, such as easy to achieve larger interaction region length scale $L$ and conveniently overlapping laser beams, that have made it the dominant experimental focus. The active vapor volume could then be straightforwardly increased by lengthening the cell to gain a factor of $O(10)$ over the result in Fig.\ \ref{fig:doppler1D}. As stated earlier, similar 2D setup gains require greatly broadening the $\sim$\,1 mm$^3$ intersecting beam volume.

\begin{figure}

\includegraphics[width=3.0in,viewport = 10 10 660 540,clip]{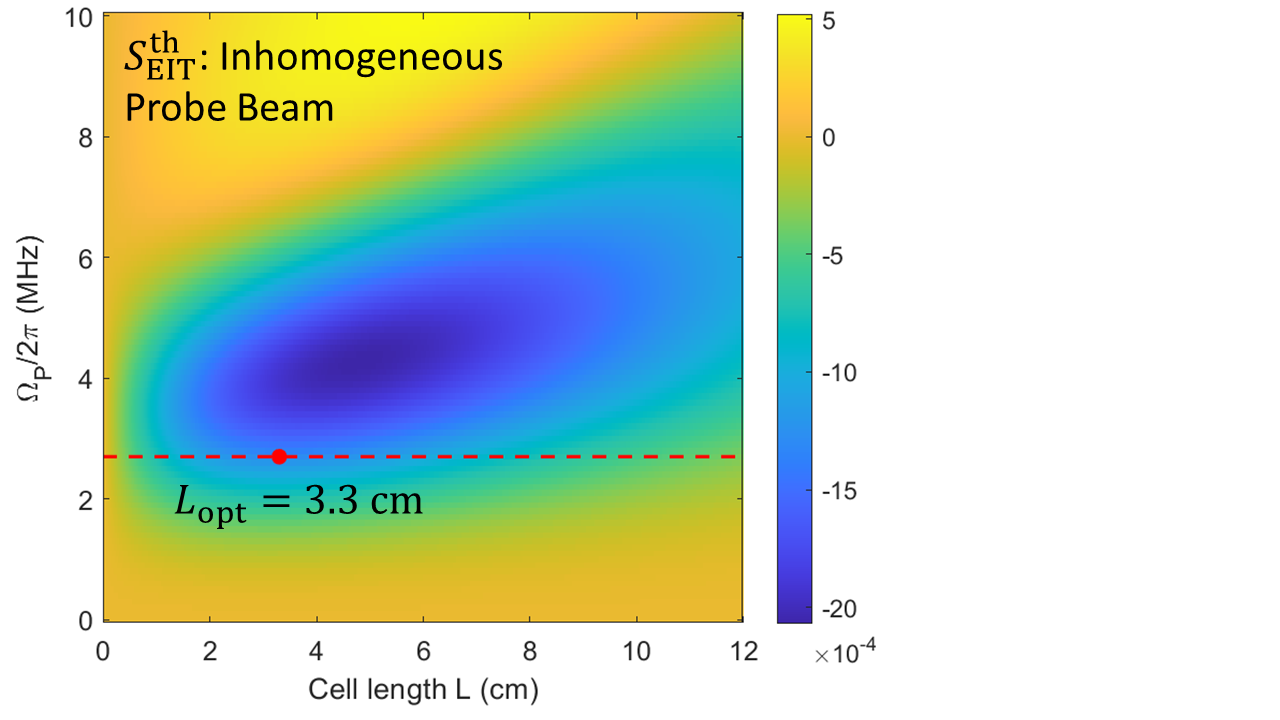}
\includegraphics[width=3.0in,viewport = 10 10 660 540,clip]{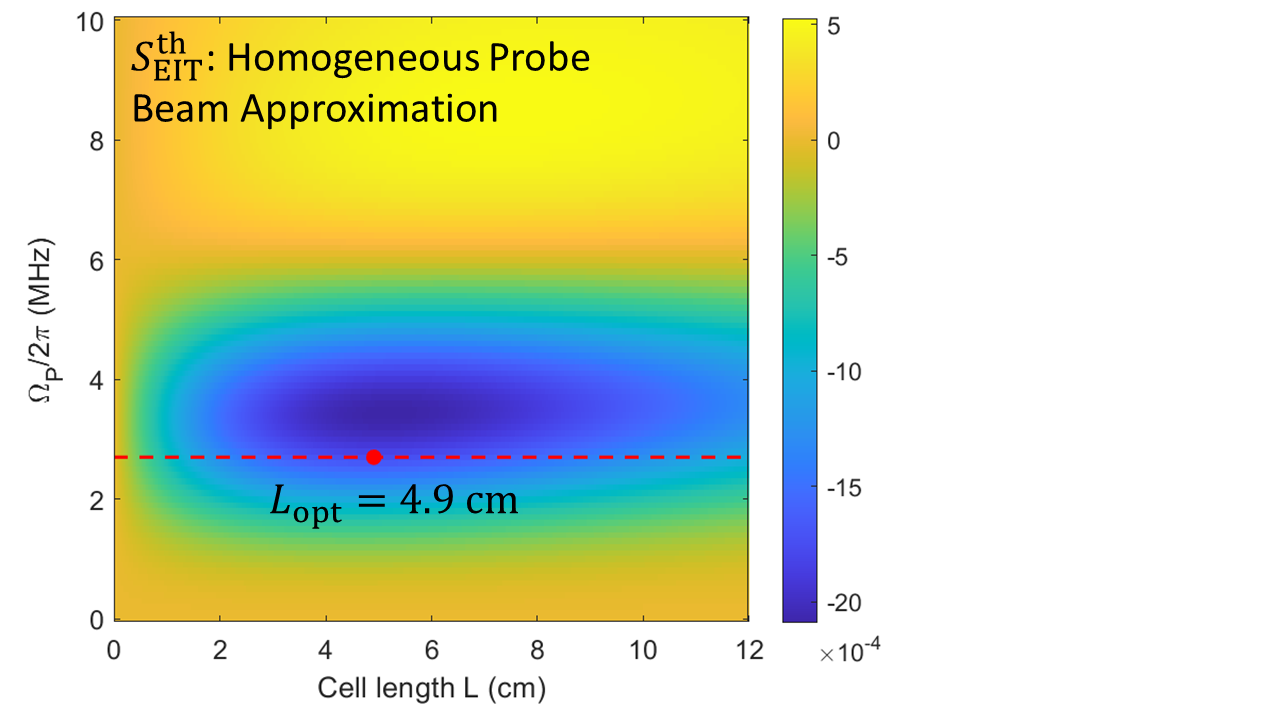}

\caption{Linear response sensitivity (Mrad/s)$^{-1}$ vs.\ Probe beam amplitude $\Omega_P \equiv \Omega_P(0)$ and cell length $L$. Parameters are otherwise as in Fig.\ \ref{fig:doppler1D}. \textbf{Above:} Full inhomogeneous solution (\ref{3.21}) using $\Omega_P(s)$ obtained from numerical solution of the coupled ODEs (\ref{3.25}). \textbf{Below:} Homogeneous approximation using uniform $\Omega_P(s) = \Omega_P$. The red dashed line highlights the result for $\Omega_P/2\pi = 2.7$ MHz for which rather different optimal cell lengths (red dots) are found.}

\label{fig:SenseInhomog}
\end{figure}

Figure \ref{fig:doppler2D_DeltaC} shows an example of the strong dependence of the Doppler-enhanced EIT signal on other parameter settings, in this case linear response sensitivity as a function of both coupling laser angle $\theta_C$ and detuning $\Delta_C$. The line shapes (horizontal transects) are extremely sensitive to the $\Delta_C$ value. Since all other detuning values are taken to vanish, the curves are even functions of $\Delta_C$. The largest peak sensitivity magnitude is at $|\Delta_C|/2\pi \simeq 0.25$ MHz, and fades rapidly for $|\Delta_C|/2\pi > 1$ MHz.

Unlike increasing $\Omega_D$ (and/or $\Omega_C$) such parameter variations do not demonstrate any strong enhancement of $|S_\mathrm{EIT}|$, only strong variability of detailed line shapes. However it is clear that there are additional $O(1)$ gain factors available through optimization within this space if desired.

\begin{figure}

\includegraphics[width=3.3in,viewport = 0 0 870 540,clip]{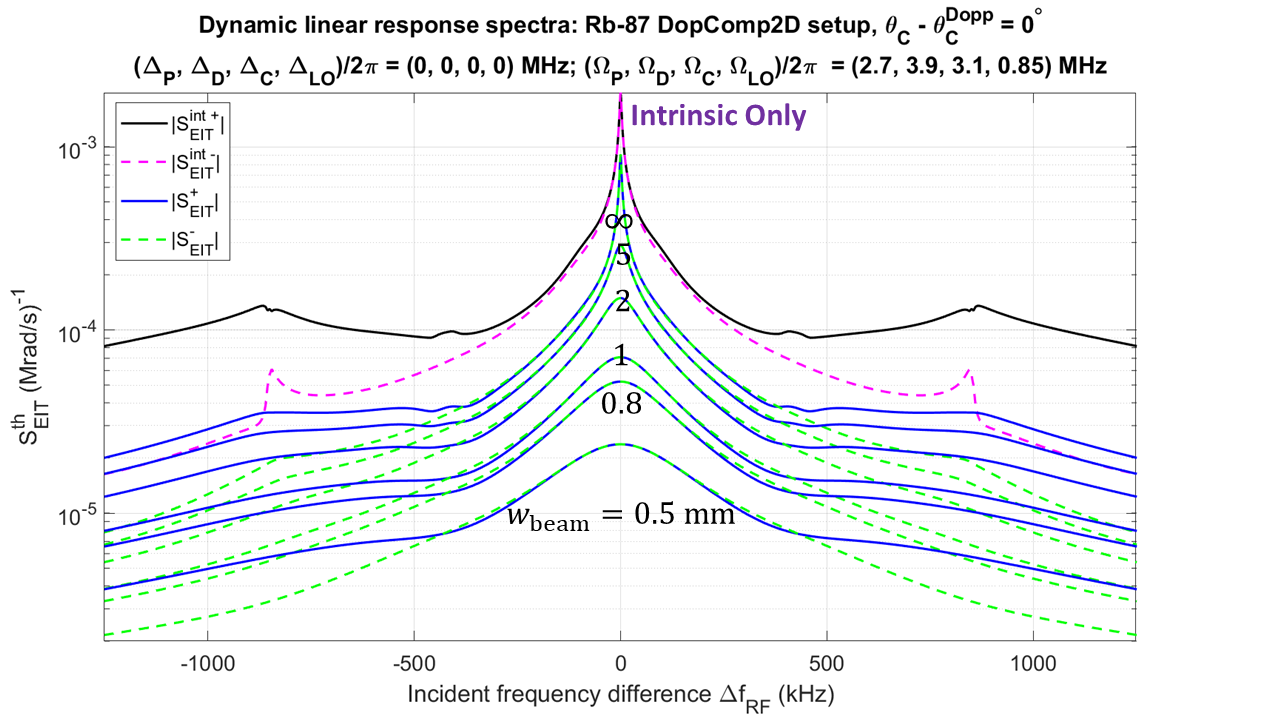}

\includegraphics[width=3.3in,viewport = 0 0 870 540,clip]{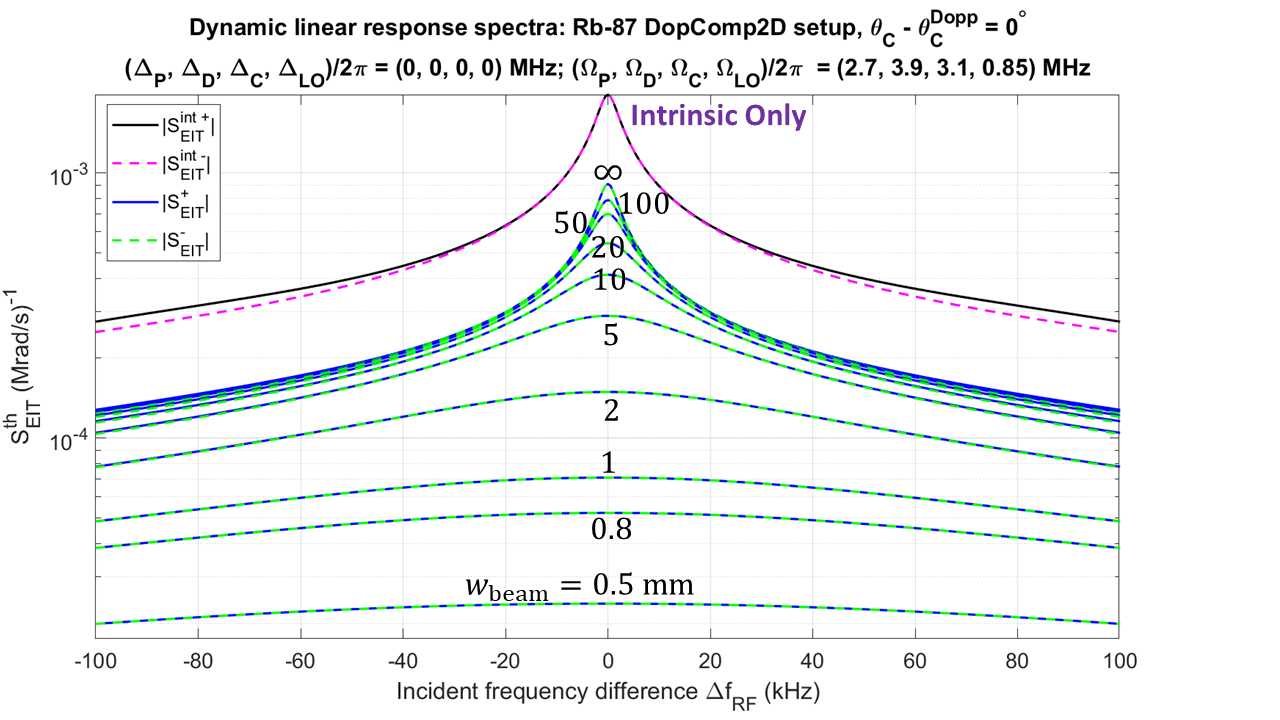}

\caption{Demonstration of the Rydberg cavity picture for the star configuration Fig.\ \ref{fig:starconfig} (hence $\theta_C = \theta_C^\mathrm{Dopp}$). Plotted are the frequency-dependent EIT linear response sensitivity magnitudes $|S_\mathrm{EIT}^{\mathrm{th},\pm}|$ (\ref{A28}) for a sequence of beam diameters 0.5 mm $\leq w \leq \infty$ using the usual illumination parameters shown in the plot titles. The adiabatic limit (\ref{2.6}) in each case corresponds to the peak at $\Delta f_\mathrm{RF} = 0$ (RF frequency identical to the transition frequency between Rydberg levels). \textbf{Top:} Expanded frequency range $|\Delta f_\mathrm{RF}| \leq 1.25$ MHz. Labeled beam diameters are all in mm, with $w = \infty$ corresponding to vanishing transit time broadening effect, $\gamma_\mathrm{Tr} = 0$. For all green and blue curves, the remaining decoherence parameters are as listed in the last three lines of Table \ref{tab:nonintrinsdecay}. For the uppermost ``intrinsic only'' curves (black, magenta) the latter are taken to vanish as well, with only the intrinsic decay rates listed in the caption to Fig.\ \ref{fig:rysetup} accounted for in the model. The latter clearly set the $\sim$\,20 kHz minimum achievable linewidth \textbf{Bottom:} Identical curves over the reduced frequency range $|\Delta f_\mathrm{RF}| \leq 100$ kHz. A few additional $w$ values are now included for clarity.}

\label{fig:lrfreq2Dstar}
\end{figure}

\begin{figure}

\includegraphics[width=3.3in,viewport = 0 10 850 530,clip]{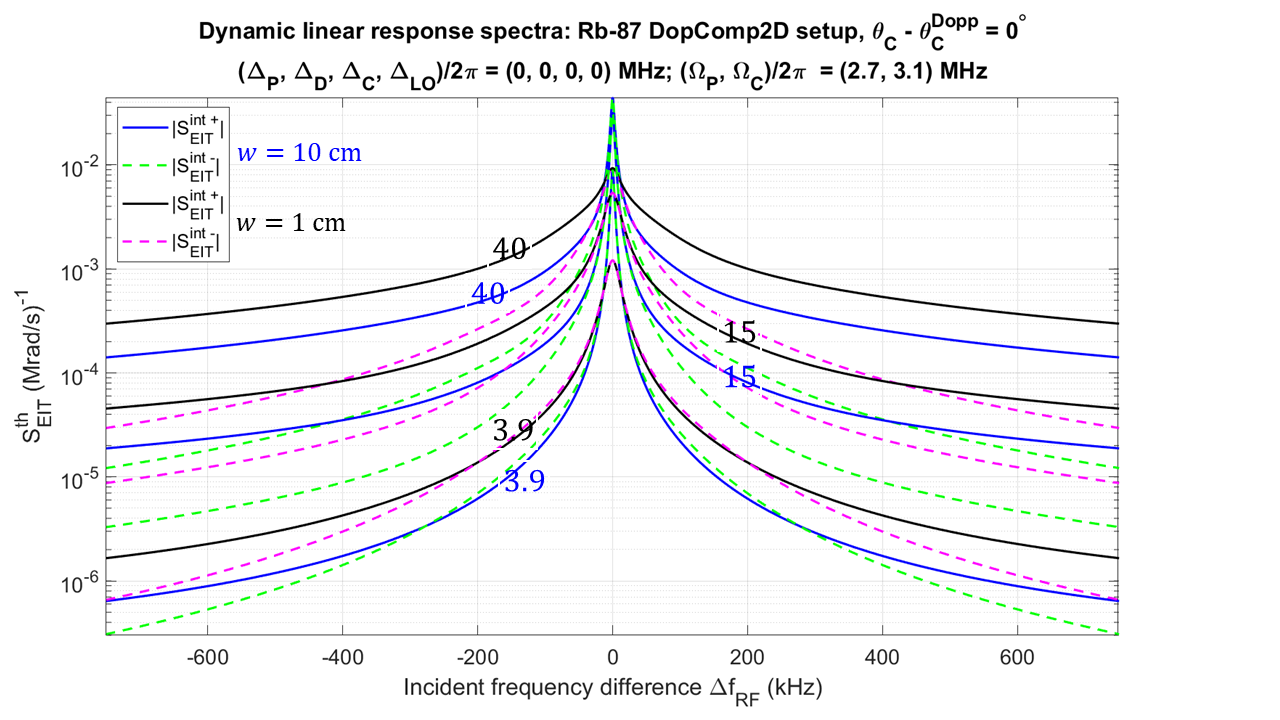}
\includegraphics[width=3.3in,viewport = 0 10 850 530,clip]{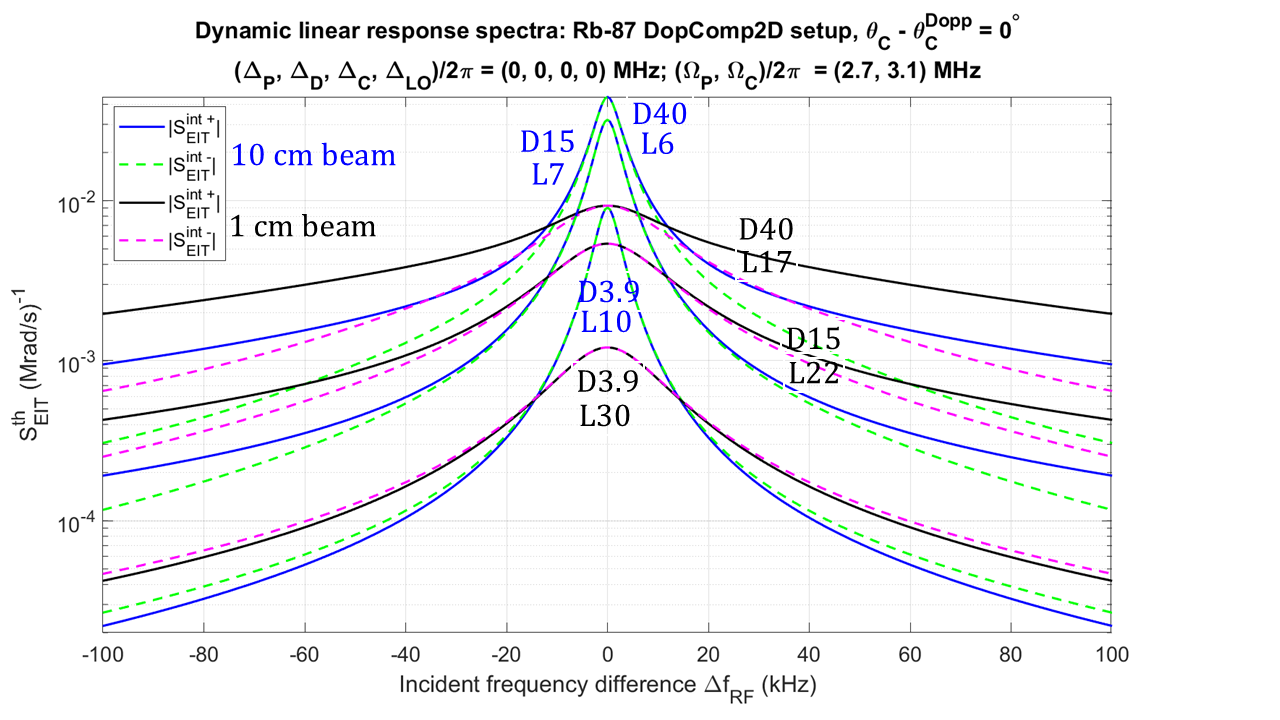}

\caption{Demonstration of the Rydberg cavity picture for the star configuration, similar to Fig.\ \ref{fig:lrfreq2Dstar}, but now in the limit of small local oscillator amplitude $\Omega_\mathrm{LO}$. Shown are selected dynamic linear response curves (frequency range $|\Delta f_\mathrm{RF}| \leq 750$ kHz in the upper panel, $|\Delta f_\mathrm{RF}| \leq 100$ kHz in the lower panel) for beam widths $w = 1$ cm (black solid and cyan dashed curves) and $w = 10$ cm (blue solid and green dashed curves). For each $w$, results are shown for dressing laser amplitudes $\Omega_D/2\pi = 3.9, 15, 40$ MHz as indicated by the equivalently colored numbers in the upper panel, and by the ``D'' label values in the lower panel. For each of the six cases, the values $\Omega_\mathrm{LO}/2\pi = 6, 7, 10, 17, 22, 30$ kHz (associated with the ``L'' label in the lower panel) is chosen near the peak of the corresponding curve in the middle and lower panels of Fig.\ \ref{fig:SensOmegaLO}. It is seen that the central peak widths ($\sim$\,40 kHz for $w = 1$ cm and $\sim$\,20 kHz for $w = 10$ cm) are essentially the same as in Fig.\ \ref{fig:lrfreq2Dstar} despite the much larger $\Omega_\mathrm{LO}/2\pi = 850$ kHz value used there.}

\label{fig:lrfreq2DstarSmallLO}
\end{figure}

\subsection{Weak local oscillator ``Rydberg cavity'' regime}
\label{sec:weakLORyCavity}

We now arrive at the second main message of this paper. Motivated by experiments, the previous results all used moderate LO amplitude values $\Omega_\mathrm{LO}/2\pi \sim 1$ MHz. However, if one focuses on the lower right $2 \times 2$ block in the Hamiltonian (\ref{3.5}), one might expect strong resonance effects for $\Omega_\mathrm{LO}$ comparable to the difference $\Delta_5 - \Delta_4 = \Delta_\mathrm{LO}$. Thus, the energy levels
\begin{eqnarray}
\epsilon_\pm &=& \frac{\Delta_5 + \Delta_4}{2}
\pm \frac{1}{2} \sqrt{\Delta_\mathrm{LO}^2 + |\Omega_\mathrm{LO}|^2}
\nonumber \\
&=& \frac{1}{2} \left(\Delta_\mathrm{LO}
\pm \sqrt{\Delta_\mathrm{LO}^2 + |\Omega_\mathrm{LO}|^2} \right),
\label{4.2}
\end{eqnarray}
in which the second line specializes to the recoil-free condition $\Delta_4 = 0$, are strongly hybridized in this regime. Of course, the core state coupling $\Omega_C$ and Rydberg level decay parameters (listed in the caption to Fig.\ \ref{fig:rysetup}) must influence any such conclusion, but to the extent that $\Omega_C$ may be viewed as an external driving of an otherwise isolated, high-q Rydberg state ``cavity,'' one may expect strong variations in the character of the full core-plus-Rydberg atomic superposition state in this regime, ultimately feeding into the Probe laser EIT response. For the sensor application, the desired sharp response will directly reflect the perturbation of $\Omega_\mathrm{LO}$ by the incident RF signal.

This expectation is fully confirmed in Fig.\ \ref{fig:SensOmegaLO}, which shows results for a much broader range of LO amplitudes, including very small values---now focused on $\Delta_4 = 0$, equivalent to the star configuration constraint $\theta_C = \theta_C^\mathrm{Dopp}$. The critical role of transit time broadening $\gamma_\mathrm{Tr}$ (see Table \ref{tab:nonintrinsdecay}) is demonstrated by comparing results for beam widths $w = 1$ mm ($\gamma_\mathrm{Tr}/2\pi = 73$ kHz), $w = 1$ cm ($\gamma_\mathrm{Tr}/2\pi = 7.3$ kHz), and $w = 10$ cm ($\gamma_\mathrm{Tr}/2\pi = 0.73$ kHz). Similar to Fig.\ \ref{fig:doppler2D}, the different curves in each panel demonstrate the advantage of increasing dressing laser amplitude $\Omega_D$, with extrema initially growing roughly linearly with $\Omega_D$, but eventually saturating at large values. For vanishing detunings $\Delta_\alpha$ used here, symmetry dictates that $P^\mathrm{th}_\mathrm{EIT}$ be an even function of $\Omega_\mathrm{LO}$, hence $S^\mathrm{th}_\mathrm{EIT} \to 0$ for $\Omega_\mathrm{LO} \to 0$.

Consider first the $w = 10$ cm case. Such a large beam width is clearly unlikely to be experimentally viable, and is shown here only to highlight the resonant structure in the limit where $\gamma_\mathrm{Tr}$ is negligible compared to all other Rydberg level decay and dephasing mechanisms listed in Table \ref{tab:nonintrinsdecay}. The very narrow $\sim$\,10 kHz wide peak centered on $\Omega_\mathrm{LO}/2\pi \simeq 7$ kHz is controlled by these other mechanisms. The peak linear response sensitivity value (for given $\Omega_D$) is enhanced by an order of magnitude compared to experimental regimes $\Omega_\mathrm{LO}/2\pi \sim 1$ MHz characteristic of typical 1D laser setups. The combination of small $\Omega_\mathrm{LO}$ and large $\Omega_D$ (compared to $\Omega_D/2\pi \sim 5$ MHz typical values) leads to a net $O(10^2)$ enhancement.

The $w = 1$ mm case shows the opposite limit in which $\gamma_\mathrm{Tr}$ completely dominates the Rydberg level linewidth. The overall (vertical) sensitivity scale of the plot is about ten times smaller than the $w = 10$ cm case. Although for large $\Omega_D/2\pi \agt 60$ MHz there is a growing sensitivity enhancement at lower $\Omega_\mathrm{LO}/2\pi \approx 200$ kHz (as well as some interesting finer scale structure below 70 kHz), for smaller $\Omega_D$ the maximum magnitude actually occurs at more conventional values $\Omega_\mathrm{LO}/2\pi \sim 700$ kHz. Note here again that we use cell length $L = 1$ cm for all three cases. As commented above, if all three beams are taken as 1 mm, the sensitivity would drop by another order of magnitude. Thus complete absence of beam broadening again leads to much worse sensing outcomes than shown in the subplot.

Finally, the $w = 1$ cm case, with $\gamma_\mathrm{Tr}$ comparable to other decay parameters, is observed to sit somewhere in between. The resonant structure is clearly evident, with somewhat broader $\sim$\,30 kHz peak widths centered on $\Omega_\mathrm{LO}/2\pi \simeq 20$ kHz. The peak linear response sensitivity values (for given $\Omega_D$) are about 5 times smaller than for the 10 cm beam, but still about 3 times larger than for larger $\Omega_\mathrm{LO}/2\pi \sim 1$ MHz. The combination of small $\Omega_\mathrm{LO}$ and large $\Omega_D$ leads to a net $O(30)$ enhancement. Interestingly, the opposite sign peaks at intermediate values $\Omega_\mathrm{LO}/2\pi \sim 0.35$ MHz have nearly the same sensitivity magnitude, and differ by only $\sim$\,20\% from the corresponding $w = 10$ cm peak values. This sign change pattern is clearly common to all beam widths, though the detailed structure varies strongly with $w$.

We have interpreted the adiabatic sensitivity limit results here in terms of effective resonant cavity behavior. More explicit support for this interpretation will be presented in Sec.\ \ref{sec:Ryfinitefresp} where the full dynamical response is computed. There the spectral linewidths will seen to be controlled by the spontaneous decay rates of the Rydberg states, with the peak occurring when the Rabi period $1/\Omega_\mathrm{LO}$ becomes comparable to the intrinsic lifetime of the resonant Rydberg cavity.

\begin{figure}

\includegraphics[width=3.3in,viewport = 0 0 880 540,clip]{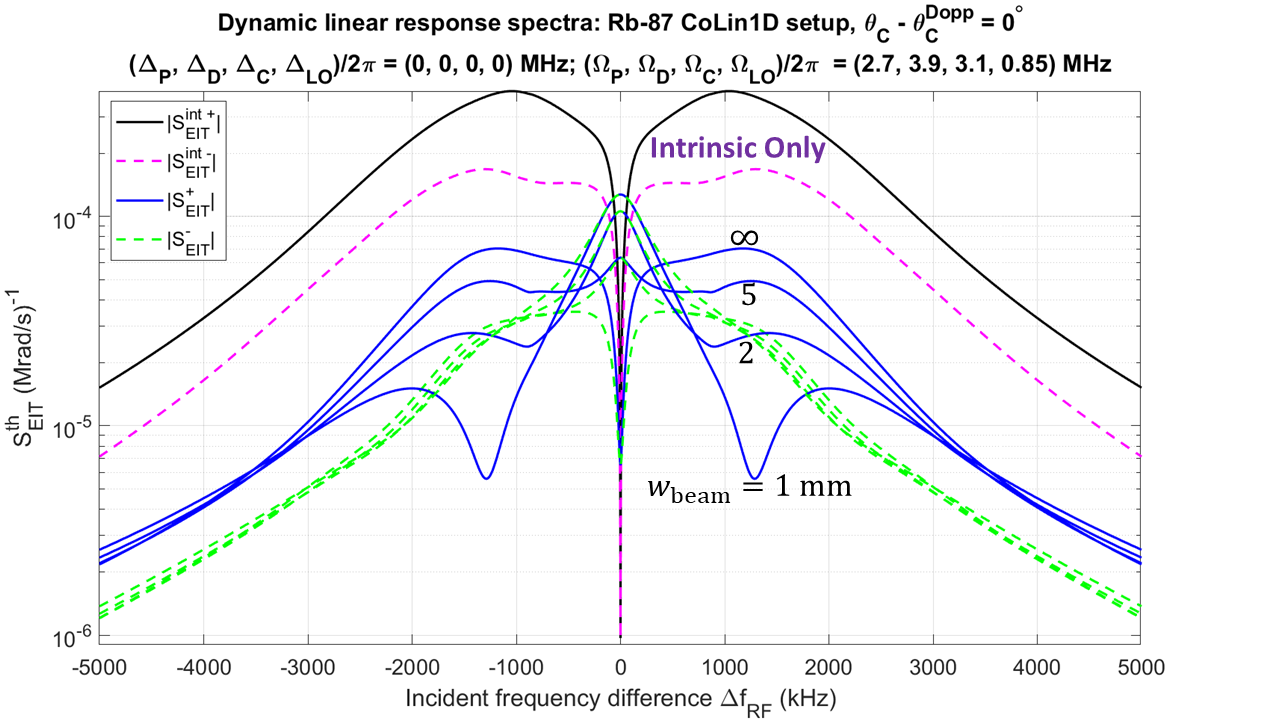}

\includegraphics[width=3.3in,viewport = 0 0 880 540,clip]{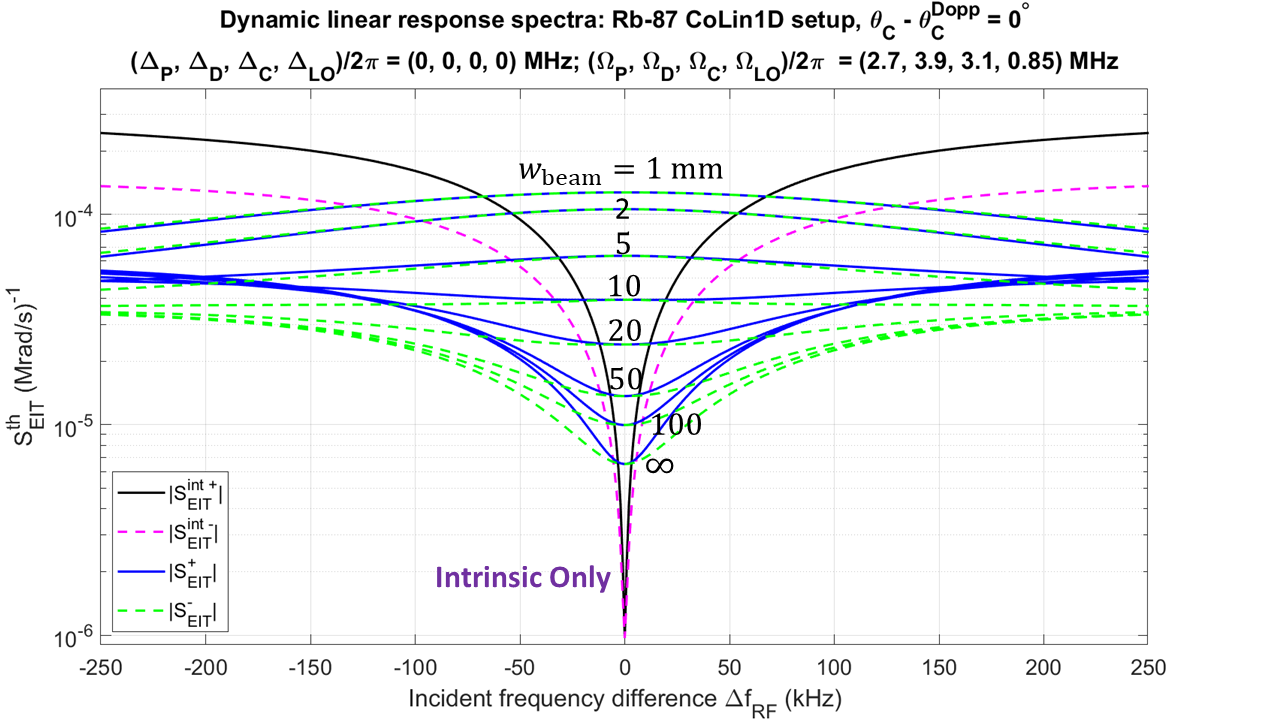}

\caption{Demonstration of the `failure' of the Rydberg cavity picture for the same linear configuration used for Fig.\ \ref{fig:doppler1D}). Plotted are the frequency-dependent EIT linear response sensitivity magnitudes $|S_\mathrm{EIT}^{\mathrm{th},\pm}|$ (\ref{A28}) for a sequence of beam diameters 1 mm $\leq w \leq \infty$ using the illumination parameters shown in the titles, and otherwise using the same parameters as in Fig.\ \ref{fig:lrfreq2Dstar}. The adiabatic limit (\ref{2.6}) again corresponds to $\Delta f_\mathrm{RF} = 0$. Although the sensitivity has strong frequency dependence, the overall linewidth is in the several MHz range. \textbf{Top:} Expanded frequency range $|\Delta f_\mathrm{RF}| \leq 5$ MHz. Labeled beam diameters are all in mm, with $w = \infty$ corresponding to vanishing transit time broadening effect. For all green and blue curves, the remaining decoherence parameters are as listed in the last three lines of Table \ref{tab:nonintrinsdecay}. For the uppermost ``intrinsic only'' curves (black, magenta) the latter are taken to vanish as well, with only the intrinsic decay rates listed in the caption to Fig.\ \ref{fig:rysetup} accounted for in the model. The sensitivity ``notch'' near zero frequency for large $w$ is due to fine-scale interference between differently moving atom sub-populations. \textbf{Bottom:} Identical curves over the reduced frequency range $|\Delta f_\mathrm{RF}| \leq 250$ kHz, exhibiting details of the notch behavior for $w > 20$ mm or so. Additional $w$ values are now included for clarity.}

\label{fig:lrfreq1Dcolin}
\end{figure}

\subsection{Inhomogeneous Probe beam}
\label{sec:inhomogprobe}

Finally we return to consider inhomogeneous probe beams. Figure \ref{fig:SenseInhomog} shows exact and approximate linear response sensitivity results vs.\ input Probe beam amplitude and cell length. The differences are not huge, but for given $\Omega_P(0)$ the two can give very different predictions for the maximum sensitivity point, highlighted here for $\Omega_P(0)/2\pi = 2.7$ MHz [in this case within the linear response regime described by (\ref{3.26}) and (\ref{3.27})]. In either case, one can more than double the sensitivity by going to considerably longer than $L = 1$ cm. Of course there are practical issues (e.g., beam diameter requirements) that may make this challenging. Although the scale of the linear response sensitivity does not change dramatically in the larger $\Omega_P$ optically nonlinear regime, the photon count, scaling as $|\Omega_P|^2$, does increase substantially. As discussed in Sec.\ \ref{sec:rysnr}, measurement SNR thereby increases linearly with $|\Omega_P|$ (to a point), so that larger values proportionately improve the minimum detectable signal.

\section{Rydberg antenna finite frequency response}
\label{sec:Ryfinitefresp}

We now turn to examples of the full finite frequency linear response function $S_\mathrm{EIT}^\mathrm{th}(\omega)$ defined in Sec.\ \ref{sec:siglinresp} and constructed in App.\ \ref{app:dynresp}. Its computation generalizes the dynamical matrix inverse relation (\ref{3.15}) and (\ref{3.17}) to the finite frequency form (\ref{A11}) and (\ref{A13}) and leading to the EIT response (\ref{A26})--(\ref{A30}).

\subsubsection{Verification of the Rydberg cavity picture for the 2D star configuration}
\label{subsec:2DRycavityverify}

We begin with example results, shown in Fig.\ \ref{fig:lrfreq2Dstar}, for the 2D star configuration setup, $\theta_C = \theta_C^\mathrm{Dopp}$, with moderate $\Omega_\mathrm{LO}/2\pi = 0.85$ MHz. The underlying adiabatic limit velocity spectrum hence corresponds to that plotted in Fig.\ \ref{fig:vspec2d}(c). Due to the structure of the Stark operator (\ref{A2}), with the operators $\hat V_0$ and $\hat V_0^\dagger$ generating transitions $|4\rangle \to |5\rangle$ and $|5\rangle \to |4\rangle$, respectively, there are corresponding positive and negative frequency contributions $S_\mathrm{EIT}^{\mathrm{th} \pm}(\omega)$, defined in (\ref{A30}), to the time-domain EIT response (\ref{A29}), plotted separately. Similar to Fig.\ \ref{fig:SensOmegaLO}, the focus is on exhibiting the dependence of $S_\mathrm{EIT}^\mathrm{th}(\omega)$ on Rydberg level line broadening effects, especially transit time broadening. The two subplots show essentially the same results on two different frequency scales. The observed multi-time-scale content reflects the spectrum of eigenvalues $\lambda_n$ of the dynamical matrix ${\bf G}$ in (\ref{3.8}), that then also characterize the linear response spectral matrix ${\bf R}(\omega)$ in (\ref{A16}).

The sharp behavior of $S_\mathrm{EIT}^\mathrm{th}(\omega)$ for small $|\omega|$ therefore reflects the smallest magnitude $\lambda_n$, in turn representing the slowest atomic processes. In the five-state model considered here, the latter are the decay and dephasing processes listed in Table \ref{tab:nonintrinsdecay}. In the plots, the effect of transit time broadening $\gamma_\mathrm{tr} \simeq 73/w[\mathrm{mm}]$ kHz is again exhibited through dependence of the spectrum on the beam width $w$. It is seen that for increasing $w \agt 2$ mm the line shapes become increasingly sharp in the regime $|f_\mathrm{RF}| \alt 30$ kHz, and the peak (adiabatic) sensitivity increasingly large. At the same time it is clear that highest sensitivity is directly limited to correspondingly low bandwidth. This limits detection of weak high bandwidth signals, but not in a way intrinsically different from the operation of finely tuned (e.g., impedance matched) classical antennas.

For $w \agt 50$ mm the central peak saturates as $\gamma_\mathrm{tr} \alt 1$ kHz falls substantially below the other decay values listed in Table \ref{tab:nonintrinsdecay}. The curves labeled ``Intrinsic Only'' correspond to zeroing out the latter values as well, leaving only the intrinsic decay rates $(\gamma_{4 \to 3}, \gamma_{5 \to 2})/2\pi = (1.64, 1.09)$ kHz quoted in the caption to Fig.\ \ref{fig:rysetup}. Interestingly, though the peak value is multiplied by a factor of $\sim$\,2, the shape of the central peak is almost unchanged, showing that the adiabatic limit density matrix ${\bm \rho}^\mathrm{ad}$ factor appearing in (\ref{A16}) is more sensitive to these parameters than are the eigenvalues themselves.

In Fig.\ \ref{fig:lrfreq2DstarSmallLO} we show results for small $\Omega_\mathrm{LO}$, close to the peaks in Fig.\ \ref{fig:SensOmegaLO}, for a representative combination of beam widths $w = 1, 10$ cm and Dressing laser amplitudes $\Omega_D/2\pi = 3.9, 15, 40$ MHz. The central peak line widths are essentially the same as seen in  Fig.\ \ref{fig:lrfreq2Dstar}. The major difference here is the much larger central peak height above the broader ``background'' peaks that are much more evident in Fig.\ \ref{fig:lrfreq2Dstar}. It follows that the larger $\Omega_\mathrm{LO}$ value may be thought of as more weakly driving the Rydberg cavity response relative to the spectrum of broader resonant responses. The small $\Omega_\mathrm{LO}$ limit therefore succeeds in greatly increasing the adiabatic response through greatly improved coupling to the Rydberg cavity. On the other hand the response now drops off much more rapidly with frequency, confirming that enhanced DC sensitivity comes at the expense of greatly diminished broad band sensitivity. This serves to further inform sensor parameter choices depending on application.

\subsubsection{`Failure' of the Rydberg cavity picture for the 1D setup}
\label{subsec:1DRycavityfail}

Finally, to again demonstrate that the 2D star configuration is required to exploit the Rydberg resonant cavity effect, we show in Fig.\ \ref{fig:lrfreq1Dcolin} dynamic linear response results for the same collinear setup used to generate Fig.\ \ref{fig:doppler1D} over a similar range of parameters. As seen, although the sensitivity has strong frequency dependence, the overall linewidth is in the several MHz range. For large beam width $w \agt 2$ cm a sensitivity ``notch'' develops near zero frequency. This is not a small eigenvalue effect, but rather a sensitivity near-cancelation due to fine-scale interference between differently moving atom sub-populations. This provides more insight into the observation in Fig.\ \ref{fig:doppler1D} that the adiabatic sensitivity actually decreases with beam width at the physical Coupling laser wavenumber. The maximum response magnitude is now actually seen to be at finite frequency $\Delta f_\mathrm{RF} \sim 1$ MHz.

\section{Rydberg receiver SNR}
\label{sec:rysnr}

We now turn to a discussion of Rydberg sensor signal-to-noise considerations, paralleling that for classical antennas in Sec.\ \ref{sec:qvsclasssense}. In the regimes of interest here the experiments are dominated by shot noise, i.e., the ability to accurately extract small changes in mean Probe beam photon count rates from discrete data, rather than instrument thermal noise. The signal-to-noise ratio in this case is
\begin{equation}
\mathrm{SNR}_\mathrm{Ry} = \frac{\delta \bar n_P}
{\sqrt{\langle (n_P - \bar n_P)^2 \rangle}}
\label{6.1}
\end{equation}
in which the numerator is the change in Probe beam mean photon count due to the incident RF signal and the denominator is the standard deviation of the total photon count. The total count is given by
\begin{equation}
\bar n_P = \frac{\epsilon_D I_P A_D \tau}{\hbar \omega_P} {\cal P}^\mathrm{th}_\mathrm{EIT}
\label{6.2}
\end{equation}
in which $\epsilon_D$ is the detector efficiency (exceeding 90\% for a good detector), $I_P$ is the probe beam intensity (power per unit area), $A_D$ is the area of the beam intersecting the detector and $\tau$ is the measurement time. Using the linear response form (\ref{2.5}) one obtains the count change
\begin{equation}
\delta \bar n_P = \bar n_P
\frac{\Omega^0_\mathrm{in} S^\mathrm{th}_\mathrm{EIT}}
{{\cal P}^\mathrm{th}_\mathrm{EIT}}.
\label{6.3}
\end{equation}
Applying Poisson statistics one obtains the variance
\begin{equation}
\langle (n_P - \bar n_P)^2 \rangle = \bar n_P,
\label{6.4}
\end{equation}
which finally yields
\begin{equation}
\mathrm{SNR}_\mathrm{Ry} = \sqrt{\bar n_P}
\frac{\Omega^0_\mathrm{in} S^\mathrm{th}_\mathrm{EIT}}
{{\cal P}^\mathrm{th}_\mathrm{EIT}}.
\label{6.5}
\end{equation}
Implicit in this formula is that the time average includes an appropriate Fourier transform filter that compensates for the $e^{-i \Delta \omega_\mathrm{in} t}$ time-dependence, which would otherwise essentially zero out the average. The primary difference between (\ref{6.5}) and the transmission perturbation $\Omega^0_\mathrm{in} S^\mathrm{th}_\mathrm{EIT}$ is the additional $\sqrt{\bar n_P}$ factor, which like the classical antenna electric field result (\ref{1.6}), yields a $\sqrt{\tau}$ SNR improvement with measurement time.

\begin{figure}

\includegraphics[width=3.0in,viewport = 10 10 670 540,clip]{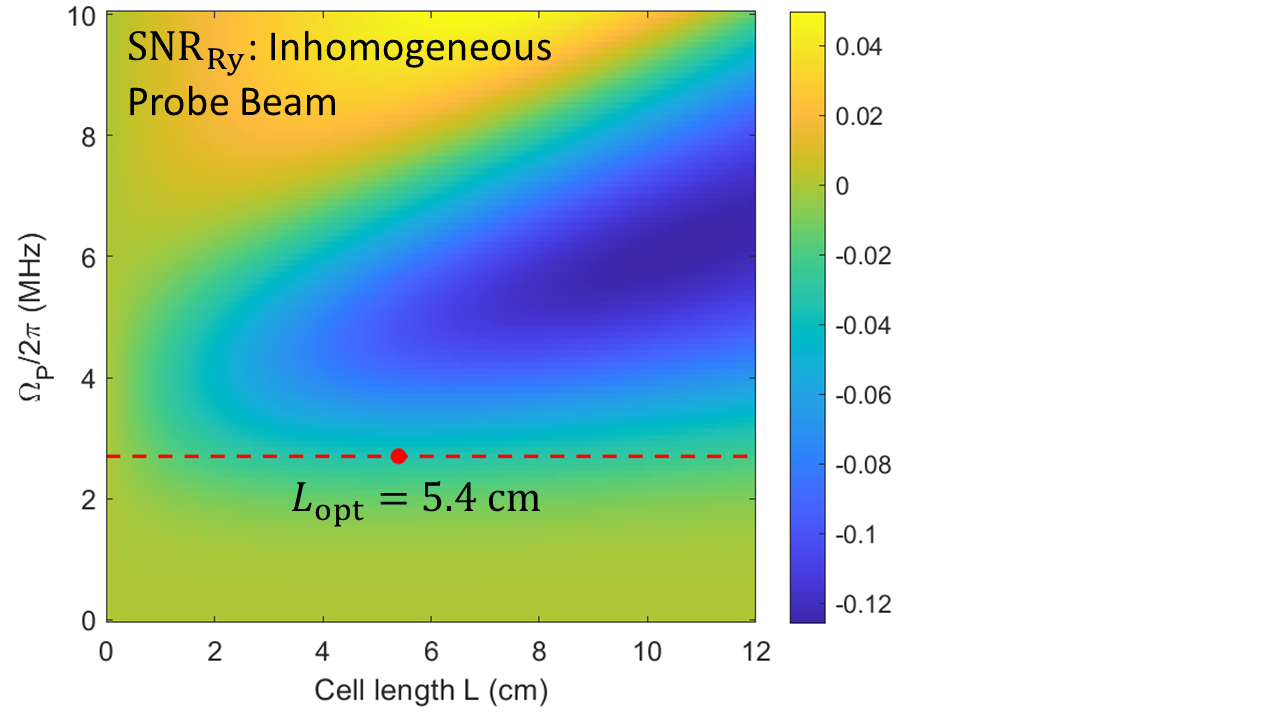}
\includegraphics[width=3.0in,viewport = 10 10 660 540,clip]{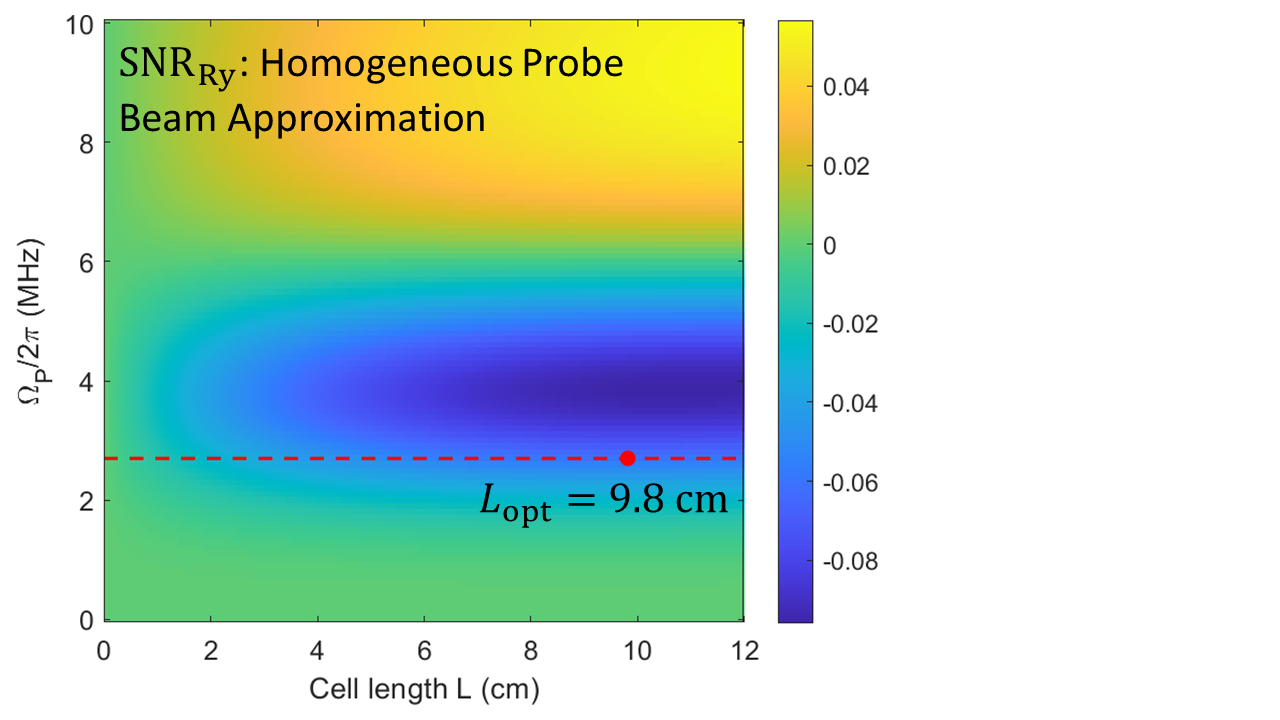}

\caption{SNR linear response sensitivity measure (\ref{6.9}) vs.\ Probe beam amplitude $\Omega_P \equiv \Omega_P(0)$ and cell length. Parameters are otherwise as in Fig.\ \ref{fig:doppler1D}. \textbf{Above:} Full inhomogeneous solution using $\Omega_P(s)$ obtained from (\ref{3.25}). \textbf{Below:} Homogeneous approximation using uniform $\Omega_P(s) = \Omega_P$. The red dashed lines highlight the result for a typical experimental value $\Omega_P/2\pi = 2.7$ MHz. Although the SNR measure here produces significantly different optimal cell lengths (red dots) compared to the sensitivity measure results shown in Fig.\ \ref{fig:SenseInhomog}, the numerical values are again rather weak functions of cell length beyond $L = 2$ cm or so.}

\label{fig:SNRInhomog}
\end{figure}

In order to express this result in physical units, we use the Poynting vector result (\ref{1.2}) and the Rabi frequency relation (\ref{2.2}) to estimate
\begin{equation}
E_P = \sqrt{2 Z_0 I_P} \approx \frac{\hbar\Omega_P}{|{\bf d}_{12}|},\ \
E_\mathrm{in} \approx \frac{\hbar \Omega^0_\mathrm{in}}{|{\bf d}_{45}|}.
\label{6.6}
\end{equation}
Scaling by physically motivated parameter values, we define first the $O(1)$ dimensionless ratio
\begin{equation}
{\cal R}_\mathrm{EIT} = \sqrt{\frac{\epsilon_D}{{\cal P}^\mathrm{th}_\mathrm{EIT}}
\frac{A_D}{1\ \mathrm{cm}^2} \frac{300\ \mathrm{THz}}{f_P}},
\label{6.7}
\end{equation}
in which we note that $\epsilon_D$ and ${\cal P}_\mathrm{EIT}$ are comparable and each slightly less than unity. One obtains finally
\begin{eqnarray}
\mathrm{SNR}_\mathrm{Ry} &=& 81.71\, {\cal R}_\mathrm{EIT}
\frac{E_\mathrm{in}}{1\ \mu\mathrm{V/cm}}
\frac{2\pi S_\mathrm{EIT}^\mathrm{th}}{1\ \mathrm{ns}}
\nonumber \\
&&\times\ \frac{\Omega_P}{2\pi(1\ \mathrm{MHz})}
\frac{|{\bf d}_{45}|}{10^3 |{\bf d}_{12}|} \sqrt{\frac{\tau}{1\ \mathrm{s}}}
\label{6.8}
\end{eqnarray}
where the linear response sensitivity factor is scaled here to be unity for $S_\mathrm{EIT}^\mathrm{th} = 10^{-3}$ (Mrad/s)$^{-1} = (2\pi)^{-1} \times 10^{-9}$ Hz$^{-1}$. The ratio $|{\bf d}_{45}|/|{\bf d}_{12}| \approx n^2$ scales quadratically with the principal quantum number, hence the corresponding factor will exceed unity for $n \agt 32$.

Example results are shown in Fig.\ \ref{fig:SNRInhomog} in which the dimensionless ``SNR sensitivity'' measure
\begin{equation}
{\cal S}_\mathrm{Ry} \equiv
\frac{\Omega_P S^\mathrm{th}_\mathrm{EIT}}{\sqrt{{\cal P}^\mathrm{th}_\mathrm{EIT}}},
\label{6.9}
\end{equation}
is plotted vs.\ Probe amplitude and cell length, and corresponds to the coefficient of $E_\mathrm{in}$ in (\ref{6.8}) with all other fixed experimental parameters dropped. The general pattern is quite similar to the unscaled linear response sensitivity result in Fig.\ \ref{fig:SenseInhomog}. The major quantitative effect is to push the optimum values to larger $\Omega_P$ [due to the latter's appearance in the numerator in (\ref{6.9})] and larger $L_\mathrm{cell}$ (which decreases ${\cal P}^\mathrm{th}_\mathrm{EIT}$ in the denominator). In both figures, the sensitivity numerical values are rather weak functions of cell length beyond roughly $L = 2$ cm.

\subsection{Comparison with classical wire antenna}
\label{sec:wirecompare}

We finally compare the predictions of (\ref{6.8}), obtained by setting $\mathrm{SNR_{Ry}} = 1$, with the classical wire antenna E-field sensitivity estimate (\ref{1.6}). Direct comparison is complicated by the fact that very different parameter sets enter the two. For specificity we set $\Omega_P/2\pi = 2.7$ MHz, and begin by setting the factor ${\cal R}_\mathrm{EIT} |{\bf d}_{45}|/10^3 |{\bf d}_{12}| = 1$.

\subsubsection{1D setup}
\label{subsec:1Dsetupsnr}

Considering first the 1D setup shown in Fig.\ \ref{fig:doppler1D}, the physical value $k_C = -k_C^\mathrm{phys}$ produces $S_\mathrm{EIT}^\mathrm{th} \approx 3 \times 10^{-4}$ (Mrad/s)$^{-1}$ and hence E-field sensitivity
\begin{equation}
{\cal E}^\mathrm{1D}_\mathrm{in} \equiv E_\mathrm{in} \sqrt{\tau}
\approx 1.4 \frac{\mu\mathrm{V}}{\mathrm{m \sqrt{Hz}}}.
\label{6.10}
\end{equation}
As discussed in Sec.\ \ref{sec:qvsclasssense}, Refs.\ \cite{Shanxi2020,foot:cqcom} quote values 2--4 times larger than this, achieved using much smaller $A_D \sim 1$ mm$^2$ (hence reducing ${\cal R}_\mathrm{EIT}$ by an order of magnitude) but then partially compensated via a combination of longer cell lengths, larger $n > 40$, and larger Probe intensity. It follows that the most direct way to further increase sensitivity for the 1D setup, improving on (\ref{6.10}) by as much as an order of magnitude, is to increase the beam area at fixed $\Omega_P$, i.e., by simultaneously broadening the beam and increasing laser power. Equivalent multi-beam assemblies are indeed under consideration \cite{foot:cqcom}.

\subsubsection{2D star configuration}
\label{subsec:2Dstarsnr}

The star configuration shown in Fig.\ \ref{fig:starconfig} offers the opportunity for significant, order of magnitude or more, Doppler compensation gains. This in combination with large $\Omega_D$ and small $\Omega_\mathrm{LO}$ (Figs.\ \ref{fig:doppler2D}, \ref{fig:SensOmegaLO}) is predicted to enable $S_\mathrm{EIT}^\mathrm{th} \approx 0.02$ (Mrad/s)$^{-1}$ (obtained from the $w = 1$ cm panel of Fig.\ \ref{fig:SensOmegaLO}), a factor $O(10^2)$ larger than the 1D result \cite{foot:other1Denhance}. All other things being equal, this improves (\ref{6.10}) to
\begin{equation}
{\cal E}^\mathrm{2D}_\mathrm{in}
\approx 20 \mathrm{\frac{nV}{m \sqrt{Hz}}},
\label{6.11}
\end{equation}
a factor of 50 smaller than the classical estimate (\ref{1.6}) if one limits consideration to small $L_\mathrm{eff} \approx 1$ cm antennas. On the other hand, $L_\mathrm{eff} \approx 50$ cm achieves the same classical E-field sensitivity as (\ref{6.11}). This improvement with wire antenna effective size is essentially equivalent to the $1/\sqrt{A_D}$ E-field sensitivity scaling implicit in (\ref{6.7}), but with a vastly lower technological barrier. Thus, increasing $A_D$ for the star configuration requires a proportionate increase in the Dress and Coupling beam areas in order to cover the required increased intersecting volume. Increasing the vapor cell length $L$, which is quite natural for the 1D setup, is even further limited for the 2D configuration.

\subsubsection{Low frequency limit}
\label{subsec:lowfreq}

One may conclude from the above discussion that the most natural application for Rydberg antennas is to low frequencies in cases where the sensor volume is highly constrained. For example, using $L_R = 1$ cm, $\lambda = 10$ m (HF band) leads, via (\ref{1.7}), to $L_\mathrm{eff} = 10^{-3}$ cm and ${\cal E}_R \approx 1$ mV/m/$\sqrt{\mathrm{Hz}}$, nearly three orders of magnitude larger even than the 1D estimate (\ref{6.10}).

One might naively predict even greater advantage in the VLF regime where $\lambda > 10$ km. However, a significant complication here is that for convenient values of $n$, the frequency difference between neighboring Rydberg levels, scaling as $1/n^3$, is in the $\sim$\,10 GHz range. An applied DC electric field ${\bf E}_0$ might be used, via the Stark effect, to tune a pair of levels much closer together. This would be in addition to the local oscillator signal, so should not require significant changes to the vapor cell design \cite{foot:stark}. Of course, the Rydberg level pairs selected in this way may have sub-optimal values of the transition dipole moment $|{\bf d}_{45}|$, and  there are line broadening mechanisms acting in same frequency range (Table \ref{tab:nonintrinsdecay}) that could severely degrade performance. There could also be other (e.g., level repulsion) issues when attempting to design such near level degeneracies, and with vapor cell operation at low frequencies. This remains very interesting work for the future.

\section{Conclusions}
\label{sec:conclude}

We have explored here several avenues for increasing Rydberg antenna sensitivity using 2D ``Doppler aware'' and ``Rydberg cavity exploitation'' setups, and quantified both the adiabatic and dynamic response of these setups. The highlight, perhaps, is demonstration of an $O(10^2)$ boost from $S_\mathrm{EIT}^\mathrm{th} \sim 10^{-4}$ for the 1D setup (Fig.\ \ref{fig:doppler1D}) to $\sim$\,$10^{-2}$ (upper curve of the $w = 1$ cm panel of Fig.\ \ref{fig:SensOmegaLO}), corresponding to optimally small $\Omega_\mathrm{LO}$ and large $\Omega_D$). The tradeoff between adiabatic and broader band sensitivity is highlighted as well by the results in Figs.\ \ref{fig:lrfreq2Dstar} and \ref{fig:lrfreq2DstarSmallLO}.

There are a number of extensions of the theory, relevant to applications, that deserve further investigation, including nonlinear response to larger amplitude signals. With regard to the low frequency (e.g, HF band) limit, the experimental parameters entering the five-state projected Hamiltonian (\ref{3.5}) required to actually obtain the required very near-degenerate Rydberg levels deserve careful investigation. Thus, the required DC Stark shift tuning may become highly nontrivial when the desired level spacing is much smaller than the original spacing.

Another exciting possibility is sensor enhancement in the vicinity of the bistable phase transition \cite{CRWAW2013,MLDGL2014,Ding2022,Wang2023}. At somewhat higher vapor densities (hence moderately warmer cells $\sim$\,$40^\circ$C) the enhanced dipole--dipole interactions between Rydberg atoms leads to a nonequilibrium phase transition that generates an enhanced many body response to the coupling laser detuning $\Delta_C$. This has already been shown to lead to enhanced electric field sensing in the single Rydberg state setup \cite{Ding2022,Wang2023}. However, the maximum demsonstrated sensitivity still lies well short that already achieved using conventional forms of the two Rydberg state setup \cite{Shanxi2020,foot:cqcom}, and it would be extremely interesting to see if the bistable phase can exploited the in the latter context. Especially intriguing is the possibility of multi-stable phases generated by different many body interactions within and between Rydberg levels, and with local oscillator perhaps capable of adjusting one's position in the phase diagram and enabling control of their relative populations.

\acknowledgments

This material is based upon work supported by the Defense Advanced Research Projects Agency (DARPA) under Contract No.\ HR001121C0122. The author also benefited from numerous conversations with Craig Price, Zak Burkley, Ying Ju Wang, Eric Bottomley, Haoquan Fan, and Shane Verploegh regarding experimental considerations.

\appendix

\section{Dynamic linear response formalism}
\label{app:dynresp}

We develop here the full dynamic (finite frequency) linear response formalism, generalizing the adiabatic limit described in Sec.\ \ref{sec:adialim} and used to generate most of the results in this paper. We now consider solutions to the density matrix equation (\ref{3.1}) when a small time-dependent perturbation is included in the Hamiltonian:
\begin{equation}
\hat H(t) = \hat H_0 + \hat V(t).
\label{A1}
\end{equation}
For example, the Rydberg level Stark form (\ref{2.4}) yields the rather sparse matrix
\begin{eqnarray}
\hat V(t) &=& -\frac{1}{2}
[\Omega_\mathrm{in}(t) \hat V_0 + \Omega_\mathrm{in}(t)^* \hat V_0^\dagger]
\nonumber \\
V_{0,mn} &\equiv& \delta_{m4} \delta_{n5}.
\label{A2}
\end{eqnarray}
Writing
\begin{equation}
\hat \rho(t) = \hat \rho^\mathrm{ad} + \delta \hat \rho(t),
\label{A3}
\end{equation}
one obtains
\begin{equation}
\partial_t \delta \hat \rho = i[\delta \hat \rho, \hat H_0]
+ i[\hat \rho^\mathrm{ad}, \hat V]  + \hat D[\delta \hat \rho]
+ i[\delta \hat \rho, \hat V]
\label{A4}
\end{equation}
in which the adiabatic equation
\begin{equation}
i[\hat \rho^\mathrm{ad}, \hat H] + \hat D[\hat \rho^\mathrm{ad}] = 0,\ \
\mathrm{tr}[\hat \rho^\mathrm{ad}] = 1
\label{A5}
\end{equation}
has been used to eliminate the zeroth order terms. The linear response solution is obtained by dropping the last term and may be solved via Fourier transform. Thus, writing
\begin{eqnarray}
\delta \hat \rho(t) &=& \int \frac{d\omega}{2\pi} \delta \tilde \rho(\omega) e^{-i\omega t}
\nonumber \\
\hat V(t) &=& \int \frac{d\omega}{2\pi} \tilde V(\omega) e^{-i\omega t}
\label{A6}
\end{eqnarray}
one obtains the linearized equation
\begin{equation}
i[\delta \tilde \rho, \hat H_0]
+ (i\omega \openone +  \hat D)[\delta \tilde \rho]
= -i[\hat \rho^\mathrm{ad}, \tilde V].
\label{A7}
\end{equation}
This equation is similar to an inhomogeneous version of (\ref{A5}), except with an additional identity matrix contribution to $\hat D$. The self-adjoint properties of $\hat \rho$ and $\hat V$ lead to
\begin{equation}
\delta \tilde \rho(\omega)^\dagger = \delta \tilde \rho(-\omega),\ \
\tilde V(\omega)^\dagger = \tilde V(-\omega)
\label{A8}
\end{equation}

\subsection{Vector formulation}
\label{app:vectorform}

Using the vector formulation (\ref{3.8}), the linear response equation (\ref{A7}) takes the form
\begin{equation}
(i\omega {\bf I} + {\bf G}) \delta \tilde {\bm \rho}(\omega)
= -i {\bf V}(\omega) {\bm \rho}^\mathrm{ad}
\label{A9}
\end{equation}
in which, similar to  (\ref{3.9}),
\begin{equation}
V_{mn,pq} =  \tilde V_{qn} \delta_{pm} - \tilde V_{mp} \delta_{qn}.
\label{A10}
\end{equation}
If (\ref{A2}) is used, this again becomes a rather sparse matrix. The formal solution is
\begin{equation}
\delta \tilde {\bm \rho}(\omega) = -i(i\omega {\bf I} + {\bf G})^{-1}
{\bf V}(\omega) {\bm \rho}^\mathrm{ad}.
\label{A11}
\end{equation}
in which, unlike for ${\bf G}$ itself with its nonempty kernel (\ref{3.14}), invertibility of the matrix is guaranteed by the presence of the identity matrix. Using the left and right eigenvector decomposition (\ref{3.16}) of ${\bf G}$ one obtains the representation
\begin{eqnarray}
{\bf G} = \sum_n \lambda_n {\bf u}_n^R {\bf u}_n^{L \dagger},\ \
{\bf I} = \sum_n {\bf u}_n^R {\bf u}_n^{L \dagger},
\label{A12}
\end{eqnarray}
in which the second line is equivalent to the normalization condition in (\ref{3.16}). Note that, via the conservation law (\ref{3.14}), one of the $\lambda_n = 0$ terms corresponds to the ``probability kernel'' ${\bm \rho}^\mathrm{ad} {\bf t}^\dagger$ (and, except under certain degeneracy conditions, will be the only $\lambda_n = 0$ term).

One obtains therefore the representation
\begin{equation}
(i\omega {\bf I} + {\bf G})^{-1}
= \sum_n \frac{{\bf u}_n^R {\bf u}_n^{L \dagger}}{i\omega + \lambda_n}
\label{A13}
\end{equation}
in which the denominator is finite even in the kernel subspace $\lambda_n = 0$. Since ${\bf t}$ is a zero eigenvalue left eigenvector, one obtains
\begin{equation}
{\bf t}^\dagger \delta \tilde {\bm \rho}
= -\frac{1}{\omega} {\bf t}^\dagger {\bf V} {\bm \rho}^\mathrm{ad} = 0
\label{A14}
\end{equation}
in which the vanishing of ${\bf t}^\dagger {\bf V}$ follows from (\ref{A9}) via
\begin{equation}
[{\bf t}^\dagger {\bf V}]_{pq}
= \sum_n V_{nn,pq} = \tilde V_{qp} - \tilde V_{qp} = 0
\label{A15}
\end{equation}
and is equivalent to the vanishing of the trace of the right hand side of (\ref{A5}). This confirms the general conservation of probability condition $\mathrm{tr}[\delta \tilde {\bm \rho}] = 0$, and (\ref{A11}) may therefore be written more explicitly in terms of the eigen-decomposition of ${\bf G}$ in the form
\begin{eqnarray}
\delta \tilde {\bm \rho}(\omega) &=& -i {\bf R}(\omega) {\bf V}(\omega){\bm \rho}^\mathrm{ad}
\nonumber \\
{\bf R}(\omega) &\equiv&
{\sum_n}' \frac{{\bf u}_n^R {\bf u}_n^{L \dagger}}{i\omega + \lambda_n}
\label{A16}
\end{eqnarray}
in which the prime indicates dropping the probability kernel term for which ${\bf u}_n^L = {\bf t}$.

\subsection{RF signal application}
\label{app:rfsigapp}

Using the form (\ref{A2}) one obtains the Fourier transform
\begin{equation}
\tilde V(\omega) = -\frac{1}{2}
[\tilde \Omega_\mathrm{in}(\omega) \hat V_0
+ \tilde \Omega_\mathrm{in}(-\omega)^* \hat V_0^\dagger].
\label{A17}
\end{equation}
In some cases $\tilde \Omega_\mathrm{in}(\omega)$ might be supported on positive frequencies so that $\tilde \Omega_\mathrm{in}(-\omega)^*$ is complementarily supported on negative frequencies. The EIT response obtained from (\ref{A11}) (specifically from $\mathrm{Im}[\delta \rho_{21}]$) is therefore a linear combination of terms containing
\begin{eqnarray}
\Omega_\mathrm{in}^{(n)}(t) &\equiv& \int \frac{d\omega}{2\pi}
\frac{\tilde \Omega_\mathrm{in}(\omega)}{i\omega + \lambda_n} e^{-i\omega t}
\nonumber \\
&=& \int_{-\infty}^t dt' e^{\lambda_n(t-t')} \Omega_\mathrm{in}(t')
\label{A18}
\end{eqnarray}
and its complex conjugate. Derivation of the last line reflects causality, and requires $\mathrm{Re}[\lambda_n] < 0$, a consequence of the decay processes encoded in ${\bf D}$. The adiabatic regime corresponds to the support of $\tilde \Omega_\mathrm{in}(\omega)$ dominated by $|\omega| \ll \mathrm{min}\{|\lambda_n|\}$. Under this condition, the solution is
\begin{eqnarray}
{\bm \rho}(t) &\simeq& {\bm \rho}^\mathrm{ad}[\Omega_\mathrm{LO} + \Omega_\mathrm{in}(t)]
\label{A19} \\
&\simeq& {\bm \rho}^\mathrm{ad}[\Omega_\mathrm{LO}]
+ {\bm \sigma}_+[\Omega_\mathrm{LO}] \Omega_\mathrm{in}(t)
+ {\bm \sigma}_-[\Omega_\mathrm{LO}] \Omega_\mathrm{in}(t)^*
\nonumber
\end{eqnarray}
in which
\begin{eqnarray}
{\bm \sigma}_+[\Omega_\mathrm{LO}]
&=& \frac{\partial {\bm \rho}^\mathrm{ad}[\Omega_\mathrm{LO}]}{\partial \Omega_\mathrm{LO}}
= \frac{i}{2} {\bf R}(0) {\bf V}_{0+} {\bm \rho}^\mathrm{ad}
\nonumber \\
{\bm \sigma}_-[\Omega_\mathrm{LO}]
&=& \frac{\partial {\bm \rho}^\mathrm{ad}[\Omega_\mathrm{LO}]}{\partial \Omega_\mathrm{LO}^*}
= \frac{i}{2} {\bf R}(0) \bar {\bf V}_{0-} {\bm \rho}^\mathrm{ad}
\label{A20}
\end{eqnarray}
Here ${\bf V}_{0\pm}$ are the matrices corresponding, respectively, to $\hat V_0$, $\hat V_0^\dagger$ in (\ref{A2}) and defined by (\ref{A10}) (note the corresponding additional factor $-\frac{1}{2}$ now appearing). The $\Omega_\mathrm{LO}$ derivatives are performed here by formally treating $\Omega_\mathrm{LO}$ and $\Omega_\mathrm{LO}^*$ (or, equivalently, the real and imaginary parts of $\Omega_\mathrm{LO}$) as separate variables through their appearance, via (\ref{2.4}), in the Hamiltonian (\ref{3.5}). This is precisely the expression previously derived for the EIT sensitivity ${\bf S}$ in the adiabatic approximation.

Another important limit is for narrow-banded signals centered on a finite frequency shift $\Delta \omega_\mathrm{in}$ [see (\ref{2.4})]. The adiabatic approximation is no longer valid, but form of the second line of (\ref{A20}) remains valid if one replaces the sensitivity expressions by
\begin{equation}
{\bm \sigma}_\pm(\Omega_\mathrm{LO},\Delta \omega_\mathrm{in})
= \frac{i}{2} {\bf R}(\Delta \omega_\mathrm{in}) {\bf V}_{0\pm} {\bm \rho}^\mathrm{ad}
\label{A21}
\end{equation}
This expression is valid not only if the signal bandwidth is much smaller than $\mathrm{Re}[\lambda_n] < 0$, but also if $|\Delta \omega_\mathrm{in}|$ is sufficiently large that the bandwidth is much smaller than $\mathrm{min}\{|-i\Delta \omega_\mathrm{in} + \lambda_n|\}$. Outside of these cases, the filtering operation (\ref{A19}) is nontrivial, at least for some values of $n$, and the signal will be significantly distorted by the exponential memory kernel.

\subsection{Thermal averaging}
\label{app:thermave}

At finite temperature motion-induced Doppler shifts (\ref{3.7}) are accounted for simply by applying the thermal average (\ref{3.17}) to the time-dependent density matrix:
\begin{equation}
\hat \rho_\mathrm{th}(t) = \langle \hat \rho(t;{\bf v}) \rangle
\equiv \int d{\bf v} P({\bf v}) \hat \rho(t;{\bf v}).
\label{A22}
\end{equation}
The incident signal response is therefore thermally averaged in the same way, leading to
\begin{eqnarray}
{\bm \rho}^\mathrm{ad}_\mathrm{th} &=& \langle {\bm \rho}^\mathrm{ad}({\bf v}) \rangle
\nonumber \\
\delta \tilde {\bm \rho}_\mathrm{th}(\omega)
&=& -i \langle {\bf R}(\omega;{\bf v}) {\bf V}(\omega) {\bm \rho}^\mathrm{ad}({\bf v}) \rangle
\label{A23}
\end{eqnarray}
in which the eigenvalues and eigenvectors of ${\bf G}({\bf v})$, arising from (\ref{A17}), are now velocity dependent through ${\bf H}({\bf v})$.

Note that since ${\bf v}$ is processed through the highly nonlinear $N^2 \times N^2$ matrix diagonalization procedure, these averages are highly nontrivial. It does not appear to be possible to avoid separately computing $\hat \rho(\omega,{\bf v})$ for each substantially thermally excited ${\bf v}$, and then evaluating the velocity integral (\ref{A22}). The linear response formalism does, however, allow one to avoid a full dynamical solution to the equation of motion (as might be needed, e.g., to obtain the nonlinear response to strong signals).

Although the matrix diagonalization does not depend on the frequency $\omega$ the thermal average does in general need to be performed separately for each. In the narrow band limit, where (\ref{A20}) and/or (\ref{A21}) are valid, one obtains the significant simplification that one need only perform a single average at $\omega = \Delta \omega_\mathrm{in}$ [which vanishes for (\ref{A20})].

The sensitivities (\ref{A22}) are now thermally averaged in the identical fashion. Using (\ref{A18}) we define the thermally averaged linear response sensitivity
\begin{equation}
{\bm \sigma}_\pm^\mathrm{th}(\omega) = \frac{i}{2} \langle
{\bf R}(\omega;{\bf v}) {\bf V}_{0\pm} {\bm \rho}^\mathrm{ad}({\bf v}) \rangle
\label{A24}
\end{equation}
in terms of which
\begin{equation}
\delta \tilde {\bm \rho}_\mathrm{th}(\omega)
= {\bm \sigma}_+^\mathrm{th}(\omega) \tilde \Omega_\mathrm{in}(\omega)
+ \bar {\bm \sigma}_-^\mathrm{th}(\omega) \tilde \Omega_\mathrm{in}(-\omega)^*.
\label{A25}
\end{equation}
Implicit dependence on $\Omega_\mathrm{LO}$, and all other setup parameters, have been dropped from the notation for clarity. The time-dependent signal is obtained from the inverse Fourier transform of this result.

Simplifying to the homogeneous limit (\ref{3.26}), the time-domain EIT response (\ref{3.21}) is obtained in the form
\begin{eqnarray}
P_\mathrm{EIT}(t) &=& e^{-\alpha L \mathrm{Im}[\rho_{21}^\mathrm{th}(t)]}
\nonumber \\
&=& P_\mathrm{EIT}^\mathrm{th}
\left\{1 - \alpha \mathrm{Im}[\delta \rho_{\mathrm{th},21}(t)]
+ O(\delta\rho^2) \right\} \ \ \ \ \ \
\label{A26}
\end{eqnarray}
in which
\begin{equation}
P_\mathrm{EIT}^\mathrm{th} = e^{-\alpha L \mathrm{Im}[\rho_{\mathrm{th},21}^\mathrm{ad}]}
\label{A27}
\end{equation}
is the result in the absence of a signal. Defining the frequency domain EIT sensitivities
\begin{equation}
S^\mathrm{th,\pm}_\mathrm{EIT}(\omega)
= -\alpha L P_\mathrm{EIT}^\mathrm{th} \sigma_{\pm,21}^\mathrm{th}(\omega)
\label{A28}
\end{equation}
one obtains the linear response form
\begin{eqnarray}
\delta P_\mathrm{EIT}(t) &\equiv& P_\mathrm{EIT}(t) - P_\mathrm{EIT}^\mathrm{th}
\nonumber \\
&=& \mathrm{Im} \int \frac{d\omega}{2\pi} \delta \tilde P_\mathrm{EIT}(\omega) e^{-i\omega t}
\label{A29}
\end{eqnarray}
with Fourier kernel
\begin{equation}
\delta \tilde P_\mathrm{EIT}(\omega)
= S_\mathrm{EIT}^\mathrm{th,+}(\omega) \tilde \Omega_\mathrm{in}(\omega)
+ S_\mathrm{EIT}^\mathrm{th,-}(\omega) \tilde \Omega_\mathrm{in}(-\omega)^*.
\label{A30}
\end{equation}


\begin{thebibliography}{}

\bibitem{HFI1990} S. E. Harris, J. E. Field, and A. Imamo$\breve{\mathrm{g}}$lu, ``Nonlinear optical processes using electromagnetically induced transparency,'' \href{https://doi.org/10.1103/PhysRevLett.64.1107}{Phys.\ Rev.\ Lett.\ \textbf{64}, 1107 (1990)}.

\bibitem{FIM2005} F. M. Fleischhauer, A. Imamo$\breve{\mathrm{g}}$lu, and J. P. Marangos, ``Electromagnetically induced transparency: Optics in coherent media,'' \href{https://doi.org/10.1103/RevModPhys.77.633}{Rev.\ Mod.\ Phys.\ \textbf{77}, 634 (2005)}.

\bibitem{RBTEY2011} I. I. Ryabtsev, I. I. Beterov, D. B. Tretyakov, V. M. Entin, and E. A. Yakshina, ``Doppler- and recoil-free laser excitation of Rydberg states via three-photon transitions,'' \href{https://doi.org/10.1103/PhysRevA.84.053409}{Phys.\ Rev.\ A \textbf{84}, 053409 (2011)}.

\bibitem{CTSSAW2012} C. Carr, M. Tanasittikosol, A. Sargsyan, D. Sarkisyan, C. S. Adams, and K. J. Weatherill, ``Three-photon electromagnetically induced transparency using Rydberg states,'' \href{http://dro.dur.ac.uk/10423/}{Opt.\ Lett. \textbf{37} 3858--3860 (2012)}.

\bibitem{Sedlacek2012} J. A. Sedlacek, A. Schwettmann, H. K\"ubler, R. L\"ow, T. Pfau, and J. P. Shaffer, ``Microwave electrometry with Rydberg atoms in a vapour cell using bright atomic resonances,'' \href{https://doi.org/10.1038/nphys2423}{Nature Phys.\ \textbf{8}, 819--824 (2012)}.

\bibitem{NIST2014} C. L. Holloway, J. A. Gordon, S. Jefferts, A. Schwarzkopf, D. A. Anderson, S. A. Miller, N. Thaicharoen, and G. Raithel, ``Broadband Rydberg atom-based electric field probe for SI-traceable, self-calibrated measurements,'' \href{https://doi.org/10.1109/TAP.2014.2360208}{IEEE Trans.\ Antennas Propag.\ \textbf{62}, 6169--6182 (2014)}.

\bibitem{SK2018} J. P. Shaffer and H. K{\"u}bler, ``A read-out enhancement for microwave electric field sensing with Rydberg atoms,'' \href{https://doi.org/10.1117/12.2309386}{Proc.\ SPIE 10674, Quantum Technologies 106740C (2018)}.

\bibitem{Michigan2019} N. Thaicharoen, K. R. Moore, D. A. Anderson, R. C. Powel, E. Peterson, and G. Raithel, ``Electromagnetically induced transparency, absorption, and microwave-field sensing in a Rb vapor cell with a three-color all-infrared laser system,'' \href{https://doi.org/10.1103/PhysRevA.100.063427}{Phys.\ Rev.\ A \textbf{100}, 063427 (2019)}.

\bibitem{Waterloo2021} J. P. Shaffer, F. Ripka, C. Liu, H. K\"ubler, J. Erskine, G. Gillett, H. Amarloo, J. Ramirez-Serrano, and J. Keaveney, ``Rydberg Atom-Based Radio Frequency Electrometry: Enhancement of the Self-Calibrated Autler-Townes Sensing Mode,'' \href{https://doi.org/10.1109/CAMA49227.2021.9703610}{2021 IEEE Conference on Antenna Measurements \& Applications 226--229 (2021)}.

\bibitem{MITRE2021} C. T. Fancher, D. R. Scherer, M. C. S. John and B. L. S. Marlow, ``Rydberg Atom Electric Field Sensors for Communications and Sensing,'' \href{https://doi.org/10.1109/TQE.2021.3065227}{IEEE Trans.\ Q. Eng. \textbf{2}, 3501313 (2021)}.

\bibitem{NIST2019a} M. T. Simons,  A. H. Haddab, J. A. Gordon, and  C. L. Holloway, ``A Rydberg atom-based mixer: Measuring the phase of a radio frequency wave,'' \href{https://doi.org/10.1063/1.5088821}{Appl.\ Phys.\ Lett.\ \textbf{114}, 114101 (2019)}.

\bibitem{NIST2019b} J. A. Gordon,  M. T. Simons,  A. H. Haddab, and  C. L. Holloway, ``Weak electric-field detection with sub-1 Hz resolution at radio frequencies using a Rydberg atom-based mixer,'' \href{https://doi.org/10.1063/1.5095633}{AIP Advances \textbf{9}, 045030 (2019)}.

\bibitem{Durham2017} C. G. Wade, N. $\breve{\mathrm{S}}$ibali\'{c}, N. R. de Melo, J. M. Kondo, C. S. Adams, and K. J. Weatherill ``Real-Time Near-Field Terahertz Imaging with Atomic Optical Fluorescence,'' \href{https://doi.org/10.1038/nphoton.2016.214}{Nature Photon \textbf{11}, 40 (2017)}.

\bibitem{JC2020} Y.-Y Jau and T. Carter ``Vapor-cell-based atomic electrometry for detection frequencies below 1 kHz,'' \href{https://doi.org/10.1103/PhysRevApplied.13.054034}{Phys.\ Rev.\ Applied \textbf{13}, 054034 (2020)}.

\bibitem{ARL2020} D. H. Meyer, Z. A. Castillo, K. C. Cox, and P. D. Kunz, ``Assessment of Rydberg atoms for wideband electric field sensing'' \href{https://doi.org/10.1088/1361-6455/ab6051}{J. Phys.\ B: At.\ Mol.\ Opt.\ Phys.\ \textbf{53}, 034001 (2020)}.

\bibitem{BVBZ2022} G. S. Botello, S. Verploegh, E. Bottomley, and Z. Popovi\'c, ``Comparison of Noise Temperature of Rydberg-Atom and Electronic Microwave Receivers,'' \href{https://doi.org/10.48550/arXiv.2209.00908}{arXiv:2209.00908 [quant-ph]}.

\bibitem{SKAW2016} N. $\breve{\mathrm{S}}$ibali\'c, J. M. Kondo, C. S. Adams, and K. J. Weatherill, ``Dressed-state electromagnetically induced transparency for light storage in uniform-phase spin waves,'' \href{https://doi.org/10.1103/PhysRevA.94.033840}{Phys.\ Rev.\ A \textbf{94}, 033840 (2016)}.

\bibitem{Balanis2016} C. A. Balanis, \href{https://www.worldcat.org/title/933291646}{\emph{Antenna theory: analysis and design (4th edition)}}  (John Wiley, Hoboken, New Jersey, 2016).

\bibitem{Shanxi2020} M. Jing, Y. Hu, J. Ma, H. Zhang, L. Zhang, L. Xiao and S. Jia, ``Atomic superheterodyne receiver based on microwave-dressed Rydberg spectroscopy,'' \href{https://doi.org/10.1038/s41567-020-0918-5}{Nature Physics \textbf{16}, 911--915 (2020)}.

\bibitem{foot:cqcom} Eric Bottomley, Haoquan Fan, private communications.

\bibitem{ARL2021} D. H. Meyer, C. O'Brien, D. P. Fahey, K. C. Cox, and P. D. Kunz ``Optimal atomic quantum sensing using EIT readout'' \href{https://doi.org/10.1103/PhysRevA.104.043103}{Phys.\ Rev.\ A \textbf{104}, 043103 (2021).}

\bibitem{foot:3laser} As discussed in Ref.\ \cite{Michigan2019} the 3-laser setup offers an added practical advantage of being based on optical frequencies (e.g., near-infrared) that can be produced by much less expensive hardware.

\bibitem{NIST2019c} C. L. Holloway, M. T. Simons, J. A. Gordon, and D. Novotny, ``Detecting and receiving phase-modulated signals with a Rydberg atom-based receiver,'' \href{https:doi.org/10.1109/LAWP.2019.2931450}{IEEE Anten.\ Wireless Prop.\ Lett.\ \textbf{18}, 1853 (2019)}.

\bibitem{NIST2022} N. Prajapati, A. Rotunno, S. Berweger, M. Simons, A. Artusio-glimpse, and C. L. Holloway, ``TV and video game streaming with a quantum receiver: a study on a Rydberg atom-based receivers bandwidth and reception clarity,'' \href{https://doi.org/10.1116/5.0098057}{AVS Quantum Sci.\ \textbf{4}, 035001 (2022)}.

\bibitem{BBNTS2022} S. M. Bohaichuk, D. Booth, K. Nickerson, H. Tai, and J. P. Shaffer ``Origins of Rydberg-atom electrometer transient response and its impact on radio-frequency pulse sensing,'' \href{https://doi.org/10.1103/PhysRevApplied.18.034030}{Phys.\ Rev.\ Applied \textbf{18}, 034030 (2022)}.

\bibitem{foot:polarize} The density matrix enters the microscopic derivation of the classical polarization density ${\bf P}({\bf x},t) = n({\bf x}) \langle \hat {\bf p}_\mathrm{at} \rangle_{\bf x}$ appearing in the macroscopic Maxwell equation. Here $n({\bf x})$ is the local atom number density and $\langle \hat {\bf p}_\mathrm{at} \rangle_{\bf x} = \mathrm{tr}[\hat \rho \hat {\bf p}_\mathrm{at}]$ is the mean atomic dipole moment of an atom centered at point ${\bf x}$ \cite{FIM2005}. The latter is in turn determined by the local electric field through the Stark coupling (\ref{2.2}), and becomes nonlinear in the vicinity of atomic resonances. Thus, in the present application ${\bf P}$ is determined by the laser illumination fields ${\bf E}_\alpha$, self-consistently varying along (and in general across) the beam paths, leading ultimately to the, in general nonlinear, solutions (\ref{3.21})--(\ref{3.23}) under the assumption only of a dilute vapor.

\bibitem{foot:other1Denhance} It is possible that the 1D setup could benefit from optimizing $\Omega_D, \Omega_\mathrm{LO}$ as well, but this remains to be studied.

\bibitem{foot:stark} The Stark effect has already been proposed by a number of groups to fine tune the Rydberg state separations to better cover the full range of UHF and higher band incident signal frequencies \cite{foot:cqcom}.

\bibitem{CRWAW2013} C. Carr, R. Ritter, C. G. Wade, C. S. Adams, and K. J. Weatherill, ``Non-equilibrium phase transition in a dilute Rydberg ensemble,'' \href{https://doi.org/10.1103/PhysRevLett.111.113901}{Phys.\ Rev.\ Lett.\ \textbf{111}, 113901 (2013)}.

\bibitem{MLDGL2014} M. Marcuzzi, E. Levi, S. Diehl, J. P. Garrahan, and I. Lesanovsky, ``Universal non-equilibrium properties of dissipative Rydberg gases,'' \href{https://doi.org/10.1103/PhysRevLett.113.210401}{Phys.\ Rev.\ Lett.\ \textbf{113}, 210401 (2014)}.

\bibitem{Ding2022} D.-S. Ding, Z.-K. Liu, B.-S. Shi, G.-C. Guo, K. M{\o}lmer, and C. S. Adams, ``Enhanced metrology at the critical point of a many-body Rydberg atomic system,'' \href{https://doi.org/10.1038/s41567-022-01777-8}{Nat.\ Phys.\ \textbf{18}, 1447 (2022)}.

\bibitem{Wang2023} Q. Wang, Z. Wang, Y. Liu, S. Guan, J. He, C.-L. Zou, P. Zhang, G. Li, and T. Zhang, ``Cavity-enhanced optical bistability of Rydberg atoms,'' \href{https://doi.org/10.1364/OL.486914}{Opt.\ Lett.\ \textbf{48}, 2865 (2023)}.

\end{thebibliography}
\end{document}